\documentclass[10pt, a4paper]{article}

\usepackage[a4paper, width=150mm,top=25mm,bottom=25mm]{geometry} % Geometry

\usepackage{fancyhdr}
\usepackage[english]{babel}
\usepackage[utf8]{inputenc}
\usepackage[T1]{fontenc}                        % 8-bit glyphs
\usepackage{authblk}                            % Adding authors and affiliations
\usepackage{bbm}                                % Mathbb for numbers
\usepackage[title, titletoc]{appendix}
\usepackage[normalem]{ulem}

\usepackage{amsmath}
\usepackage{amsthm}
\usepackage{amsbsy}
\usepackage{amssymb}
\usepackage{bm}                                 % Bold math symbols
\usepackage{mathtools}                          % Some additional command like dcases etc.
\usepackage{physics}
\allowdisplaybreaks                             % Allow pagebreaks in math environments
\numberwithin{equation}{section}                % Number equations in the sections

\usepackage{graphicx}
\usepackage[dvipsnames]{xcolor}
\usepackage[section]{placeins}

\usepackage{cite}
\usepackage{hyperref}
\hypersetup{hidelinks, colorlinks=true, allcolors=[RGB]{0,84,159}}

\usepackage[capitalise]{cleveref}             
\crefname{section}{Sect.}{Sects.}
\crefname{fig_a}{Fig.}{Fig.}                    % Reference figures with (a) etc.
\crefname{fig_b}{Fig.}{Fig.}                    % Reference figures with (b) etc.
\Crefname{fig_a}{Figure}{Figure}                % At begining of sentence
\Crefname{fig_b}{Figure}{Figure} 
\creflabelformat{fig_a}{#2{#1(a)}#3}
\creflabelformat{fig_b}{#2{#1(b)}#3}
\usepackage[perpage]{footmisc}
\usepackage{doi}
\newcommand*\subtxt[1]{_{\textnormal{#1}}}
\DeclareRobustCommand\_{\ifmmode\expandafter\subtxt\else\textunderscore\fi}

\newcommand*\supertxt[1]{^{\textnormal{#1}}}
\DeclareRobustCommand\^{\ifmmode\expandafter\supertxt\else\textasciicircum\fi}

\newtheorem*{theorem}{Theorem}
\newcommand{\LG}[1]{\textcolor{black}{{#1}}} % Lukas
\title{$\mathcal{PT}$-symmetric, non-Hermitian quantum many-body physics---a methodological perspective}

\author[1]{V.~Meden$^\dagger$}
\author[1,2]{L.~Grunwald}
\author[1,2]{D.~M.~Kennes}

\affil[1]{Institut f{\"u}r Theorie der Statistischen Physik, RWTH Aachen University, 52056 Aachen, Germany}
\affil[2]{Max Planck Institute for the Structure and Dynamics of Matter, Center for Free Electron Laser Science, 22761 Hamburg, Germany}
\affil[$^\dagger$]{\textit meden@physik.rwth-aachen.de}

\begin{document}

\maketitle

\begin{abstract}
We review the methodology to theoretically treat parity-time- ($\mathcal{PT}$-) symmetric, non-Hermitian quantum many-body systems. They are realized as open quantum systems with $\mathcal{PT}$ symmetry and couplings to the environment which are compatible. $\mathcal{PT}$-symmetric non-Hermitian quantum systems show a variety of fascinating properties which single them out among generic open systems. The study of the latter has a long history in quantum theory. These studies are based on the Hermiticity of the combined system-reservoir setup and were developed by the atomic, molecular, and optical physics as well as the condensed matter physics communities. The interest of the mathematical physics community in $\mathcal{PT}$-symmetric, non-Hermitian systems led to a new perspective and the development of the elegant mathematical formalisms of $\mathcal{PT}$-symmetric and biorthogonal quantum mechanics, which do not make any reference to the environment. In the mathematical physics research, the focus is mainly on the remarkable spectral properties of the Hamiltonians and the characteristics of the corresponding single-particle eigenstates. Despite being non-Hermitian, the Hamiltonians can show parameter regimes, in which all eigenvalues are real. To investigate emergent quantum many-body phenomena in condensed matter physics and to make contact to experiments one, however, needs to study expectation values of observables and correlation functions. One furthermore, has to investigate statistical ensembles and not only eigenstates. The adoption of the concepts of  $\mathcal{PT}$-symmetric and biorthogonal quantum mechanics by parts of the condensed matter community led to a controversial status of the methodology. There is no consensus on fundamental issues, such as, what a proper observable is, how expectation values are supposed to be computed, and what adequate equilibrium statistical ensembles and their corresponding density matrices are. With the technological progress in engineering and controlling open quantum many-body systems it is high time to reconcile the Hermitian with the $\mathcal{PT}$-symmetric and biorthogonal perspectives. We comprehensively review the different approaches, including the overreaching idea of pseudo-Hermiticity. To motivate the Hermitian perspective, which we propagate here, we mainly focus on the ancilla approach. It allows to embed a non-Hermitian system into a larger, Hermitian one. In contrast to other techniques, e.g., master equations, it does not rely on any approximations. We discuss the peculiarities  of $\mathcal{PT}$-symmetric and biorthogonal quantum mechanics. In these, what is considered to be an observable depends on the Hamiltonian or the selected (biorthonormal) basis. Crucially in addition, what is denoted as an ``expectation value'' lacks a direct probabilistic interpretation, and what is viewed as the canonical density matrix is non-stationary and non-Hermitian. Furthermore, the non-unitarity of the time evolution is hidden within the formalism. We pick up several model Hamiltonians, which so far were either investigated from the Hermitian perspective or from the $\mathcal{PT}$-symmetric and biorthogonal one, and study them within the respective alternative framework. This includes a simple two-level, single-particle problem but also a many-body lattice model showing quantum critical behavior. Comparing the outcome of the two types of computations shows that the Hermitian approach, which, admittedly, is in parts clumsy, always leads to results which are physically sensible. In the rare cases, in which a comparison to experimental data is possible, they furthermore agree to these. In contrast, the mathematically elegant $\mathcal{PT}$-symmetric and biorthogonal approaches lead to results which, are partly difficult to interpret physically. We thus conclude that the Hermitian methodology should be employed. However, to fully appreciate the physics of $\mathcal{PT}$-symmetric, non-Hermitian quantum many-body systems, it is also important to be aware of the main concepts of $\mathcal{PT}$-symmetric and biorthogonal quantum mechanics. Our conclusion has far reaching consequences for the application of Green function methods, functional integrals, and generating functionals, which are at the heart of a large number of many-body methods. They cannot be  transferred in their established forms to treat $\mathcal{PT}$-symmetric, non-Hermitian quantum systems. It can be considered as an irony of fate that these methods are available  only within the mathematical formalisms of  $\mathcal{PT}$-symmetric and biorthogonal quantum mechanics.   

\end{abstract}

\tableofcontents % Zum einfacheren navigieren vom dokument

\section{Introduction}
\label{sec:introduction}

Let us be clear about this from the outset: We trust that the reader consulting this review already has a personal motivation to learn more about the theoretical and, in particular, methodological aspects of parity-time-symmetric, non-Hermitian quantum systems. We thus avoid the standard introductory paragraphs emphasizing the growing interest in such systems and do not give an overview of very recent experimental breakthroughs. We also restrict our discussion of the novel and surprising physical effects already identified in these systems, to a few examples. A comprehensive account of this can be found in several very recent topical reviews \cite{El-Ganainy2018-xn,Miri2019,Ashida2020-vp,Hurst2022}. Instead, we quickly dive into the methodological questions of parity-time-symmetric, non-Hermitian quantum systems. 

We entered the field of non-Hermitian quantum systems with the goal to theoretically study emergent, collective many-body phenomena of correlated systems in the presence of terms breaking Hermiticity. In the Hermitian case, effects such as, e.g., phase transitions, quantum critical behavior, Tomonaga-Luttinger physics, etc.,
%and such%
are well understood \cite{Bruus2003,Sachdev2011}.  To study these phenomena requires the computation of time-dependent expectation values of observables, time-dependent correlation functions, and thermodynamical properties \cite{Bruus2003,Sachdev2011}. Focusing on the spectra and  eigenstates of the Hamiltonian, as it is often done in the context of other topics involving non-Hermitian Hamiltonians \cite{Bender2005-py,Bender2007-tf,Bender2019,Ghatak2019,Bergholtz2019-ss}, is, in this case, not sufficient. 

We were, in particular, aiming at non-Hermitian models showing a combined parity and time-reversal symmetry---in short, $\mathcal{PT}$-symmetry \cite{Bender2005-py,Bender2007-tf,Bender2019}. They are popular as the corresponding Hamiltonians have the peculiar property that their eigenvalues are either entirely real or are partly real and come partly in complex conjugate pairs. This leads to surprising physical properties. Furthermore, this symmetry has a clear physical interpretation.

We think of the non-Hermitian terms in the Hamiltonian to arise from the coupling of the quantum system under investigation to the outer world \cite{Bender2019,Ashida2020-vp}. Non-Hermitian systems are thus open quantum systems \cite{Carmichael1993,Breuer2007,Weiss2012}. The $\mathcal{PT}$ symmetry is only realized in special open systems and, on top, requires a fine tuning of the system-reservoir couplings. Hamiltonians of this type are easily written down but need sophisticated techniques for an experimental realization; see \cite{El-Ganainy2018-xn,Miri2019,Ashida2020-vp,Hurst2022} and references therein. However, such effort is rewarded by the observation of unique effects. If one aims at a bold characterization of the dynamics of non-Hermitian, $\mathcal{PT}$-symmetric quantum systems, one could say that it is located in between the one of closed systems and that of generic open systems. Below we will exemplify this.  

Studying the literature on $\mathcal{PT}$-symmetric, non-Hermitian quantum many-body systems, we quickly realized that many methodological questions are partly controversial and partly open. Different groups of authors use different incompatible definitions of fundamental objects such as quantum mechanical expectation values or correlation functions. Some employ the (grand-) canonical partition function to derive equilibrium thermodynamic observables while others argue, that the underlying statistical ensemble does not have any meaning if the Hamiltonian is non-Hermitian. Also the question of the use of generating functionals, Green functions, response functions, and other standard tools of quantum many-body theory is controversial. We believe that this unsatisfying state of affairs is partly due to the lack of a review focusing on the methodological aspects of $\mathcal{PT}$-symmetric or, more generally (see below), pseudo-Hermitian \cite{Ashida2020-vp} quantum many-body theory. We intend to close this gap.         

The systems investigated in most of the experiments aiming at $\mathcal{PT}$-symmetric, non-Hermitian quantum physics were, in fact, classical. What is exploited there is the analogy to optics (optical metamaterials with gain and loss) and other wave phenomena such as microwaves and even to (classical) mechanics. In that sense $\mathcal{PT}$-symmetric, non-Hermitian quantum effects were mostly  only emulated \cite{El-Ganainy2018-xn,Miri2019,Xiao2019,Ashida2020-vp,xueNonHermitianKibbleZurek2021}. Experiments in genuine  quantum systems are still rare \cite{Wu2019,Li2019,Wang2021}. In a very new trend experiments on non-Hermitian, $\mathcal{PT}$-symmetric systems are performed on quantum processors \cite{Dogra2021,Gautam2022}.  To explain the results of the experiments emulating this physics does not require a full-fledged, consistent methodology including many-body concepts such as, e.g., Green, response, and correlation functions. Understanding the properties of the spectrum and the single-particle eigenfunctions of the Hamiltonian is largely sufficient \cite{Bender2016,El-Ganainy2018-xn,Miri2019,Hurst2022,Ashida2020-vp}. We mention this here to show that the pressure from the experimental side to converge to a commonly accepted quantum mechanical formalism to treat $\mathcal{PT}$-symmetric, non-Hermitian quantum many-body systems was up to now rather moderate. However, with the increasing progress in the control of few- and many-body quantum systems, this is about to change. 
%DMK: This should be evident by the work
%We trust, that our review is timely.

In this manuscript, we provide strong evidence that the notion of observables and expectation values should be carried over from Hermitian quantum mechanics. Alternative concepts, such as those of $\mathcal{PT}$-symmetric \cite{Bender2019} or biorthogonal quantum mechanics \cite{Brody2013-br}, should not be used beyond the study of spectra. A minimal reading selection elucidating this, is provided by \cref{sec:introduction,sec:non-herm_ham,subsec:basic_idea,sec:obexp}, whereby \cref{sec:obexp} contains the main points of the argument. However, a word of warning is in order. The reader will only fully benefit from this methodological review if carefully studying all parts. It should thus not merely be used as a reference guide. Other reviews on non-Hermitian and $\mathcal{PT}$-symmetric quantum systems, with less of a theoretical and methodological
%mathematical
focus, are more suitable in this respect \cite{El-Ganainy2018-xn,Miri2019,Ashida2020-vp,Hurst2022}. In Sect.~\ref{sec:sum} we summarize the formalism which proved to be most reasonable on physical grounds in a compact form. This way we hope to make it accessible for direct use. 

%A word of warning is in order. This methodological review is written in a way, that the reader will most likely only benefit from it, if carefully studying all parts. It should thus not merely be used as a reference guide. Other reviews on non-Hermitian and $\mathcal{PT}$-symmetric quantum systems are more suitable in this respect \cite{El-Ganainy2018-xn,Miri2019,Ashida2020-vp}. Only towards the end we summarize the formalism which proved to be most reasonable on physical grounds in a compact form.  

% \begin{figure}
%     \centering
%     \includegraphics{figures/model_general2.pdf}
%     \caption{\LG{Shall we keep such a figure at the beginnig of the paper as an illustration of balanced-gain and loss. (Mental picture with water filled boxes)}}
% \end{figure}

\subsection{Hermitian quantum theory}

In quantum mechanics, as it was developed about a century ago, the dynamics of a system of particles is described by the Schr\"odinger equation ($\hbar=1$)
\begin{equation}
i \partial_t \left| \psi(t) \right> = H_t    \left| \psi(t) \right>, \quad  \left| \psi(0) \right> =  \left| \psi_0 \right> 
\label{eq:SG}
\end{equation}
with a Hermitian, in general time dependent, Hamiltonian $H_t$ and a time dependent  state vector $ \left| \psi(t) \right>$ from an underlying Hilbert space ${\mathcal{H}}$ \cite{Sakurai2017}. We use standard Dirac notation. The initial state is denoted as $\left| \psi_0 \right>$ and is assumed to be normalized $\left< \psi_0 \right. \left| \psi_0 \right> =1$, according to the canonical
%standard
inner product. We select $t=0$ as the (arbitrary) initial time. The Hermiticity of $H_t$ ensures that the time evolution operator 
\begin{equation}
U(t) = T \exp{-i \int_0^t dt' H_{t'}},
\label{eq:timeevol}
\end{equation}
with the time-ordering symbol $T$, is unitary, i.e., the dynamics is unitary, 
%DMK: I Ddont understand the e.g. here
% e.g., 
\begin{align}
\left< \psi(t) \right. \left| \psi(t) \right> = \left< \psi_0\right| U^\dag(t) U(t) \left| \psi_0 \right> = \left< \psi_0 \right. \left| \psi_0 \right> =1 . 
\label{eq:unitarity}
\end{align}

The Hermiticity of $H_t$ is also a sufficient condition for its instantaneous eigenvalues  $E_\nu^t$, fulfilling $H_t \left| \mbox{R}_\nu^t \right> = E_\nu^t  \left|\mbox{R}_\nu^t \right>$, with the instantaneous right  eigenstates $ \left|\mbox{R}_\nu^t \right>$ (which explains the label R), to be real.  In case of degeneracy, $\nu$ denotes a multi-index. The set of eigenstates is pairwise orthonormal as well as complete and forms a basis of ${\mathcal{H}}$.
The Hamiltonian is, in addition, the observable of the energy and the real eigenvalues are the possible outcomes of an energy measurement. Also all other observables in quantum mechanics are represented by Hermitian operators with real eigenvalues and complete, pairwise orthonormal sets of eigenstates. All this lies at the heart of our theoretical understanding of quantum mechanics as well as its probabilistic interpretation \cite{Sakurai2017}. 

\subsection{Open quantum systems}

Bearing in mind the above, one might wonder why at all non-Hermitian Hamiltonians are considered as the generator of the time evolution of a quantum system and as the observable of its energy \cite{Bender2005-py,Bender2007-tf,Bender2019,Rotter2009-ar,Ashida2020-vp}. However, if the system investigated is only one part of a larger setup, i.e., if one is interested in an open quantum system, studying (effectively) non-Hermitian Hamiltonians becomes reasonable \cite{Carmichael1993,Breuer2007,Weiss2012,Rotter2009-ar,Ashida2020-vp}. 

In fact, the idea that an energy eigenvalue of an open system can effectively become complex dates back at least to 1928 when Gamow studied radioactive decay of a nucleus (the system) using the, back then, new quantum theory \cite{Gamow1928}. Physically, the imaginary part of the energy is associated to the lifetime of the state. The goal to better understand nuclear reactions also led Feshbach and others to introduce the idea of complex (real-space) single-particle potentials and thus of non-Hermitian Hamiltonians \cite{Feshbach1954,Feshbach1958}. Effective (phenomenological) models with non-Hermitian Hamiltonians were thus considered from the early days of modern quantum theory on. 

Mainly three theoretical schemes leading to (effectively) non-Hermitian Hamiltonians of open quantum systems are popular: Feshbach projection, quantum master equations, and the ancilla approach. They were developed for specific classes of physical problems encountered by different communities. In this review we only give the basic ideas of the first two approaches. More can be found in the reviews \cite{Rotter2009-ar}, \cite{Daley2014}, \cite{Ashida2020-vp}, and \cite{Roccati2022}. %VM: added Roccati2022
Although, both approaches are important to derive generic non-Hermitian Hamiltonians of experimental relevance, we believe that for a theoretical analysis of $\mathcal{PT}$-symmetric, open quantum systems, the third framework, the ancilla approach \cite{Guenther2008,Kawabata2017,Wu2019}, is the most transparent one. Often it is also referred to as the dilation method, in particular, in the context of quantum information theory \cite{Guenther2008}. The ancilla approach has the additional advantage that it does not rely on any approximations or phenomenological reasoning. Therefore, our considerations will build mainly upon this. Surprisingly, the ancilla approach was, up to now, not reviewed in a comprehensive way in the quantum many-body context. We give a brief overview of this scheme in the present introduction and a full account, including all technical details, in Sect.~\ref{sec:ancilla}.

Open quantum systems \cite{Carmichael1993,Breuer2007,Weiss2012} are at the heart of the timely and vivid field of quantum engineering which includes quantum information processing. This is one of the reasons why the study of non-Hermitian systems experiences a recent revival \cite{El-Ganainy2018-xn,Ashida2020-vp}.  $\mathcal{PT}$-symmetric, open systems are of particular interest. Their dynamical properties can, e.g., show coherence features which differ strongly from those of closed systems but, at the same time, lack a decay, as it does occur in generic open systems \cite{Wu2019,Dogra2021,Wang2021,Gautam2022}. For a discussion of this, see Sect.~\ref{subsec:further_examples_2_2}. One often refers to this phenomenon as a balance of gain and loss of the open system \cite{Bender2019}. 

\subsubsection{Feshbach projection}
\label{subsec:Feshbach}

We first present the main ideas of Feshbach projection \cite{Rotter2009-ar,Ashida2020-vp}. The total, Hermitian Hamiltonian $H=H_{\rm s} + H_{\rm r} + H_{\rm sr}$ consists of a system part $H_{\rm s}$ with a few degrees of freedom, a reservoir part $H_{\rm r}$ with many degrees of freedom, and the coupling $H_{\rm sr}$. One takes the thermodynamic limit for the reservoir part for which its spectrum becomes continuous. One way of deriving an effective Hamiltonian of the system part is to consider the resolvent or Green function $G(z)=\left\{z - H \right\}^{-1}$, $z \in {\mathbbm C}$, of the total Hamiltonian $H$. The total Green function is projected onto the system leading to the (exact) system Green function $G_{\rm s}(z)=\left\{z - \left[H_{\rm s} + \Sigma_{\rm r}(z)\right]\right\}^{-1}$, with the operator valued reservoir self-energy $\Sigma_{\rm r}(z)$. Note that $ \Sigma_{\rm r}(z)$ only contains system degrees of freedom but depends on the spectrum of the reservoir as well as the details of the system-reservoir coupling. Now $H_{\rm s} + \Sigma_{\rm r}(z)$ is interpreted as the effective Hamiltonian at energy $z$. In general, $\Sigma_{\rm r}(z)$ will not be Hermitian and the effective Hamiltonian will be non-Hermitian.  In addition, $\Sigma_{\rm r}(z)$ depends on the (complex) energy variable $z$. Generically, the effective Hamiltonian will not have $\mathcal{PT}$ symmetry. This will be realized only in special systems involving fine-tuning of the system-reservoir coupling. 

The Feshbach projection is often employed in scattering theory and in mesoscopic physics to study resonances with a finite broadening in the quantum regime \cite{Feshbach1958,Rotter2009-ar,Ashida2020-vp}. In many-body physics it is used to integrate out the non-interacting parts (reservoirs) of a setup such that computationally demanding quantum many-body methods must only be employed to the system with fewer degrees of freedom; see, e.g., \cite{Enss2005}. 

\subsubsection{Master equations and the quantum trajectory approach}
\label{subsec:master}

The second scheme to treat open quantum systems which leads to non-Hermitian Hamiltonians are master equations complemented by the quantum trajectory approach, which is heavily used in the  atomic, molecular, and optical (AMO) community, including quantum optics, as well as in quantum information theory \cite{Daley2014}. Employing the rotating wave approximation, the Born approximation, and the Markov approximation leads to a Markovian master equation of Lindblad form 
\begin{equation}
i \partial_t \rho_{\rm s}(t) \! = \! \left[H_{\rm s},\rho_{\rm s}(t)\right] \! - \! \frac{i}{2} \! \sum_{l} \! \gamma_l  \! \left\{ a_l^\dag a_l \rho_{\rm s}(t) + \rho_{\rm s}(t) a_l^\dag a_l - 2 a_l \rho_{\rm s}(t) a_l^\dag \right\}  
\label{eq:lindblad}
\end{equation}
for the time evolution of the systems' reduced density matrix $\rho_{\rm s} = \mbox{Tr}_{\rm r} \rho$, where $\rho$ is the density matrix of the combined system and reservoir setup. The trace is taken over the reservoir degrees of freedom. The $a_l^{(\dag)}$ are called jump operators and describe the systems loss processes with corresponding (real) rates $\gamma_l$. Often, the jump operators and the rates can only be constructed based on phenomenological reasoning. Note that Eq.~(\ref{eq:lindblad}) ensures that the reduced density matrix $\rho_{\rm s}(t)$ is Hermitian for all $t$ if it is Hermitian at $t=0$. We emphasize that in contrast to Feshbach projection the reservoir is not projected but traced out.  Due to an appropriate hierarchy of the system and reservoir energy scales and a weak system-reservoir coupling, the criteria underlying the above three approximations are often met in AMO systems \cite{Daley2014,Ashida2020-vp}. A typical example is a few-level atom coupled to a quantized light field.

Equation (\ref{eq:lindblad}) can be rewritten as 
\begin{equation}
i \partial_t \rho_{\rm s}(t)= H_{\rm eff} \rho_{\rm s}(t) -  \rho_{\rm s}(t) H^\dag_{\rm eff} - \frac{i}{2} \sum_{l} \gamma_l a_l \rho_{\rm s}(t) a_l^\dag , 
\label{eq:lindblad_non_herm}
\end{equation}
with
\begin{equation}
H_{\rm eff} = H_{\rm s} - \frac{i}{2} \sum_{l} \gamma_l a_l^\dag a_l .
 \label{eq:H_eff}    
\end{equation}
Up to the so-called recycling term $ \sum_{l} \gamma_l a_l \rho_{\rm s} a_l^\dag $, Eq.~(\ref{eq:lindblad_non_herm}) has the form of a generalized von Neumann equation known from quantum statistical mechanics \cite{Sakurai2017}, with a non-Hermitian Hamiltonian $H_{\rm eff}$. As $H_{\rm eff}^\dag \neq H_{\rm eff}$ the first two terms cannot be combined to a commutator. Generically, $H_{\rm eff}$ will not be $\mathcal{PT}$-symmetric. This requires a special system and a fine-tuned system-reservoir coupling, which is encoded in the jump operators and the rates. 

The master equation in the form of Eq.~(\ref{eq:lindblad}) or equivalently (\ref{eq:lindblad_non_herm}) approximately describes the dynamics of the reduced system density matrix. A further reduction to a Hamiltonian dynamics can be obtained along the following line. If one 
%if no information about measurement outcomes is available. The idea is now to 
performs a continuous measurement on the system and disregards all instances in which losses via jumps, described by the addends $\gamma_l a_l \rho_{\rm s}(t) a_l^\dag$, occurred, the dynamics is captured by only the first two terms on the right hand side of Eq.~(\ref{eq:lindblad_non_herm}).  Therefore, understanding the dynamics of $\rho_{\rm s}(t)$ resulting out of the generalized von Neumann equation without the last addend in Eq.~(\ref{eq:lindblad_non_herm}), but the non-Hermitian Hamiltonian $H_{\rm eff}$, is very useful. Note that also in this case the equation of motion guarantees that the systems reduced density matrix remains Hermitian at all times if it is Hermitian initially. More on this can be found in the reviews \cite{Ashida2020-vp}, \cite{Daley2014}, and \cite{Roccati2022}. %VM: added Roccati2022   

\subsubsection{The ancilla approach}

We will extensively use the ancilla approach which is the third scheme leading to equations involving non-Hermitian Hamiltonians \cite{Guenther2008,Kawabata2017,Wu2019}. For our purpose of reconciling the open quantum system view on non-Hermitian many-body systems with that of $\mathcal{PT}$-symmetric systems, we judge this to be the most transparent method. This holds, in particular, as the ancilla approach does not rely on any approximations or an effective Hamiltonian picture. In Sect.~\ref{sec:ancilla} we thus give a comprehensive account of this.

In the ancilla approach one assumes that a non-Hermitian Hamiltonian $H_{\rm s}$ of a system, acting on a Hilbert space $\mathcal{H}_{\rm s}$, is given. It might be $\mathcal{PT}$-symmetric. The system is supplemented by a single spin-1/2 ancilla. One can construct a Hermitian Hamiltonian $H_{\rm sa}$ for the combined system-ancilla setup, i.e., a linear, Hermitian operator acting on $\mathcal{H}_{\rm sa} =  \mathbbm{C}^2 \otimes \mathcal{H}_{\rm s} $, which leaves a certain subset $\mathcal{H}_{\rm sa}^{\rm sub}$ of states from $\mathcal{H}_{\rm sa}$ invariant (see Sect.~\ref{sec:ancilla} for details). The non-Hermitian system is hence embedded in a Hermitian one. In this sense the logic in the ancilla approach is inverse to that of  Feshbach projection and the master equation technique, in which effective, non-Hermitian Hamiltonians of the subsystem are derived from Hermitian ones. 
%If an initial, normalized state $\left| \psi _{\rm sa}^{\rm sub} (0) \right>$ from $\mathcal{H}_{\rm sa}^{\rm sub}$ is time-evolved according to the Schr\"odinger equation with  $H_{\rm sa}$ up to time $t$ (unitary dynamics), a measurement of the ancilla spin on (the normalized) $\left| \psi _{\rm sa}^{\rm sub} (t) \right>$ is performed, and only the instances which gave spin-up are kept, then the unnormalized, time-dependent state after the measurement can also be reached by the non-unitary time evolution with $H_{\rm s}$. In that sense the state after the measurement evolved according to the Schr\"odinger equation with a non-Hermitian Hamlitonian. 

% One then performs a gedanken experiment
The non-unitary dynamics in the ancilla approach results as follows. A normalized state $\left| \psi _{\rm sa}^{\rm sub} (0) \right>$ from $\mathcal{H}_{\rm sa}^{\rm sub}$ is time-evolved with the unitary time evolution operator corresponding to $H_{\rm sa}$ up to time $t$. At that time a measurement of the ancilla spin is performed, and only the instances which give spin-up are kept.
% One then notices that the un-normalized, time-dependent state after the measurement can also be reached by the non-unitary time evolution with the time evolution operator associated to the non-Hermitian $H_{\rm s}$.
One then notices that, up to a normalization, the time evolution of the above outlined protocol, can also be described by the non-unitary time evolution with the time evolution operator associated to the non-Hermitian $H_{\rm s}$. In this sense a continuous measurement of the ancilla spin leads to a state that evolves according to the Schr\"odinger equation with a non-Hermitian Hamiltonian. No approximations are required. Details are give in Sect.~\ref{sec:ancilla}. 

In a very impressive real world experiment with a single nitrogen-vacancy center in diamond the ancilla scheme was directly implemented and non-Hermitian dynamics in a ${\mathcal{PT}}$-symmetric system was observed \cite{Wu2019}. This shows, that the ancilla idea is not only theoretically appealing but can be brought to life in real systems.

\subsection{\texorpdfstring{$\mathcal{PT}$}{PT}-symmetric Hamiltonians}

\subsubsection{\texorpdfstring{$\mathcal{PT}$}{PT}-symmetry}
\label{subsec:PT}

Despite the early appearance of non-Hermitian Hamiltonians in open system quantum theory, their mathematical structure \cite{Ashida2020-vp} was analyzed in detail only later. This was substantially triggered by the observation of Bender \textit{et al.}, that the non-Hermitian single-particle Hamiltonian 
\begin{equation}
H=\hat p^2 + \hat x^2+ (i \hat x)^N ,
\label{eq:toy_ham}
\end{equation}
with $N\geq 2$, has an entirely real spectrum and that the set of right eigenstates forms a (non-orthogonal) basis of the Hilbert space  \cite{benderRealSpectra1998,doreySpectralEquivalences2001}.\footnote{Note also footnote [1] of \cite{benderRealSpectra1998}} This Hamiltonian was not studied in relation to open system quantum mechanics, but emerged as a toy model in the context of quantum field theory \cite{Bender2005-py,Bender2007-tf}. Further model Hamiltonians with these properties were found \cite{Bender2005-py,Bender2007-tf,Bender2019,Ashida2020-vp}.

Bender and collaborators speculated that the reality of the spectrum of the toy Hamiltonian Eq.~(\ref{eq:toy_ham}) originates from its combined ${\mathcal{PT}}$ symmetry. For a (continuous) translational degree of freedom ${\mathcal{PT}}$ symmetry implies that the Hamiltonian is invariant under simultaneous spatial reflection $x \to -x$ (${\mathcal{P}}$) and time reversal  (${\mathcal{T}}$); the latter also changes $i \to -i$ (anti-linearity of ${\mathcal{T}}$ \cite{Sakurai2017}). Due to this, the combined ${\mathcal{PT}}$ transformation is anti-linear. In mathematical terms a Hamiltonian $H$ is $\mathcal{PT}$-symmetric if
\begin{equation}
\left[ H, \mathcal{PT} \right]=0.
\label{eq:H_PT_com}    
\end{equation}
It would be very appealing if the reality of the spectrum of a Hamiltonian could be ensured based on a physical symmetry principle rather than the mathematical requirement of Hermiticity \cite{Bender2002}. In particular, as the Hermiticity of a Hamiltonian is sufficient, but, apparently not necessary for the spectrum to be entirely real; see the above toy model Eq.~(\ref{eq:toy_ham}) for an example. The ultimate goal was to set up a quantum theory in which the mathematical axiom of Hermiticity of the Hamiltonian and, for that matter, all other observables, would be replaced by a more physical one associated to a symmetry. This spurred a tremendous research effort, originally in the field of mathematical physics \cite{Bender2019}. For an overview, we merely summarize some crucial insights
into the structure of the theory of ${\mathcal{PT}}$-symmetric, non-Hermitian Hamiltonians,
following the historical development in this introduction.
% %and, if needed,
If needed, we provide more technical details and, in particular, further examples in Sect.~\ref{sec:non-herm_ham}. 

This line of research opened up an independent second route towards the current popularity of the field of non-Hermitian,  $\mathcal{PT}$-symmetric quantum systems. The investigation of $\mathcal{PT}$-symmetric Hamiltonians led to two formalisms which are nowadays subsumed under the terms ``$\mathcal{PT}$-symmetric quantum mechanics'' \cite{Bender2019} and ``biorthogonal quantum mechanics''  \cite{Brody2013-br}, respectively. They have a common ground but there are also specific differences which we will discuss in this review. This, in particular, concerns the definition of the concept of an observable. The focus of this research program is more on the spectral properties of the Hamiltonian and not so much on its role as the generator of the dynamics. It is crucial to keep in mind that even if the spectrum of a non-Hermitian Hamiltonian is entirely real, the dynamics of the underlying quantum system is still non-unitary if the standard formalism of quantum mechanics is used. We emphasize this already at this stage, as it is very likely, that the typical reader of this review will be familiar with or, at least, will have heard of the idea of a biorthogonal inner product \cite{Brody2013-br}.\footnote{Within $\mathcal{PT}$-symmetric quantum mechanics this is also denoted as the $\mathcal{CPT}$ inner product \cite{Bender2019}. We here use this synonymously.} Postulating a modified inner product between two quantum states, that is, working in a modified Hilbert space, the dynamics can become effectively unitary if the spectrum is real. Reconciling the biorthogonal and $\mathcal{PT}$-symmetric quantum  mechanics view on $\mathcal{PT}$-symmetric non-Hermitian Hamiltonians with the standard formalism, familiar from Hermitian quantum theory, is one of the main goals of this review. Note that only the standard formalism of Hermitian quantum mechanics is employed for non-$\mathcal{PT}$-symmetric, that is, generic open quantum systems \cite{Carmichael1993,Breuer2007,Weiss2012}. We will return to this observation.

${\mathcal{PT}}$ symmetry is obviously not a necessary condition for the spectrum to be real. Hermitian Hamiltonians can be given which are not ${\mathcal{PT}}$-symmetric but have an entirely real spectrum. Think, e.g., of the Hamiltonian of a single particle confined to move in one spatial dimension, with mass $m$, and kinetic energy $\hat p^2/(2m)$ subject to a real potential $V(\hat x)$ which is not an even function. It is important to emphasize that this only holds, if the parity transformation is understood in its direct sense, namely, a as spatial reflection. Employing more general definitions of the parity transformation \cite{Mostafazadeh2003,Bender2003}, every Hermitian Hamiltonian turns out to have a  ``generalized'' $\mathcal{P}$ symmetry and the above one, $H=\hat p^2/(2m) + V(\hat x)$, would also be  ``generalized'' ${\mathcal{PT}}$-symmetric \cite{Bender2003}. This generalized parity transformation, not referring to spatial reflection, is, however, not very intuitive. 

The study of the toy model Eq.~(\ref{eq:toy_ham}) also indicates that the ${\mathcal{PT}}$ symmetry is not sufficient for the reality of the spectrum. For $1<N<2$ the Hamiltonian Eq.~(\ref{eq:toy_ham}) has partly real eigenvalues and partly complex conjugate pairs of eigenvalues. For $N\leq 1$ no real eigenvalues are left \cite{benderRealSpectra1998,doreySpectralEquivalences2001,Bender2005-py,Bender2007-tf}. 

It was shown, that \cite{Bender2005-py,Bender2007-tf,Bender2019}:
\begin{theorem}[$\mbox{T}^{\mathcal{PT}}_{1}$]
A ${\mathcal{PT}}$-symmetric Hamiltonian has an entirely real spectrum if and only if all of its right eigenvectors are also eigenvectors of ${\mathcal{PT}}$.
\end{theorem}
\noindent If this is not the case one speaks of a spontaneously broken ${\mathcal{PT}}$ symmetry. One might wonder why it is even an option that the eigenvectors of $H$ are not simultaneously eigenvectors of $\mathcal{PT}$ even though $H$ and $\mathcal{PT}$ commute; see Eq.~(\ref{eq:H_PT_com}). This is possible as $\mathcal{PT}$ is an anti-linear operator and not a linear one \cite{Bender2007-tf}. In the toy Hamiltonian Eq.~(\ref{eq:toy_ham}) the phase of broken $\mathcal{PT}$ symmetry is associated to the appearance of the complex conjugate pairs of eigenvalues mentioned above. Exactly at the point of the ${\mathcal{PT}}$ transition, a  minimum of two eigenenergies, which do not repel each other, are degenerate and the set of right eigenvectors does no longer form a basis; the eigenvectors coalesce. One denotes this as an exceptional point \cite{Heiss2012-nj,Ashida2020-vp}; there are too few eigenvectors of $H$ to span the entire Hilbert space.
%Exactly at the point of the ${\mathcal{PT}}$ transition, at minimum two eigenenergies are degenerate (they do not repel each other when approaching the critical parameter value), but the set of right eigenvectors does no longer form a basis and one speaks of an exceptional point \cite{Ashida2020-vp}; there are too few eigenvectors of $H$ to span the entire Hilbert space.
As in the toy example Eq.~(\ref{eq:toy_ham}), it is often the case that the system can be driven across the ${\mathcal{PT}}$ transition by tuning a parameter of the Hamiltonian (here $N$). When doing so, further exceptional points might occur, which, however, are not necessarily related to a ${\mathcal{PT}}$ transition. In short ${\mathcal{PT}}$ transitions occur at an exceptional point but not every exceptional point is related to a ${\mathcal{PT}}$ transition. 
%There appears to be some confusion about this in the many-body literature \cite{Zhang2022}. We already now emphasize that it is advisable to be strict in the nomenclature to prevent any misconceptions. 
Going beyond ${\mathcal{PT}}$-symmetric systems the role of exceptional points of non-Hermitian Hamiltonians in many interesting physical effects was reviewed in \cite{Heiss2012-nj,Miri2019,Ashida2020-vp,Ghatak2019,Bergholtz2019-ss}. 

The quest for a necessary and sufficient condition for the spectrum of a Hamiltonian to be entirely real led to a very fruitful academic dispute between Bender and his collaborators on the one side and Mostafazadeh on the other. Mostafazadeh showed \cite{Mostafazadeh2003a}: 
\begin{theorem}[$\mbox{T}^{\mathcal{PT}}_{2}$]
For every ${\mathcal{PT}}$-symmetric Hamiltonian $H$ with unbroken ${\mathcal{PT}}$ symmetry, there exists a Hermitian Hamiltonian $h$, which is isospectral to $H$ and is related to it by a similarity transformation: $h=S H S^{-1}$, where $S$ denotes an invertible linear operator.
\end{theorem}
\noindent In this sense ${\mathcal{PT}}$-symmetric Hamiltonians with real spectra do not extend the class of Hermitian Hamiltonians. Bender and others reacted by emphasizing that for most ${\mathcal{PT}}$-symmetric $H$ with unbroken ${\mathcal{PT}}$ symmetry of physical interest it is neither possible to exactly construct $S$ nor $h$ \cite{Jones2005,Bender2005-ys}. In fact, even for the simple ${\mathcal{PT}}$-symmetric model 
\begin{equation}
H= (\hat p^2 + \hat x^2)/2 + ig \hat x^3
\label{eq:sci_post_ham}
\end{equation}
with real spectrum, $S$ and $h$ can only be constructed in perturbation theory in $g$. Already to $\mathcal{O}(g)$, $h$ has very unfamiliar terms $\sim \hat p^3$ and $\sim \hat x \hat p \hat x$ \cite{Jones2005,Mostafazadeh2005}. This becomes worse in higher orders. Only in a few exceptional cases, closed exact expressions for $S$ or $h$ can be given \cite{Geyer2004,Jones2005,Jones2006,Bender2006,Jones2006a,Ashida2016-ll,Dora2022}. We return to this in Sect.~\ref{subsec:examples_2_2}. 

It was furthermore shown that ${\mathcal{PT}}$ symmetry is not special and that Hamiltonians being invariant under other anti-linear symmetry transformations can also have entirely real spectra \cite{Bender2002a,Mostafazadeh2003,Mannheim2018}.  

\subsubsection{Pseudo-Hermiticity}
\label{subsec:pseudoherm}

In his studies, Mostafazadeh emphasized that the underlying principle which might render even the spectrum of non-Hermitian Hamiltonians to be entirely real is pseudo-Hermiticity \cite{Pauli1943,Mostafazadeh2002-gk,Mostafazadeh2003,Mostafazadeh2003a}. An operator---here the Hamiltonian $H$---is denoted as pseudo-Hermitian if a linear, Hermitian, and invertible operator $\eta$ exists, such that $H^\dag=\eta H \eta^{-1}$. For $\eta=\mathbbm{1}$ the condition of pseudo-Hermiticity reduces to Hermiticity. The class of pseudo-Hermitian Hamiltonians thus includes the Hermitian ones. In general, the operator $\eta$ depends on the Hamiltonian considered and might not even be unique \cite{Mostafazadeh2002-gk}. One thus also speaks of $\eta$-pseudo-Hermiticity of a given Hamiltonian. 
\begin{theorem}[$\mbox{T}^{\eta}_{1}$]
A direct relation between ${\mathcal{PT}}$ symmetry and pseudo-Hermiticity can be made when the ${\mathcal{PT}}$-symmetric Hamiltonian can be represented by a complex, symmetric matrix. In this case $H$ is pseudo-Hermitian with $\eta = \mathcal{P}$ \cite{Mostafazadeh2003a}. 
\end{theorem}
\noindent Furthermore, several of the $\mathcal{PT}$-symmetric one-dimensional real space continuum models studied by Bender et al.~\cite{Bender2019} have a Hamiltonian of the form $H=\hat p^2 + V_{\rm r}(\hat x) + i V_{\rm i}(\hat x)$ with two real-valued potentials $V_{\rm r/i}(x)$, with $V_{\rm r}$ being an even function and $V_{\rm i}$ being an odd one. Also for this type of $\mathcal{PT}$-symmetric, non-Hermitian Hamiltonians it is straightforward to show that they are $\mathcal{P}$-pseudo-Hermitian \cite{Mostafazadeh2002-gk}. 

To fully understand a crucial theorem on the spectral properties of non-Hermitian operators, we first have to introduce the notion of left eigenvectors. As $H^\dag \neq H$, the left eigenvectors $\left|\mbox{L}_\nu \right>$, with 
%\begin{equation}
%H^\dag \left|\mbox{L}_\nu \right> = \textcolor{red}{\tilde E_\nu \to  E_\nu^\dagger} \left|\mbox{L}_\nu \right>,
%\label{eq:leftevdef}
%\end{equation}
\begin{equation}
H^\dag \left|\mbox{L}_\nu \right> = 
\tilde E_\nu  \left|\mbox{L}_\nu \right>,
\label{eq:leftevdef}
\end{equation}
are no longer equal to the right ones. Furthermore, the set of right eigenvectors cannot be chosen as (pairwise) orthonormal: $\left< \mbox{R}_\nu \right. \left| \mbox{R}_\mu \right> \neq \delta_{\nu,\mu}$. Instead one can normalize (and label) the left and right eigenstates such that $\left< \mbox{L}_\nu \right. \left| \mbox{R}_\mu \right> = \delta_{\nu,\mu}$ \cite{Brody2013-br}. Mostafazadeh proved that \cite{Mostafazadeh2002-gk,Wigner1960}:
\begin{theorem}[$\mbox{T}^{\eta}_{2}$]
If $H$ is a non-Hermitian Hamiltonian with a discrete spectrum and the set $\left\{ \left| \mbox{R}_\nu \right> ,\left|\mbox{L}_\nu \right> \right\}$ is complete, i.e., $\sum_{\nu}\left| \mbox{R}_\nu \right> \left<\mbox{L}_\nu \right|= {\mathbbm{1}}$ (complete biorthonormal basis), then $H$ is pseudo-Hermitian if and only if the spectrum of $H$ is either entirely real or complex eigenvalues come in complex conjugate pairs.
\end{theorem}
\noindent While, as discussed above, ${\mathcal{PT}}$ symmetry of a Hamiltonian is neither a necessary nor a sufficient condition for its entire spectrum to be real, we conclude that pseudo-Hermiticity (which includes Hermiticity) is at least necessary.  Mostafazadeh was also able to prove \cite{Mostafazadeh2002}:
\begin{theorem}[$\mbox{T}^{\eta}_{3}$]
 A necessary and sufficient condition for the spectrum of an $\eta$-pseudo-Hermitian Hamiltonian to be real,  is the existence of the square root of $\eta$, i.e., one can write $\eta=\eta^{1/2} \eta^{1/2}$ with a linear, Hermitian, and invertible operator $\eta^{1/2}$ . 
\end{theorem}
\noindent In fact, $\eta^{1/2}$ is related to the operator $S$ of theorem $\mbox{T}^{\mathcal{PT}}_{2}$ which transforms a ${\mathcal{PT}}$-symmetric Hamiltonian $H$ with entirely real spectrum to the isospectral Hermitian one $h$; for more on this, see Sect.~\ref{subsec:examples_2_2}. Finally, it was shown that \cite{Mostafazadeh2002-gk}: 
\begin{theorem}[$\mbox{T}^{\eta}_{4}$]
Every $\mathcal{PT}$-symmetric Hamiltonian $H$ with a discrete spectrum and a complete, biorthogonal basis of eigenstates is pseudo-Hermitian. Its eigenvalues are either entirely real or come in complex conjugate pairs.     
\end{theorem} 
$\mathcal{PT}$-symmetric Hamiltonians form a subset of the larger set of pseudo-Hermitian ones. If, in the following, we speak about properties of pseudo-Hermitian Hamiltonians with entirely real spectra, one can think of a $\mathcal{PT}$-symmetric Hamiltonian in its symmetry unbroken phase. For a pseudo-Hermitian Hamiltonian with partly real eigenvalues and partly complex conjugate pairs of eigenvalues one can think of a $\mathcal{PT}$-symmetric Hamiltonian in its symmetry broken phase.

All this is, unfortunately, pretty far away from Benders original idea to directly relate ${\mathcal{PT}}$ symmetry of a Hamiltonian to the reality of its spectrum. In particular, the above theorems do not allow for a simple, physically intuitive understanding of the conditions under which a non-Hermitian Hamiltonian is guaranteed to have a real spectrum. We will return to the spectral theory of pseudo-Hermitian Hamiltonians as well as to what is denoted as biorthogonal \cite{Brody2013-br} and $\mathcal{PT}$-symmetric \cite{Bender2019} quantum mechanics in Sects.~\ref{sec:non-herm_ham}-\ref{sec:obexp}.

\subsection{Outline}

As an introduction, we summarized a few notions of relevance for the understanding of the theory of $\mathcal{PT}$-symmetric, non-Hermitian, and open quantum systems. We also gave a historical account of these. 
%We next dive deeper in our quest to provide a comprehensive review of a proper formalism to treat non-Hermitian quantum many-body systems. 

%In Sect.~\ref{sec:non-herm_ham} we analyze the properties of non-Hermitian, pseudo-Hermitian, and  $\mathcal{PT}$-symmetric Hamiltonians in more detail. Doing so we (first) ignore the question why it is meaningful to  study the Schr\"odinger equation with a non-Hermitian Hamiltonian on the right hand side as well as the spectral properties of such an operator. The situation with entirely real spectra (e.g., for an unbroken $\mathcal{PT}$ symmetry in the case of a  $\mathcal{PT}$-symmetric system) as well as that with complex spectra (e.g., for a broken $\mathcal{PT}$ symmetry) is considered. We introduce the concept of a biorthonormal basis and a corresponding resolution of unity as well as the spectral representation of the Hamiltonian. This paves the way towards biorthogonal and $\mathcal{PT}$-symmetric quantum mechanics. We give more details on pseudo-Hermiticity and construct $\eta$ operators. 

In Sect.~\ref{sec:non-herm_ham} we analyze the properties of non-Hermitian, pseudo-Hermitian, and  $\mathcal{PT}$-symmetric Hamiltonians in more detail. Doing so we (first) ignore the question why it is meaningful to  study these Hamiltonians. We introduce the concept of a biorthonormal basis and a corresponding resolution of unity. This paves the way towards biorthogonal and $\mathcal{PT}$-symmetric quantum mechanics.
In Sect.~\ref{sec:non-herm_ham} we also introduce our three example systems employed to illustrate the reviewed methodologies. The first is a $\mathcal{PT}$-symmetric, non-Hermitian $2 \times 2$ matrix corresponding to a quantum particle hopping between two lattice sites (or a single spin-1/2 degree of freedom). We refer to it as the two-level problem. This is heavily used in the literature as one of the simplest systems to convey some of the major differences between Hermitian and non-Hermitian quantum mechanics \cite{Bender2005-py,Bender2007-tf,Brody2013-br,Ashida2020-vp,Tetling2022}. The second is the $\mathcal{PT}$-symmetric resonant level model with complex hybridization between the single (quantum dot) level and the two leads which destroys Hermiticity \cite{Yoshimura2020}. The Hermitian resonant level model is a simple toy model of mesoscopic physics to study quantum transport through a quantum dot.  The third is a  $\mathcal{PT}$-symmetric one-dimensional tight-binding model with staggered complex hopping and staggered onsite energy very recently introduced as a  $\mathcal{PT}$-symmetric toy model for quantum critical behavior in the presence of non-Hermitian terms \cite{Dora2022a}. All these models show a phase of unbroken $\mathcal{PT}$ symmetry as well as a symmetry broken phase.      

In Sect.~\ref{sec:ancilla} we review the ancilla approach which is a transparent way to show how the dynamics of a quantum system can become effectively non-unitary. In the ancilla approach the system is complemented by a single spin-1/2 and a measurement on this is performed. The combined setup of system and ancilla spin evolves unitarily, i.e., according to the Schr\"odinger equation with a Hermitian Hamiltonian. This framework does not rely on any approximations or an effective Hamiltonian picture and was brought to life in an experiment \cite{Wu2019}. We illustrate the ancilla idea for the example of the non-Hermitian two-level problem. Further, we discuss one of the reasons why biorthogonal and $\mathcal{PT}$-symmetric quantum mechanics are popular in mathematical physics and quantum field theory. Within these frameworks the time evolution is (effectively) unitary with respect to a modified inner product as long as the entire spectrum is real.

Building on the ideas which evolved out of Sects.~\ref{sec:non-herm_ham} and \ref{sec:ancilla} we introduce the concepts of observables and quantum mechanical expectation values, as postulated within the different formalisms in Sect.~\ref{sec:obexp}. We argue that a definition of observables, just like in Hermitian quantum mechanics, is physically more compelling on general grounds as compared to the ones used in biorthogonal and $\mathcal{PT}$-symmetric quantum mechanics. The same holds for the concept of an expectation value. This is further substantiated by the explicit comparison of results obtained for our models within the two alternative frameworks. It is still very useful to be aware of the main concepts of biorthogonal and $\mathcal{PT}$-symmetric quantum mechanics when analyzing non-Hermitian Hamiltonians, since both are mathematically powerful.

The considerations of Sects.~\ref{sec:ancilla} and \ref{sec:obexp} naturally lead to an extension towards statistical ensembles which we present in Sect.~\ref{sec:statens}. As the dynamics is non-unitary a crucial question is to clarify if ensembles with non-trivial statistical operators exist, which are stationary and can thus be employed to set up equilibrium thermodynamics. We show that for pseudo-Hermitian Hamiltonians with entirely real spectra, such exist but that the standard (grand) canonical ensemble of Hermitian quantum statistical mechanics does not fall into this class. For the case of pseudo-Hermitian Hamiltonians with pairs of complex conjugate eigenvalues we were unable to identify a mixed state, stationary statistical operator. We employ the ancilla approach to make contact with what is known for the Hermitian system-ancilla Hamiltonian. Quantum statistical expectation values are introduced and we review linear response theory for non-Hermitian Hamiltonians. This leads us to conclude that response functions, which are central in Hermitian quantum many-body theory when making contact to experiments, are less practical in non-Hermitian systems.  

In Sect.~\ref{sec:genfun} we show, that crucial concepts of quantum many-body theory, such as generating functionals, functional integrals, and Green functions, can be mathematically defined for non-Hermitian systems within the frameworks of biorthogonal and $\mathcal{PT}$-symmetric quantum mechanics. However, these concepts turn out to be less useful and even difficult to define in the physical scheme we propagate here. As they are at the heart of most analytical quantum many-body methods and also at some numerical ones, their usage for non-Hermitian systems has to be reconsidered. In short, for non-Hermitian systems one cannot remove the explicit dependence of expectation values and correlation functions on many-body states. That these drop out is one of the strengths of Green function techniques in Hermitian many-body theory. We identify this as a major roadblock for the application of a plethora of quantum many-body methods. We, in addition, discuss the use of correlation functions to investigate quantum critical behavior in one of our model systems.          

We conclude our review by presenting a recap of the methodological aspects of non-Hermitian quantum many-body theory in Sect.~\ref{sec:sum}. It summarizes the formalism which proved to be most reasonable on physical grounds in a compact form. 

The Appendices contain some technical details, in particular, those on our numerical computations.  

\section{\texorpdfstring{$\mathcal{PT}$}{PT}-symmetric and pseudo-Hermitian Ha\-miltonians}
\label{sec:non-herm_ham}

In this section we take it for granted that it is meaningful to study both the Schr\"odinger equation (\ref{eq:SG}) with a non-Hermitian Hamiltonian $H_t$ on the right hand side as well as the spectral properties of this operator. For now we suppress the index $t$.  

For a non-Hermitian Hamiltonian the time evolution operator $U(t)$ Eq.~(\ref{eq:timeevol}) is no longer unitary. This implies that even if the evolution starts in a state $ \left| \psi_0 \right>$  which is normalized to one, $\left< \psi_0 \right.  \left| \psi_0 \right> =1$, the time-evolved state will not be normalized for all $t>0$, $\left< \psi(t) \right. \left| \psi(t) \right> \neq 1$. Here $\left< \ldots \right.  \left| \ldots \right>$ denotes the (canonical) inner product of Hermitian quantum mechanics.  

%Not much of tangible value can be said about the 
Generic statements about the
spectral properties of a general non-Hermitian (Hamilton) operator on a general Hilbert space are difficult \cite{Ashida2020-vp}. We thus have to further specify the non-Hermitian Hamiltonians of interest to us. Furthermore, we want to avoid any mathematical subtleties related to infinite dimensional Hilbert spaces when it comes to statements about separability and completeness. Such are already known from the Hermitian case and one would not expect matters to improve for non-Hermitian operators \cite{Brody2013-br}. Also our mathematical statements in Sects.~\ref{subsec:PT} and \ref{subsec:pseudoherm} must be understood with this caveat. Investigating such refinements is a challenge for mathematical physics or even pure mathematics. Generically they do not play any role for the operators of interest in quantum many-body theory. Therefore, we  
%restrict our considerations. We 
only consider finite dimensional Hilbert spaces of dimension $n$ and thus have to deal with $ n \times n$-dimensional matrices \cite{Brody2013-br,Ashida2020-vp}. For models, for which this does not apply, such as, e.g., the toy model Eq.~(\ref{eq:toy_ham}), we take a pragmatic approach. We first restrict the size of the Hilbert space based on physical considerations, e.g., by considering the low-energy sector,\footnote{Note that for complex energy eigenvalues one would first have to define what is meant by this.} and later on relax this restriction. This can also be used as a strategy for numerically determining the eigenvalues and eigenvectors. For an example see \cite{Grunwald2022} in which the non-Hermitian but $\mathcal{PT}$-symmetric Hamiltonian Eq.~(\ref{eq:sci_post_ham}) is investigated. With this restriction the spectrum is also guaranteed to be discrete, which is an additional asset when it comes to avoiding mathematical subtleties \cite{Mostafazadeh2002-gk,Mostafazadeh2003,Mostafazadeh2003a,Brody2013-br,Ashida2020-vp}. In many-body problems this can always be ensured by studying a finite system and taking the thermodynamic limit only at the end of the calculation.      

\subsection{The biorthonormal basis}
\label{subsec:bio}

To set up a proper extension of Hermitian quantum mechanics we assume that, for a given $H$, the set of right eigenvectors $\left\{ \left| \mbox{R}_\nu \right> \right\}$ and the set of left ones (i.e., the right eigenvectors of $H^\dag$) $\left\{ \left| \mbox{L}_\nu \right> \right\}$ each form a basis in $\mathcal{H}$. Or, put differently, we only accept operators as Hamiltonians which have this property. Needless to say that our example Hamiltonians introduced in Sect.~\ref{subsec:all_models} fall into this class. In the Hermitian case the existence of a basis consisting of eigenstates of $H$ is always ensured and the sets of right and left eigenvectors are identical.  As emphasized above, for non-Hermitian operators the states of each of the two sets are not guaranteed to be pairwise orthogonal \cite{Brody2013-br}. In case the Hamiltonian depends on parameters which can be tuned such that isolated exceptional points occur, the corresponding parameter sets leading to these must be excluded.

For non-degenerate spectra it is now straightforward to show that \cite{Brody2013-br}:
\begin{theorem}[$\mbox{T}^{\rm bio}_{1}$]
With a proper labeling and normalization of the eigenstates one can always ensure the
%has $\left< \mbox{L}_\nu \right.  \left| \mbox{R}_\mu \right> = 0$ if $\nu \neq \mu$ and  $\left< \mbox{L}_\nu \right.  \left| \mbox{R}_\nu \right> \neq 0$. After normalization we thus have the 
biorthonormality relation 
\begin{equation}
  \left< \mbox{L}_\nu \right.  \left| \mbox{R}_\mu \right>  = \delta_{\nu,\mu}.
\label{eq:biorthogonal}
\end{equation}
% From now on we will always assume that the eigenstates of the Hamiltonian are normalized in this way. 
In addition, one finds $\tilde E_\nu = E_\nu^\ast$; see Eq.~(\ref{eq:leftevdef}). \end{theorem}
\noindent From now on we will assume that the eigenstates of the Hamiltonian are normalized in this way. Employing the standard Gram-Schmidt process this can be generalized to the case of degenerate eigenvalues \cite{Mostafazadeh2002-gk}. With all this we end up with the completeness relation
\begin{equation}
\sum_{\nu}\left| \mbox{R}_\nu \right> \left<\mbox{L}_\nu \right|= {\mathbbm{1}} 
%\quad \Rightarrow \quad  \sum_{\nu}\left| \mbox{L}_\nu \right> \left<\mbox{R}_\nu %\right| = {\mathbbm{1}} 
\label{eq:complete}    
\end{equation}
and the $\left\{ \left| \mbox{R}_\nu \right>, \left| \mbox{L}_\nu \right>\right\} $ form a complete biorthonormal basis. In addition, we have the spectral representation
\begin{equation}
H = \sum_{\nu} E_\nu  \left| \mbox{R}_\nu \right> \left<\mbox{L}_\nu \right|
\quad \Rightarrow \quad H^\dag = \sum_{\nu} E_\nu^\ast  \left| \mbox{L}_\nu \right> \left<\mbox{R}_\nu \right| . 
\label{eq:specrepr}    
\end{equation}
Remember that if degeneracy matters, $\nu$ is a multi-index. Identifying $\left| \mbox{R}_\nu \right> = \left| \mbox{L}_\nu \right>$ we get back to the well known expressions in the Hermitian case.

\subsection{Pseudo-Hermiticity and \texorpdfstring{$\eta$}{ETA} operators}
\label{subsec:eta_op}

To determine, whether a given non-Hermitian Hamiltonian $H$ is pseudo-Hermitian, we have to find a linear, Hermitian operator $\eta$ which fulfills $\eta H \eta^{-1} = H^\dag$. As an ansatz for this let us consider\footnote{The reason for the index r will become clear soon.} \cite{Mostafazadeh2002-gk}
\begin{equation}
\eta_{\rm r} = \sum_{\nu}  \left| \mbox{L}_\nu \right>  \left< \mbox{L}_\nu \right| .
\label{eq:ansatz_eta}    
\end{equation}
This operator is obviously Hermitian. Employing the biorthonormality relation Eq.~(\ref{eq:biorthogonal}) and the completeness relation Eq.~(\ref{eq:complete}) one can show that
\begin{equation}
\eta_{\rm r}^{-1} = \sum_{\nu}  \left| \mbox{R}_\nu \right>  \left< \mbox{R}_\nu \right| .
\label{eq:ansatz_eta_inv}    
\end{equation}
With this and the spectral representation Eq.~(\ref{eq:specrepr})  we obtain
\begin{equation}
 \eta_{\rm r} H \eta_{\rm r}^{-1} =  \sum_{\nu,\mu,\kappa}  E_\mu \left| \mbox{L}_\nu \right>  \left< \mbox{L}_\nu \right. \left| \mbox{R}_\mu \right> \left<\mbox{L}_\mu \right.  \left| \mbox{R}_\kappa \right>  \left< \mbox{R}_\kappa \right| = \sum_{\nu} E_\nu \left| \mbox{L}_\nu \right> \left<\mbox{R}_\nu \right| .
\label{eq:anastz_eta_calc}    
\end{equation}
If the spectrum is real, the expression to the right of the last equal sign, is equal to $H^\dag$; see Eq.~(\ref{eq:specrepr}). For the case of an entirely real spectrum (explaining the index r of $\eta$) $H$ is therefore $\eta_{\rm r}$-pseudo-Hermitian
\begin{equation}
\eta_{\rm r} H \eta_{\rm r}^{-1} = H^\dag .
\label{eq:eta_doesit}    
\end{equation}
 As emphasized in Sect.~\ref{subsec:pseudoherm} the $\eta$ operator of a pseudo-Hermitian Hamiltonian depends in general on the Hamitonian; in the above ansatz via the eigenvectors. Note that Bender and coworkers avoid to introduce the operator $\eta_{\rm r}$. However, they introduce a linear operator $\mathcal C$ instead, which, in a proper combination with $\mathcal P$ and $\mathcal T$, plays the role of $\eta_{\rm r}$. In particular, also $\mathcal C$ depends on the Hamiltonian under investigation. We here do not further elaborate on this and refer the interested reader to \cite{Bender2019} and \cite{Brody2013-br}. 

We so far constructed a feasible $\eta$ for the case of an entirely real spectrum ($\mathcal{PT}$-symmetric phase in case of a $\mathcal{PT}$-symmetric Hamiltonian). In accordance with the theorem $\mbox{T}^{\eta}_{2}$, it is also possible to give a   $\eta_{\rm cp}$ for the case in which the eigenvalues of the non-Hermitian Hamiltonian are partly real and partly come in complex conjugate pairs ($\mathcal{PT}$ symmetry broken phase in case of a $\mathcal{PT}$-symmetric Hamiltonian).  Here the index cp stands for  ``complex pairs''. $\eta_{\rm cp}$ is constructed in close analogy to $\eta_{\rm r}$ but requires a slight extension of our notation. Let us denote the left eigenvectors to the real eigenvalues $E_{\nu_{\rm r}}$  as $ \left| \mbox{L}^{\rm r}_{\nu_{\rm r}} \right> $. We write the left eigenvectors of one pair $(E_{\nu_{c}}, E^\ast_{\nu_{c}})$ of complex conjugate eigenvalues as $ \left| \mbox{L}^{{\rm c}}_{\nu_{\rm c}} \right> $ and  $ \left| \mbox{L}^{{\rm c},\ast}_{\nu_{\rm c}} \right> $, respectively. Then \cite{Mostafazadeh2002-gk}
\begin{equation}
\eta_{\rm cp} = \sum_{\nu_{\rm r}}  \left| \mbox{L}^{\rm r}_{\nu_{\rm r}} \right>  \left< \mbox{L}^{\rm r}_{\nu_{\rm r}} \right| +  \sum_{\nu_{\rm c}}  \left( \left| \mbox{L}^{\rm c}_{\nu_{\rm c}} \right>  \left< \mbox{L}^{{\rm c},\ast}_{\nu_{\rm c}} \right| +  \left| \mbox{L}^{{\rm c},\ast}_{\nu_{\rm c}} \right>  \left< \mbox{L}^{{\rm c}}_{\nu_{\rm c}} \right|  \right).
\label{eq:ansatz_eta_cp}    
\end{equation}
The inverse $\eta_{\rm cp}^{-1}$ is obtained by replacing L $\to$ R, just as in going from Eq.~(\ref{eq:ansatz_eta}) to (\ref{eq:ansatz_eta_inv}).
After rewriting the spectral representation of $H$  Eq.~(\ref{eq:specrepr}) in the new notation, it is straightforward to see that Eq.~(\ref{eq:eta_doesit}) holds with $\eta_{\rm r} \to \eta_{\rm cp}$.

Let us assume that a general biorthonormal basis $\left\{\left|{\rm r}_\nu \right> , \left|{\rm l}_\nu \right>   \right\}$ is given which fulfills the orthonormalization relation Eq.~(\ref{eq:biorthogonal}) as well as the completeness relation Eq.~(\ref{eq:complete}). As these basis vectors are not necessarily the right and left eigenvectors of a non-Hermitian Hamiltonian $H$, we denote them with small letters instead of capital ones. In analogy to Eq.~(\ref{eq:ansatz_eta})  we define the linear operator \cite{Brody2013-br}
\begin{equation}
\hat g = \sum_{\nu}  \left| \mbox{l}_\nu \right>  \left< \mbox{l}_\nu \right| .
\label{eq:ansatz_hat_g}    
\end{equation}
If one applies $\hat g$ to a right basis state, it is mapped onto the corresponding left one
\begin{equation}
\hat g \left| \mbox{r}_\nu \right> = \sum_{\mu}  \left| \mbox{l}_\mu \right>  \left< \mbox{l}_\mu \right. \left| \mbox{r}_\nu \right> =  \left| \mbox{l}_\nu \right> ,
\label{eq:right_left_map}    
\end{equation}
were we employed the biorthonormality relation Eq.~(\ref{eq:biorthogonal}). With this and the property that $\left\{ \left| \mbox{r}_\nu \right> \right\}$ forms a (non-orthogonal) basis, i.e., $\left| \psi \right> = \sum_{\nu} a_\nu  \left| \mbox{r}_\nu \right>$, $a_\nu \in {\mathbbm{C}}$, we obtain
\begin{equation}
\left< \psi \right| \hat g \left| \psi \right> =  \sum_{\nu,\mu}  a_\nu^\ast a_{\mu} \left< \mbox{r}_\nu \right| \hat g \left| \mbox{r}_\mu \right> = \sum_{\nu} \left| a_\nu \right|^2 > 0 
\label{eq:eta_pos}    
\end{equation}
for an arbitrary state $ \left| \psi \right> \neq 0$ from $\mathcal{H}$. Thus $\hat g$ is a positive definite operator. 

For the special case that the biorthonormal basis vectors are eigenvectors of a non-Hermitian Hamiltonian we have $\eta_{\rm r} = \hat g$. Thus $\eta_{\rm r}$ is positive definite and the square root $\eta_{\rm r}^{1/2}$ can be taken. This Hermitian operator $\eta_{\rm r}^{1/2}$ is the one of theorem $\mbox{T}^{\eta}_{3}$. In contrast, $\eta_{\rm cp}$ is not a positive definite operator (for an example, see Sect.~\ref{subsec:examples_2_2})  \cite{Mostafazadeh2002-gk,Brody2013-br} . The positivity of $\eta_{\rm r}$ (or $\hat g$) can be exploited to define an alternative inner product \cite{Pauli1943,Scholtz1992,Bender2019,Brody2013-br}. Further down we will return to this.

We next introduce our three model systems. The first is the non-Hermitian two-level problem represented by a simple non-Hermitian $2 \times 2$ matrix. This we use to illustrate all the general concepts and theorems of $\mathcal{PT}$-symmetric and pseudo-Hermitian non-Hermitian quantum theory discussed so far. We also discuss its interesting dynamical properties. For the other two models, the resonant level model with complex hybridization and the staggered tight-binding chain with complex hopping, we focus primarily on the physics. However, we partly also use them to demonstrate some of the general ideas in a more complex setting. When introducing these two models, we first aim at their single-particle properties. In Sects.~\ref{sec:obexp} and \ref{sec:genfun} we will extend this and study the non-Hermitian many-body physics of those. 

\subsection{Models}
\label{subsec:all_models}

\subsubsection{A non-Hermitian two-level problem}
\label{subsec:examples_2_2}

\begin{figure}
    \centering
    \includegraphics{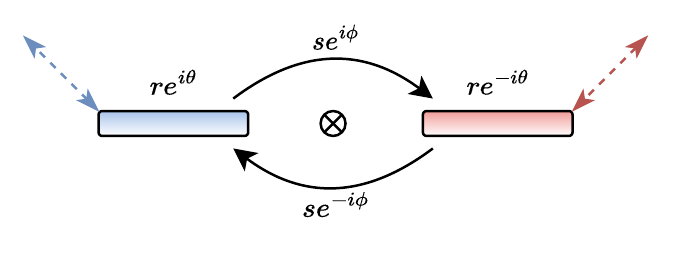}
    \caption{\textbf{Sketch of the $\bm{\mathcal{PT}}$-symmetric two-level system} with non-Hermitian Hamiltonian \cref{eq:spin_1_2}. It describes two sites with complex ``energy'' $r e^{\pm i \theta}$ (red and blue) that are connected by a complex hopping $s e^{\pm i \phi}$. The on-site matrix elements describe balanced gain and loss via the coupling to an environment, indicated by the dashed arrows and leading to the phase $e^{\pm i \theta}$. The phase of the hopping $e^{\pm i \phi}$ originates from a perpendicular magnetic flux indicated by the ``$\otimes$''.}
    \label{fig:two_level_hop_mod}
\end{figure}

As our simplest toy problem we study a single quantum particle hopping between two lattice sites. The orthonormal basis states in which the particle occupies the left or the right lattice site are denoted by $\left| \uparrow \right>$ and $\left| \downarrow \right>$, respectively. This notation is reminiscent of the often employed interpretation of this model in terms of a single spin-1/2 degree of freedom. The  onsite ``energies'' are set to $re^{\pm i \theta}$ with the complex phase resulting out of a gain and loss to a reservoir \cite{Bender2019,Ashida2020-vp}. The hopping amplitude $s e^{\pm i \phi}$ is assumed to be complex as well. Its phase can be understood as originating from a magnetic vector potential in the Peierls substitution \cite{Peierls1933}. The model is sketched in Fig.~\ref{fig:two_level_hop_mod}.
In the up-down-basis the Hamiltonian reads
\begin{equation}
H(\phi) \doteq \left( 
\begin{array}{cc}
r e^{i \theta} & s e^{i \phi} \\
s e^{-i \phi} & r e^{-i \theta}  
\end{array} 
\right) 
%\quad  \Rightarrow \quad H^\dag(\phi) \doteq \left( 
%\begin{array}{cc}
%r e^{-i \theta} & s e^{i \phi} \\
%s e^{-i \phi} & r e^{i \theta}  
%\end{array} 
%\right) .
\label{eq:spin_1_2}    
\end{equation}
It depends on the real variable $\phi$ and the three real parameters $r$, $s$, and $\theta$. We distinguish between a variable and parameters as we will further down investigate $H(\phi)$ at the special point $\phi=0$ but for arbitrary $r$, $s$, and $\theta$. For $r \neq 0$ and $\theta \neq n \pi$, with $n \in {\mathbbm{N}}$ the Hamiltonian is non-Hermitian. The eigenvalues are given by
\begin{equation}
\frac{E_{\pm}}{s} = \frac{r}{s} \cos{\theta} \pm \left\{ \begin{array}{ll}
\sqrt{1- \left(  \frac{r}{s}\right)^2 \sin^2{\theta}} & \mbox{for}\,  \left|  \frac{r}{s} \sin{\theta} \right| < 1  \\
i \sqrt{\left(  \frac{r}{s}\right)^2 \sin^2{\theta} -1 } &  \mbox{for}\,  \left| \frac{r}{s} \sin{\theta} \right| > 1.  \end{array} \right.
\label{eq:spin_1_2_eigenvalues}    
\end{equation}
They are independent of the variable $\phi$. The overall energy scale is set by $s$. The remaining two dimensionless parameters are $r/s$ and $\theta$. We introduce
\begin{align}
z = \frac{r}{s} \sin{\theta} \in {\mathbb R}.
\label{eq:spin_1_2_zdef}
\end{align}
For $|z| < 1$ the spectrum is real, while a pair of complex conjugate eigenvalues is found for $|z| >1$. For $|z|=1$ the two real eigenvalues are degenerate and for $z=0$ the Hamiltonian is Hermitian. The right and left eigenvectors are given by
\begin{align}
\left| \mbox{R}_\pm \right> &=
\frac{ e^{i \phi} \left(i z \pm \sqrt{1-z^2} \right)  \left| \uparrow \right> + \left| \downarrow \right>}{\left[ 1+ \left(i z \pm \sqrt{1-z^2} \right)^2 \right]^{1/2} } ,
\label{eq:spin_1_2_eigenvectors_1_R}   \\
\left| \mbox{L}_\pm \right> 
&=
\frac{ e^{i \phi} \left(-i z \pm \sqrt{1-z^2} \right)  \left| \uparrow \right> + \left| \downarrow \right>}{\left[ 1+ \left(-i z \pm \sqrt{1-z^2} \right)^2 \right]^{1/2} } 
\label{eq:spin_1_2_eigenvectors_1_L}    
\end{align}
for $|z|<1$ and
\begin{align}
\left| \mbox{R}_\pm \right> 
&=
\frac{i e^{i \phi} \left(z \pm \sqrt{z^2-1} \right)  \left| \uparrow \right> + \left| \downarrow \right>}{\left[1-\left(z \pm \sqrt{z^2-1} \right)^2 \right]^{1/2} } , 
\label{eq:spin_1_2_eigenvectors_2_R}  \\
\left| \mbox{L}_\pm \right> 
&=
\frac{-i   e^{i \phi} \left(z \pm \sqrt{z^2-1} \right)  \left| \uparrow \right> + \left| \downarrow \right>}{\left\{\left[1-\left(z \pm \sqrt{z^2-1} \right)^2\right]^{1/2} \right\}^\ast } 
\label{eq:spin_1_2_eigenvectors_2_L}    
\end{align}
for  $|z|>1$. It is easy to verify that $\left| \mbox{R}_+ \right> $ and $\left| \mbox{R}_- \right> $ are linearly independent and thus form a basis of  $\mathbbm{C}^2$, but that they are not orthogonal. The same holds for  $\left| \mbox{L}_+ \right> $ and $\left| \mbox{L}_- \right> $. However, $\left< \mbox{L}_\nu \right. \left| \mbox{R}_\mu \right> = \delta_{\nu,\mu}$ as in Eq.~(\ref{eq:biorthogonal}) due to a proper normalization. For $|z|=1$ the eigenvectors for  $\nu=+$ and $\nu=-$ are equal and cannot be used to span the Hilbert space $\mathbbm{C}^2$, i.e., the eigenvectors of the Hamiltonian do no longer form a basis. This corresponds to an exceptional point. The resolution of unity Eq.~(\ref{eq:complete}) and the spectral representation of the Hamiltonian Eq.~(\ref{eq:specrepr}) can also be verified explicitly. In the Hermitian case $z=0$ the right and left eigenvectors Eqs.~(\ref{eq:spin_1_2_eigenvectors_1_R})  and (\ref{eq:spin_1_2_eigenvectors_1_L}) are equal.

To analyze the spectral properties of the Hamiltonian Eq.~(\ref{eq:spin_1_2}) from the perspective of the theory of $\mathcal{PT}$-symmetric quantum systems, we first have to investigate if $H(\phi)$ is $\mathcal{PT}$-symmetric. With the above interpretation of the Hamiltonian as that of a particle hopping between a left and a right lattice site, it is evident, that the parity operator ${\mathcal{P}}$, which is supposed to interchange left and right, is represented by \cite{Bender2005-py}
\begin{equation}
{\mathcal{P}} \doteq  \left( 
\begin{array}{cc}
0 & 1 \\
1 & 0   
\end{array} 
\right) = \sigma_x ,
\label{eq:spin_1_2_parity}   
\end{equation}
with the Pauli matrix $\sigma_x$. With $\mathcal{T}$ performing complex conjugation it is straightforward to show that
\begin{equation}
\left[\mathcal{PT},H\right] =0
\label{eq:spin_1_2_PTinvar}    
\end{equation}
and thus the Hamiltonian Eq.~(\ref{eq:spin_1_2}) is ${\mathcal{PT}}$-symmetric. 

It is easy to show that for $|z|<1$ the right eigenvectors $\left| \mbox{R}_\pm \right>$ Eq.~(\ref{eq:spin_1_2_eigenvectors_1_R}) are also eigenvectors of ${\mathcal{PT}}$ with eigenvalues $e^{-i \phi} \left(i z \pm \sqrt{1-z^2}\right)$.\footnote{Note that the absolute value of the eigenvalue is 1. This holds on general grounds \cite{Bender2007-tf}.} This does no longer hold for $|z|>1$. These findings are in accordance with the theorem $\mbox{T}^{\mathcal{PT}}_{1}$. For $|z|>1$,  ${\mathcal{PT}}$ symmetry is spontaneously broken and the exceptional point $|z|=1$ is the point of a ${\mathcal{PT}}$ transition.  

For our two-level toy model it is also possible to determine an exact expression for the similarity transformation $S$ which, in the ${\mathcal{PT}}$ symmetry unbroken phase $|z|<1$, maps $H$ to an isospectral Hermitian Hamiltonian $h = S H S^{-1}$; see theorem $\mbox{T}^{\mathcal{PT}}_{2}$. We postpone the explicit construction of $S$ and $h$ until we analyzed the spectral properties of the Hamiltonian Eq.~(\ref{eq:spin_1_2}) within the framework of pseudo-Hermiticity; this will significantly simplify the construction. Employing the result that the eigenvalues Eq.~(\ref{eq:spin_1_2_eigenvalues}) of the Hamiltonian Eq.~(\ref{eq:spin_1_2}) are either real or form  complex conjugate pairs and the theorem $\mbox{T}^{\eta}_{2}$ or, alternatively, theorem  $\mbox{T}^{\eta}_{4}$, which states that every $\mathcal{PT}$-symmetric Hamiltonian is also pseudo-Hermitian, we know that the Hamiltonian $H(\phi)$ is pseudo-Hermitian (in both phases). We next aim to determine appropriate $\eta$'s. 

For $\phi=0$, Eq.~(\ref{eq:spin_1_2}) is a complex, symmetric, ${\mathcal{PT}}$-symmetric matrix and the theorem $\mbox{T}^{\eta}_{1}$ ensures that $H(0)$ is $\mathcal{P}$-pseudo-Hermitian. It is indeed straightforward to show that with
\begin{equation}
\eta = \mathcal{P} \doteq   \left( 
\begin{array}{cc}
0 & 1 \\
1 & 0   
\end{array} 
\right) \doteq \eta^\dag = \eta^{-1}   
\label{eq:spin_1_2_eta_1}    
\end{equation}
one finds
\begin{equation}
\eta H(0) \eta^{-1} = H^\dag(0)  .  
\label{eq:spin_1_2_pseudo_1}    
\end{equation}
Note that this holds in both phases $|z|<1$ and $|z|>1$.
%If the two off-diagonal (hopping) elements of Eq.~(\ref{eq:spin_1_2}) are replaced by $s e^{i \phi}$ and $s e^{- i \phi}$, respectively,\footnote{In the interpretation of the model as presenting a single quantum particle hopping between two lattice sites.} the resulting Hamiltonian 
%\begin{equation}
%\tilde H \doteq \left( 
%\begin{array}{cc}
%r e^{i \theta} & s e^{i \phi} \\
%s e^{- i \phi} & r e^{-i \theta}  
%\end{array} 
%\right)
%\label{eq:spin_1_2_tilde}    
%\end{equation}
%is still $\mathcal{PT}$-symmetric---Eq.~(\ref{eq:spin_1_2_PTinvar}) also holds with $H$ replaced by $\tilde H$---but no longer $\mathcal{P}$-pseudo-Hermitian; the matrix is no longer symmetric. Remarkably, $\tilde H$ has the same eigenvalues Eq.~(\ref{eq:spin_1_2_eigenvalues}) as $H$ and applying theorem $\mbox{T}^{\eta}_{2}$ we know that $\tilde H$ is pseudo-Hermitian. The eigenvectors of $\tilde H$ follow from Eqs.~(\ref{eq:spin_1_2_eigenvectors_1_R})-(\ref{eq:spin_1_2_eigenvectors_2_L}) by multiplying the prefactor of the up-spin component by $e^{i\phi}$.

To explicitly determine a $\eta$ for $H(\phi)$ and all $\phi$ we first consider the $\mathcal{PT}$-symmetric phase $|z|<1$ with real eigenvalues and employ Eq.~(\ref{eq:ansatz_eta}). Using Eq.~(\ref{eq:spin_1_2_eigenvectors_1_L}) for the left eigenvectors we obtain in the orthonormal basis of spin-up and -down states 
\begin{equation}
\eta_{\rm r} \doteq 
 \frac{1}{\sqrt{1-z^2}} \left( 
\begin{array}{cc}
1 & -i z e^{i \phi} \\
i z e^{- i \phi} & 1  
\end{array} 
\right) .
\label{eq:spin_1_2_eta_a}    
\end{equation}
This matrix is obviously Hermitian: $\eta_{\rm r}=\eta_{\rm r}^\dag$. It becomes the unity matrix in the Hermitian limit $z=0$. It is easy to check explicitly that for $|z| <1$ the pseudo-Hermiticity relation $\eta_{\rm r} H(\phi) \eta_{\rm r}^{-1} = H^\dag(\phi)$ holds. For $\phi=0$, $\eta_{\rm r}$ Eq.~(\ref{eq:spin_1_2_eta_a}) and $\eta$ Eq.~(\ref{eq:spin_1_2_eta_1}) do not coincide. This gives an explicit example that the $\eta$ operator of a pseudo-Hermitian Hamiltonian is not necessarily unique. We can now take the square root of the positive operator $\eta_{\rm r}$ and obtain (see theorem $\mbox{T}^{\eta}_{3}$)
\begin{equation}
\eta_{\rm r}^{1/2} \doteq 
 \frac{1}{\left[ 1-z^2 \right]^{1/4}} \left( 
\begin{array}{cc}
\left[ \frac{1}{2} + \frac{1}{2} \sqrt{1-z^2} \right]^{1/2}& 
-i \left[ \frac{1}{2} - \frac{1}{2} \sqrt{1-z^2} \right]^{1/2}  e^{i \phi} \\
i \left[ \frac{1}{2} - \frac{1}{2} \sqrt{1-z^2} \right]^{1/2} e^{- i \phi} &  \left[ \frac{1}{2} + \frac{1}{2} \sqrt{1-z^2} \right]^{1/2}
\end{array} 
\right) \!\! .
\label{eq:spin_1_2_O}    
\end{equation}
Note that in contrast to $\eta_{\rm r}$, $\eta=\mathcal{P}=\sigma_x$ is not a positive operator (remember that the eigenvalues of $\sigma_x$ are $\pm 1$) and the square root cannot be taken. We next compute 
\begin{equation}
\eta_{\rm r}^{1/2} H(\phi) \eta_{\rm r}^{-1/2} \doteq  \left( 
\begin{array}{cc}
r \cos{\theta} & 
s \sqrt{1-z^2} e^{i \phi} \\ s \sqrt{1-z^2} e^{-i \phi} & r \cos{\theta} \end{array} 
\right) \doteq  h(\phi).
\label{eq:spin_1_2_h}    
\end{equation}
$h(\phi)$ is represented by a Hermitian matrix which has the same eigenvalues Eq.~(\ref{eq:spin_1_2_eigenvalues}) as $H(\phi)$. This illustrates that the operator $\eta^{1/2}$ of theorem $\mbox{T}^{\eta}_{3}$ coincides with $S$ of theorem $\mbox{T}^{\mathcal{PT}}_{2}$. As announced above for the present model it was thus easy to exactly determine $S$ and $h$ of $\mbox{T}^{\mathcal{PT}}_{2}$ by employing pseudo-Hermiticity.

We now turn to the $\mathcal{PT}$ symmetry broken phase with $|z|>1$. Using the eigenvectors Eq.~(\ref{eq:spin_1_2_eigenvectors_2_L}) and the general expression  Eq.~(\ref{eq:ansatz_eta_cp}) of $\eta_{\rm cp}$ for the case of complex conjugate pairs of eigenvalues, the matrix form is
\begin{equation}
\eta_{\rm cp} \doteq 
 \mbox{sgn}(z) \left( 
\begin{array}{cc}
0 & e^{i \phi} \\
e^{- i \phi} & 0  
\end{array} 
\right) .
\label{eq:spin_1_2_eta_cp}    
\end{equation}
Straightforward matrix multiplication gives $\eta_{\rm cp} H \eta_{\rm cp}^{-1} = H^\dag$ for $|z|>1$. Also in this phase $\eta=\mathcal{P}$ and $\eta_{\rm cp}$ for $\phi=0$ do not coincide if $z<-1$. However, they coincide for $z>1$. As $\eta_{\rm cp}$ has eigenvalues $\pm 1$, it is not positive definite, as stated at the end of Sect.~\ref{subsec:eta_op} for the general case.    

\subsubsection{The resonant level model with complex hybridization}
\label{subsec:examples_RL}

\begin{figure}
    \centering
    \includegraphics[width = 13cm]{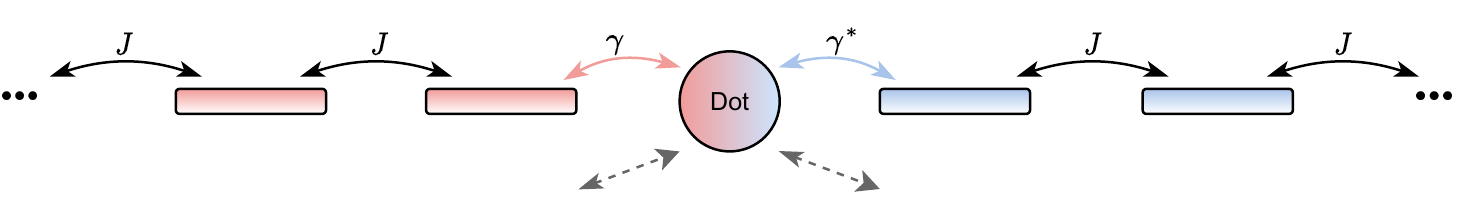}
    \caption{\textbf{Sketch of the $\bm{\mathcal{PT}}$-symmetric resonant level model} with non-Hermitian Hamiltonian \cref{eq:RL_ham}. The quantum dot couples to leads with complex hopping amplitude $\gamma$ (left) and $\gamma^\ast$ (right). The leads themselves are Hermitian and modeled as a fermionic chain with hopping $J$. The non-Hermiticity is induced by a coupling of the dot to the environment, which is indicated by the dashed arrows.}
    \label{fig:RL_model}
\end{figure}

The resonant level model describes a single quantum level which is tunnel coupled to two leads. The fermionic particle(s) residing on the level and in the leads are assumed to be spinless. In mesoscopic physics the model is often considered as a simple toy model to study resonant transport through a quantum dot. We here use a version of the resonant level model in which the leads are given by tight-binding chains with hopping amplitude $J>0$ and lattice constant $a=1$. The Hamiltonian is
\begin{equation}
H =  -J \sum_{j=-N/2}^{-2} \left( c_j^\dag c_{j+1} + \mbox{H.c.} \right) - \gamma \left( c_{-1}^\dag c_0 + \mbox{H.c.} \right)  
- \gamma^\ast \left( c_{0}^\dag c_1 + \mbox{H.c.} \right)  -J    \sum_{j=1}^{N/2-1} \left( c_j^\dag c_{j+1} + \mbox{H.c.} \right)  . 
\label{eq:RL_ham} 
\end{equation}
Here $j$ denotes the lattice site index and $c_j^{(\dag)}$ the annihilation (creation) operator of a particle on site $j$ in standard second quantization notation.  The Wannier states $\left\{ \left| j \right> \right\}$ form an orthonormal basis of the single-particle Hilbert space. In this section we discuss the problem of a single particle and we could avoid second quantization. In Sect.~\ref{sec:obexp} we will, however, be interested in the many-particle problem. This explains why we already now use this notation. Each lead has $N/2$ lattice sites and open boundary conditions at the end opposing the dot site. Combined with the dot site at $j=0$ the system has $N+1$ lattice sites in total. The tunnel coupling of the dot level to the left lead is given by $\gamma$. In the Hermitian resonant level model $\gamma$ is taken to be real and one usually assumes $\gamma \ll J$ to describe resonance phenomena. We obtain a non-Hermitian model if $\gamma = \gamma_{\rm r} + i \gamma_{\rm i} \in {\mathbbm{C}}$ is assumed (with $\gamma_{\rm r/i} \geq 0$). Such complex hopping amplitude could result from the local coupling of the lattice model to an additional environment (not to be confused with the two leads of the model) and a corresponding gain and loss. If the tunnel coupling between the level and the right lead is taken to be $\gamma^\ast$, the non-Hermitian Hamiltonian Eq.~(\ref{eq:RL_ham}) is obviously $\mathcal{PT}$-symmetric (reflection at site $j=0$ and complex conjugation). The model is sketched in \cref{fig:RL_model}.   

The single-particle left and right eigenstates of $H$ Eq.~(\ref{eq:RL_ham})  fall into two classes. The first is made of the scattering states which are spatially delocalized and have real eigenvalues. They appear for all $\gamma_{\rm r/i}$. The second class are bound states exponentially localized around the quantum dot, which appear in certain parameter regimes. For $\gamma_{\rm r} > \sqrt{J^2+\gamma_{\rm i}^2}$ 
% they
we find two bound states with
% have a
real eigenvalues while for $\gamma_{\rm i} > \gamma_{\rm r}$ their eigenvalues are purely imaginary and complex conjugate to each other. In this parameter regime the ${\mathcal PT}$ symmetry is spontaneously broken. In the following we elaborate on the results for the eigenvalues and eigenvectors which are necessary for our later considerations. Details on the derivation can be found in \cite{Yoshimura2020}.

Using a standard standing wave ansatz (open boundary conditions) including a scattering phase shift, for the right eigenstates in the Wannier (position) basis representation 
\begin{equation}
  \left< j_{\phantom{k}} \!\!\!\!  \right. \left| {\rm R}_k \right> = \left\{ \begin{array}{ll} 
  N_{\rm r} \left( e^{ikj} + e^{i 2 k \delta(k)} e^{-i kj}  \right) & j=1,2,3,\ldots \\
   N_{\rm l} \left( e^{ikj} + e^{-i 2 k \delta(k)} e^{-i kj}  \right) & j=-1,-2,-3,\ldots \end{array} \right. ,
\label{eq:scat_ansatz}    
\end{equation}
one obtains for the eigenvalues of the scattering states the usual real band energies of a one-dimensional tight-binding chain
\begin{equation}
E_k = -2 J \cos{k} ,
\label{eq:RL_ss_disp}    
\end{equation}
where we replaced the general (quantum number multi-) index $\nu$ used so far by the momentum $k$. The quantization condition for $k$ follows from the boundary conditions at the end of the two leads opposing the dot site. These manifest in the implicit equation
\begin{equation}
e^{ikN} = \frac{\Gamma_\lambda -J^2 e^{-2ik}}{\Gamma_\lambda - J^2 e^{2ik}} ,
\label{eq:RL_ss_k_quant}    
\end{equation}
with $\lambda = \pm 1$ and
\begin{equation}
\Gamma_\lambda = (1+\lambda) \tilde \Gamma -J^2 , \quad \tilde \Gamma = \frac{\gamma^2 + (\gamma^\ast)^2}{2} .
\label{eq:RL_ss_Gamdef}    
\end{equation}
For fixed $N$, $\gamma$, and $J$ one thus finds two sets of allowed momenta. To avoid an overloading of the notation, we suppress the index $\lambda$ as well as one for the discreetness of $k$. The right eigenstates are 
\begin{equation}
\left< j^{\phantom{\lambda}}_{\phantom{k}} \!\!\!\!  \right. \left| {\rm R}_k^\lambda\right> = N_{\rm R}^\lambda \times \left\{ \begin{array}{ll} 
\cos{\left( k  \left[ j-\delta_\lambda(k) \right]\right)} & j=1,2,3,\ldots \\
M_\lambda \cos{\left( k \left[ j+\delta_\lambda(k)  \right]\right)} & j=-1,-2,-3,\ldots \end{array} \right. , 
\label{eq:RL_ss_rstates}    
\end{equation}
with the two phase shifts $\delta_\lambda(k)$ determined by
\begin{equation}
e^{i 2 k \delta_{\lambda}(k)} = \frac{J^2-\Gamma_\lambda e^{i2k}}{\Gamma_\lambda - J^2 e^{2ik}} 
\label{eq:RL_ss_phaseshift}    
\end{equation}
and $M_{+}= \gamma^2/|\gamma|^2$ as well as $M_{-}=- (\gamma^\ast)^2/|\gamma|^2$. For notational ease we here and in the following suppressed the $1$ when replacing $\lambda$ in an index by one of the two options $\pm 1$. 
For $\lambda=-1$ one finds $\Gamma_{-} = - J^2$ and $e^{i 2 k \delta_{-}(k)} = -1$. The phase shift is thus $\delta_{-}(k)=\pi/(2 k)$. Inserting this in Eq.~(\ref{eq:RL_ss_rstates}) the $j$ dependence is given by $\sin{(kj)}$. This is the odd solution of the scattering problem with a vanishing amplitude on the dot site 
\begin{equation}
\left< 0^{\phantom{\lambda}}_{\phantom{k}} \!\!\!\!  \right. \left| {\rm R}_k^{-}\right> =0 .
\label{eq:RL_ss_odd0} 
\end{equation}
For the $\lambda=+1$ solution the phase shift is non-trivial and must be determined by solving Eq.~(\ref{eq:RL_ss_phaseshift}). The amplitude on the dot site is give by 
\begin{equation}
\left< 0^{\phantom{\lambda}}_{\phantom{k}} \!\!\!\!  \right. \left| {\rm R}_k^{+}\right> = -  N_{\rm R}^{+} \cos{\left(k \left[ 1- \delta_{+}\right] \right)} \left( \gamma M_{+} + \gamma^\ast  \right) / E_k .
\label{eq:RL_ss_even0} 
\end{equation}
With this the right scattering eigenstates are fully determined up to the overall normalization constant $N_{\rm R}^\lambda$. This can be used to achieve Eq.~(\ref{eq:biorthogonal}); see below. The scattering state solutions of the time-independent Schr\"odinger equation (eigenvalue problem) occur for arbitrary $(\gamma_{\rm r},\gamma_{\rm i})$. 

As in Eq.~(\ref{eq:RL_ham}), $H^\dag = H(\gamma \to \gamma^\ast)$ (in obvious notation), the left scattering eigenstates follow from the above expressions by interchanging $\gamma \leftrightarrow \gamma^\ast$ and replacing $N_{\rm R}^\lambda \to N_{\rm L}^\lambda$. Biorthonormalization  Eq.~(\ref{eq:biorthogonal}) is achieved for 
\begin{equation}
N_{\rm R}^{-} = \frac{\gamma}{\sqrt{2 \tilde \Gamma}} \left[ \frac{1}{2} + \frac{N}{4} \right]^{-1/2}, \quad N_{\rm R}^{+} =  \frac{\gamma^\ast}{ \sqrt{ 2 \tilde \Gamma}}  \left[ \frac{1}{2} + \frac{N}{4} + \frac{2 \tilde \Gamma (J^2 - \tilde \Gamma)}{\Delta_{+}(k)} \right]^{-1/2} ,
\label{eq:RL_ss_biortho}    
\end{equation}
with 
\begin{equation}
\Delta_{+}(k) = \Gamma_{+}^2 + J^4 - 2 \Gamma_{+} J^2 \cos{(2 k)}
\label{eq:RL_ss_Delta+def}    
\end{equation}
and $N_{\rm L}^{\lambda} = \left(N_{\rm R}^{\lambda} \right)^\ast $. Employing the latter we obtain $\left< j^{\phantom{\lambda}}_{\phantom{k}} \!\!\!\!  \right. \left| {\rm L}_k^{\lambda}\right> = \left< j^{\phantom{\lambda}}_{\phantom{k}} \!\!\!\!  \right. \left| {\rm R}_k^{\lambda}\right>^\ast$. In the absence of bound states---see the next paragraph---the $\left\{ \left| {\rm R}_k^{\lambda}\right>,  \left| {\rm L}_k^{\lambda}\right>\right\} $ form a biorthonormal basis.

Next we elaborate on bound state solutions of the  time-independent Schr\"odinger equation. As it is difficult to give their analytical form in the presence of the open boundaries of the leads at $j=\pm N/2$, we take the thermodynamic limit $N \to \infty$. By doing so, we relax the constraints we discussed in the beginning of the present section. However, this does not lead to any mathematical problems as we are only doing it for the discrete bound states. In Sect.~\ref{sec:obexp} when computing expectation values of observables we will, in addition, compute the bound states numerically for finite $N$. For $\gamma_{\rm r} > \sqrt{J^2 + \gamma_{\rm i}^2}$ the ansatz of an exponentially decaying wave function for increasing $|j|$ gives the real eigenenergies\footnote{Note the different sign as compared to Eq.~(19) of \cite{Yoshimura2020}.} 
\begin{equation} 
E_{\pm}^{\rm r} = \mp \frac{\gamma^2 + (\gamma^\ast)^2}{\sqrt{\gamma^2 + (\gamma^\ast)^2 -J^2}}  . 
\label{eq:RL_br_spec}    
\end{equation}
These are solutions which are also present if $\gamma$ is real ($\gamma_{\rm i} =0$) in case the tunnel coupling is larger than the hopping in the leads $\gamma_{\rm r} > J$.  In mesoscopic physics this parameter regime is usually of no interest as the resonance width $\sim \gamma_{\rm r}^2/J$ is larger than the band width $\sim J$ (and one can no longer speak of a resonance). The energies of the two bound states are located above and below the band edges $\pm 2 J$. The right eigenstate wave function is given by
\begin{equation}
\left< j^{\phantom{\lambda}}_{\phantom{k}} \!\!\!\!  \right. \left| {\rm R}_\pm^{\rm r}\right> = N_{\rm R}^{\rm r} \times \left\{ \begin{array}{ll} 
\left( \pm 1\right)^j e^{-j/\xi} & j=1,2,3,\ldots \\
\frac{J}{\gamma^\ast} & j=0 \\
M_+ \left( \pm 1\right)^j e^{j/\xi}  & j=-1,-2,-3,\ldots \end{array} \right. ,
\label{eq:RL_br_Rstates}    
\end{equation}
with the localization length $\xi = 2 / \ln{|\Gamma_+/J^2|}$.
For the left eigenstates it holds $\left< j^{\phantom{\lambda}}_{\phantom{k}} \!\!\!\!  \right. \left| {\rm L}_\pm^{\rm r}\right> = \left< j^{\phantom{\lambda}}_{\phantom{k}} \!\!\!\!  \right. \left| {\rm R}_\pm^{\rm r}\right>^\ast$.\footnote{Note the typo in Eq.~(18) of \cite{Yoshimura2020}.} Biorthonormalization is achieved for
\begin{equation}
N_{\rm R}^{\rm r} = \frac{\gamma^\ast}{J} \sqrt{\frac{J^2 -\tilde \Gamma}{J^2 - 2 \tilde \Gamma}} = \left(N_{\rm L}^{\rm i} \right)^\ast
\label{eq:RL_bi_biortho}    
\end{equation}
To obtain a biorthonormal basis in this parameter regime, the set $\left\{ \left| {\rm R}_k^{\lambda}\right>,  \left| {\rm L}_k^{\lambda}\right>\right\} $ has to be supplemented by these bound states.

A second type of bound state solutions is found for $\gamma_{\rm i} > \gamma_{\rm r}$. These have purely imaginary energies
\begin{equation} 
E_{\pm}^{\rm i} = \mp i \frac{\gamma^2 + (\gamma^\ast)^2}{\sqrt{J^2 - \gamma^2 - (\gamma^\ast)^2}}   
\label{eq:RL_bi_spec}    
\end{equation}
and right eigenstate wave functions
\begin{equation}
\left< j^{\phantom{\lambda}}_{\phantom{k}} \!\!\!\!  \right. \left| {\rm R}_\pm^{\rm i}\right> = N_{\rm R}^{\rm i} \times \left\{ \begin{array}{ll} 
\left( \pm i \right)^j e^{-j/\xi} & j=1,2,3,\ldots \\
\frac{J}{\gamma^\ast} & j=0 \\
M_+ \left( \mp i\right)^j e^{j/\xi}  & j=-1,-2,-3,\ldots \end{array} \right. .
\label{eq:RL_bi_Rstates}    
\end{equation}
The left eigenstates in position representation again follow from complex conjugation: $\left< j^{\phantom{\lambda}}_{\phantom{k}} \!\!\!\!  \right. \left| {\rm L}_\pm^{\rm i}\right> = \left< j^{\phantom{\lambda}}_{\phantom{k}} \!\!\!\!  \right. \left| {\rm R}_\pm^{\rm i}\right>^\ast$ and biorthonormalization requires 
\begin{equation}
N_{\rm R}^{\rm i} = \frac{\gamma^\ast}{J} \sqrt{\frac{\tilde \Gamma - J^2}{2 \tilde \Gamma - J^2}} = \left(N_{\rm L}^{\rm r} \right)^\ast .
\label{eq:RL_br_biortho}    
\end{equation}
The biorthonormal basis for $\gamma_{\rm i} > \gamma_{\rm r}$ is obtained by adding the  bound states with imaginary energy to  $\left\{ \left| {\rm R}_k^{\lambda}\right>,  \left| {\rm L}_k^{\lambda}\right>\right\} $.

For the single-particle $\mathcal{PT}$-symmetric resonant level model with complex hybridization $\gamma= \gamma_{\rm r} + i \gamma_{\rm i}$ one thus finds a spontaneous breaking of the $\mathcal{PT}$ symmetry for $\gamma_{\rm i} > \gamma_{\rm r}$. In the symmetry broken phase a (purely imaginary) pair of complex conjugate energies is found. To apply the spatial reflection $\mathcal P$ to an eigenstate one simply has to replace $\gamma \to \gamma^\ast$. Applying $\mathcal{T}$ leads to complex conjugation. Using the above explicit expressions Eqs.~(\ref{eq:RL_ss_rstates}) and (\ref{eq:RL_br_Rstates}) for the scattering states and the bound states with real energies, respectively, it is straightforward to see that they are eigenstates of $\mathcal{PT}$; compare to theorem $\mbox{T}^{\mathcal{PT}}_{1}$. This does no longer hold for the bound states with imaginary energy Eq.~(\ref{eq:RL_bi_Rstates}) as $i$ explicitly appears in the wave function.

According to theorems $\mbox{T}^{\eta}_{4}$ or $\mbox{T}^{\eta}_{2}$ the Hamiltonian is also pseudo-Hermitian. Corresponding $\eta$ operators can be obtained from the general expressions Eqs.~(\ref{eq:ansatz_eta}) and (\ref{eq:ansatz_eta_cp}) but are not constructed explicitly here. We neither attempt to explicitly determine the operators $\eta_{\rm r}^{1/2}$ of theorem  $\mbox{T}^{\eta}_{3}$ nor $S$ and the associated Hermitian Hamiltonian $h$ (with the same spectrum as $H$) of theorem $\mbox{T}^{\mathcal{PT}}_{2}$ in the symmetry unbroken parameter regime $\gamma_{\rm i} < \gamma_{\rm r}$. However, from these theorems we know that they must exist.  

% The staggered tight-binding chain with complex hopping
\subsubsection{The staggered tight-binding chain with complex hopping}
\label{subsec:examples_stag}

\begin{figure}
    \centering
    \includegraphics[width = 13cm]{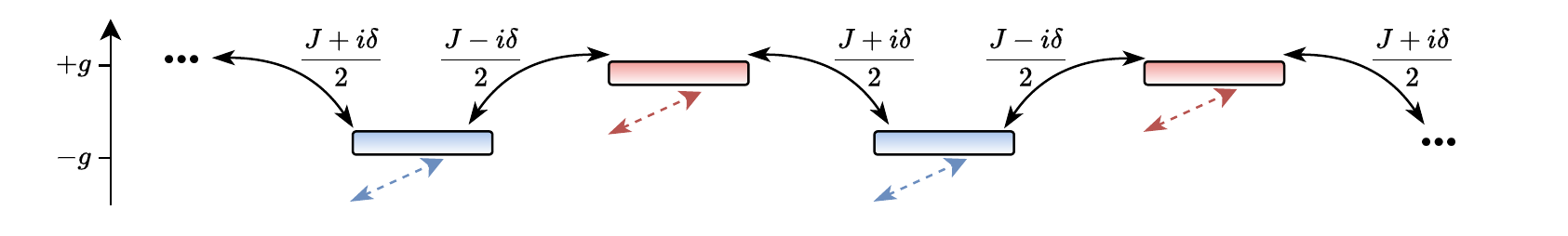}
    \caption{\textbf{Sketch of the $\bm{\mathcal{PT}}$-symmetric staggered tight-binding chain} with non-Hermitian Hamiltonian \cref{eq:SC_ham}. Lattice sites with alternating on site potential $\pm g$ are coupled by a uniform hopping $J$ and a imaginary alternating hopping $\delta$. The latter is induced by a coupling to the environment, which differs on even and odd sides, leading to the alternation.}
    \label{fig:stag}
\end{figure}

The lattice model (lattice constant $a=1$) with a staggered onsite energy $g \geq 0$ and a staggered complex hopping $\delta \geq 0$ with non-Hermitian Hamiltonian
\begin{equation}
 H=\sum_{j=1}^N \left[ \frac{J+ i \delta (-1)^j}{2} \left( c_j^\dag c_{j+1} + \mbox{H.c.} \right) + g (-1)^j c_j^\dag c_j \right]
\label{eq:SC_ham}    
\end{equation}
was  recently introduced as a toy model to study non-Hermitian quantum critical behavior \cite{Dora2022a}. Note that the uniform hopping has amplitude $J/2$, not $J$ as in Eq.~(\ref{eq:RL_ham}), and the opposite sign. Using this convention we follow the authors of  \cite{Dora2022a}. We take an even number of lattice sites $N$ and consider periodic boundary conditions; we identify the single-particle Wannier states $\left|j=N+1\right>$ and $\left|j=1 \right>$. The model is sketched in Fig.~\ref{fig:stag}. As the authors of \cite{Dora2022a} emphasize the non-Hermitian term can be thought of as arising in $H_{\rm eff}$ of the Lindblad equation (\ref{eq:lindblad_non_herm}) when using jump operators $\sqrt{\delta} \left( c_j \pm c_{j+1} \right)$ on even and odd bonds, respectively, and dropping a term proportional to the particle number operator. As the spatial reflection at any lattice site ($\mathcal P$) in combination with complex conjugation ($\mathcal T$) leaves $H$ Eq.~(\ref{eq:SC_ham}) invariant, the model is indeed ${\mathcal{PT}}$-symmetric; with theorem  $\mbox{T}^{\eta}_{4}$ it is also pseudo-Hermitian.

The system is translationally invariant and hence best treated in momentum space. We  transform from the single-particle Wannier basis $\left\{ \left| j \right>  \right\}$ to momentum space states $\left\{ \left| k \right>  \right\}$. In this, the momenta $k$ and $k-\pi$, with $0 \leq k < \pi$ are coupled, that is, the single-particle (1p) Hamiltonian is block-diagonal with non-Hermitian sub-blocks
\begin{equation}
 H_k^{\rm 1p} \doteq   \left( 
\begin{array}{cc}
J \cos{k}  &  g + \delta \sin(k) \\
g - \delta \sin{k}& - J \cos{k}  
\end{array} 
\right)   .
\label{eq:SC_ham_k}    
\end{equation}
To ease notation we again suppress an index on $k$, indicating that the momenta are discrete $k=n 2\pi/N$, $n \in \mathbbm{Z}$. The single-particle dispersion (eigenvalues) can now be obtained by diagonalizing Eq.~(\ref{eq:SC_ham_k}). This gives 
\begin{equation}
E_k^\pm = \pm \sqrt{\left( J^2 + \delta^2 \right) \cos^2{k} + g^2 - \delta^2} .
\label{eq:SC_ham_eigenvalues}    
\end{equation}
For $g \geq \delta$ all single-particle eigenvalues are real, that is, the $\mathcal{PT}$ symmetry is unbroken. When studying quantum critical behavior, we will only be interested in this regime. For $g < \delta$ there is always a region of momenta $k$ close to $\pi/2$ with complex conjugate pairs of complex eigenvalues and the $\mathcal{PT}$ symmetry is spontaneously broken. 

\begin{figure}
    \centering
    \includegraphics{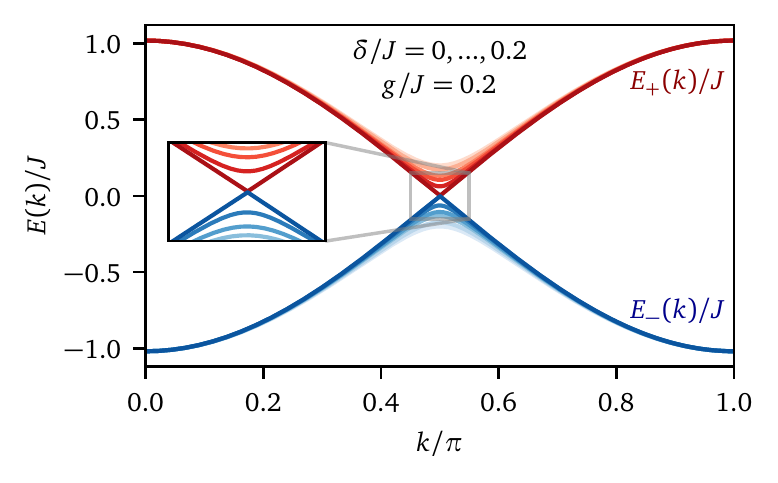}
    \caption{\textbf{Single-particle dispersion of the quantum critical model} \cref{eq:SC_ham} for $g/J = 0.2$ and $\delta/J = 0.0, 0.1, 0.15, 0.17, 0.19, 0.2$ from desaturated to saturated lines. For all parameters shown the $\mathcal{PT}$ symmetry is unbroken. As $\delta \nearrow g$, the direct band-gap at $k=\pi/2$ closes and the system becomes quantum critical  \cite{Sachdev2011,Dora2022a}.}
    \label{fig:critical_dispersion}
\end{figure}

In Fig.~\ref{fig:critical_dispersion} the single-particle dispersion is shown for $g/J=0.2$ and different $\delta/J$ from the phase of unbroken $\mathcal{PT}$ symmetry. For $g > \delta$ the dispersion is gapped with a minimal value of the gap at $k=\pi/2$, while it becomes gapless for $g=\delta$. For this case the dispersion is linear close to $k = \pi/2 $. In the many-body case and for half-filling, with Fermi momentum $k_{\rm F} = \pi/2$, this is the parameter regime in which the model becomes quantum critical \cite{Sachdev2011}. If, for $g=\delta$, one expands the elements of the matrix $H_k^{\rm 1p}$ of Eq.~(\ref{eq:SC_ham_k}) close to $k=\pi/2$ to leading order, one obtains
\begin{equation}
 H_{k}^{\rm 1p} \dot \approx   \left( 
\begin{array}{cc}
-J \left[ k - \frac{\pi}{2} \right]&  2g \\
0&   J \left[ k - \frac{\pi}{2} \right]
\end{array} 
\right).
\label{eq:SC_ham_k_exp}    
\end{equation}
This single-particle momentum space block is also realized in a field theoretical continuum model with Hamiltonian 
\begin{equation}
 H = \int dx \left\{ J \left[ \hat \psi_{\rm r}^\dag(x) i \partial_x \hat \psi_{\rm r}(x) -  \hat \psi_{\rm l}^\dag(x) i \partial_x \hat \psi_{\rm l}(x) \right]  + 2 g \hat \psi_{\rm r}^\dag(x)  \hat \psi_{\rm l}(x) \right\} ,
\label{eq:SC_ham_field}    
\end{equation}
which describes a critical system with dynamical critical exponent $z=1$ \cite{Sachdev2011}. Here $\hat \psi_{\rm r/l}(x)$ denotes right and left moving fermionic fields.

We do not compute the single-particle eigenstates of the lattice model Eq.~(\ref{eq:SC_ham}) analytically. Instead when studying correlation functions of this model in Sect.~\ref{sec:genfun}, we will numerically determine these. For our purposes this is sufficient. We also do not aim to compute any $\eta$ operator.   

\section{Non-unitary dynamics and the ancilla approach}
\label{sec:ancilla}

In this section we discuss in which sense the dynamics induced by a non-Hermitian Hamiltonian, and thus by a non-unitary time evolution operator, might be of relevance for physical systems. The non-Hermitian system is embedded into a larger one described by a Hermitian Hamiltonian. We, in particular, focus on the ancilla approach \cite{Guenther2008,Kawabata2017,Wu2019}. Some readers familiar with Feshbach projection or Lindblad master equations and the quantum trajectory approach, might find these methods of deriving non-Hermitian Hamiltonians physically more appealing; see \cref{subsec:Feshbach} and \cref{subsec:master} for a brief summary of these. However, we hope that at the end of this section all readers can agree, that the ancilla approach is conceptionally transparent and formally attractive. Crucially, it does not rely on any approximations or an effective Hamiltonian picture. It was so far not reviewed in the context of quantum many-body theory. 

We show that, within the ancilla approach, one can always find an embedding, which, in combination with a measurement, yields the non-Hermitian physics. We use this embedding approach in subsequent sections for deriving a consistent methodological framework for theoretically studying $\mathcal{PT}$-symmetric, non-Hermitian systems. Note that considerations of this type, or its converse, of deriving a non-Hermitian subsystem Hamiltonian from a Hermitian one, are rarely put forward in the mathematical physics research which led to the concepts of pseudo-Hermiticity as well as biorthogonal and $\mathcal{PT}$-symmetric quantum mechanics \cite{Bender2005-py,Bender2007-tf,Bender2019,Mostafazadeh2002-gk,Mostafazadeh2002,Mostafazadeh2003a,Brody2013-br}, at least beyond cartoons; see Chapter 1 of \cite{Bender2019}. On the one hand, we believe that it is important to keep this in mind to better appreciate the nature of these approaches. On the other hand, we judge this embedding into a Hermitian system to be crucial for the  development of a methodology which can be used to study emergent quantum many-body phenomena.

To introduce the ancilla approach we start out with the time-dependent Schr\"odinger equation 
\begin{equation}
i \partial_t \left| \psi_{\rm s}(t) \right> = H_{\rm s}(t)    \left| \psi_{\rm s}(t) \right>, \quad  \left| \psi_{\rm s}(0) \right> =  \left| \psi_{{\rm s,}0} \right> .
\label{eq:ancilla_SG}
\end{equation}
We slightly adopted the notation; compare Eq.~(\ref{eq:SG}). The non-Hermitian Hamiltonian $H_{\rm s}$ and the state now carry a label s; denoting ``system''. The possible time dependence is indicated as the argument instead of by an index. We often suppress the argument and only restore it if time dependence is important. The state $\left| \psi_{\rm s}(t) \right> $ is an element of the systems Hilbert space $\mathcal{H}_{\rm s}$ and formally given by
\begin{equation}
\left| \psi_{\rm s}(t) \right> = U_{\rm s}(t) \left| \psi_{\rm s}(0) \right> = T \exp{-i \int_0^t dt' H_{\rm s}(t') } \left| \psi_{\rm s}(0) \right> .
\label{eq:ancilla_timeevol}
\end{equation}
We now pose the question if we can find an enlarged quantum system, with Hilbert space $\mathcal{H}_{\rm sa}$ and a dynamics given by the Schr\"odinger equation with a Hermitian Hamiltonian $H_{\rm sa}$ which results in the state $\left| \psi_{\rm s}(t) \right>$ ``if one focusses'' on the system part. Here the index a denotes ancilla. One of the crucial issues is to decipher the meaning of the phrase in the quotation marks. It will turn out that complementing the system by a single spin-1/2 degree of freedom is sufficient and ``focusing on the system'' corresponds to performing a measurement on the ancilla spin.   

\subsection{The basic idea}
\label{subsec:basic_idea}

To present the basic idea we first consider a general non-Hermitian Hamiltonian $H_{\rm s}(t)$. Below, we will focus on pseudo-Hermitian and $\mathcal{PT}$-symmetric Hamiltonians. We complement the systems Hilbert space $\mathcal{H}_{\rm s}$ by $\mathbbm{C}^2$ and consider $\mathcal{H}_{\rm sa} = \mathbbm{C}^2 \otimes \mathcal{H}_{\rm s}$ as the combined system-ancilla Hilbert space. In $\mathbbm{C}^2$ we take the orthonormal basis states $\left| \uparrow \right>$ and $\left| \downarrow \right>$. 
%Do not confuse these with the orthonormal basis states of our $2 \times 2$ matrix example Eq.~(\ref{eq:spin_1_2}). 
We now focus on time-dependent states from a subset $\mathcal{H}_{\rm sa}^{\rm sub} \subset \mathcal{H}_{\rm sa}$ of the special form
\begin{equation}
\left| \psi_{\rm sa}^{\rm sub}(t) \right> = K \left[ \left| \uparrow \right> \otimes \left| \psi_{\rm s}(t) \right> + \left| \downarrow \right> \otimes g(t) \left| \psi_{\rm s}(t) \right> \right],
\label{eq:ancilla_state_form}
\end{equation}
with a, for now, unspecified linear, time-dependent operator $g(t)$ and a normalization constant $K$. Assuming that $\left| \psi_{\rm s}(t) \right>$ solves Eq.~(\ref{eq:ancilla_SG}) with the given non-Hermitian $H_{\rm s}(t)$ we ask:  Can one construct a Hermitian Hamiltonian $H_{\rm sa}(t)$, an initial state $\left| \psi_{\rm sa}^{\rm sub}(0) \right> \in \mathcal{H}_{\rm sa}^{\rm sub}$, and an operator $g(t)$  such that, for times $0 \leq t \leq \tau$, $\left| \psi_{\rm sa}^{\rm sub}(t) \right>$ of Eq.~(\ref{eq:ancilla_state_form}) fulfills the Schr\"odinger equation with this $H_{\rm sa}(t)$? Here $\tau$ denotes the largest time up to which the time evolution is performed. Ideally, we can consider the limit $\tau \to \infty$; see below. Furthermore, $\left| \psi_{\rm sa}^{\rm sub}(0) \right>$ is supposed to be normalized $\left< \psi_{\rm sa}^{\rm sub}(0) \right. \left| \psi_{\rm sa}^{\rm sub}(0) \right>=1$. To achieve this we can use $K$ which allows to normalize $ \left| \psi_{\rm sa}^{\rm sub}(0) \right>$ independent of what is assumed for $\left< \psi_{\rm s}(0) \right. \left| \psi_{\rm s}(0) \right>$.\footnote{Note that $\left< \psi_{\rm sa}^{\rm sub}(0) \right. \left| \psi_{\rm sa}^{\rm sub}(0) \right>$ and $\left< \psi_{\rm s}(0) \right. \left| \psi_{\rm s}(0) \right>$ denote canonical inner products on the two different Hilbert spaces $\mathcal{H}_{\rm sa}$ and  $\mathcal{H}_{\rm s}$, respectively.} If we can find a Hermitian $H_{\rm sa}(t)$ the normalization of  $ \left| \psi_{\rm sa}^{\rm sub}(0) \right>$ would persist for $0 < t \leq \tau$ as the dynamics of the combined system-ancilla setup is unitary.

To conclude the reasoning, let us first assume that we succeeded and postpone the construction of $H_{\rm sa}$, $\left| \psi_{\rm sa}^{\rm sub}(0) \right>$, and $g$. We time-evolve the initial state $\left| \psi_{\rm sa}^{\rm sub}(0) \right>$ up to time $t \leq \tau$, employing the unitary time evolution operator $U_{\rm sa}(t)$, obtained from Eq.~(\ref{eq:ancilla_timeevol}) by replacing $H_{\rm s}(t')$ by $H_{\rm sa}(t')$,  leading to $\left| \psi_{\rm sa}^{\rm sub}(t) \right>$ of Eq.~(\ref{eq:ancilla_state_form}). Next we perform a measurement on the ancilla spin, and only keep the instances in which the ancilla spin pointed upwards. This process is often referred to as post-selection. According to the principles of Hermitian quantum mechanics, after the measurement the system is in the state 
\begin{align}
\left| \psi_{\rm sa}^{\uparrow} (t) \right> 
&=
N(t) \left( P_\uparrow \otimes \mathbbm{1}_{\rm s} \right) \left| \psi_{\rm sa}^{\rm sub}(t) \right> \nonumber \\
&=
N(t) \left( \left| \uparrow \right> \left< \uparrow \right| \otimes \mathbbm{1}_{\rm s} \right) \left( \left| \uparrow \right> \otimes \left| \psi_{\rm s}(t) \right> + \left| \downarrow \right> \otimes g(t) \left| \psi_{\rm s}(t) \right> \right) \nonumber \\
&=
N(t)   \left| \uparrow \right> \otimes \left| \psi_{\rm s}(t) \right> ,
\label{eq:ancilla_proj}    
\end{align}
with the normalization\footnote{This holds as $\left< \uparrow \right. \left| \uparrow \right> =1$, with the inner product in $\mathbbm{C}^2$.}  
\begin{equation}
\left| N(t) \right|^2 = \left<  \psi_{\rm s}(t) \right. \left| \psi_{\rm s}(t) \right>^{-1}  .
\label{eq:ancilla_norm}    
\end{equation}
Remember that because of the non-Hermiticity of $H_{\rm s}$,  $\left<  \psi_{\rm s}(t) \right. \left| \psi_{\rm s}(t) \right> \neq 1$ even if we would have taken  $\left<  \psi_{\rm s}(0) \right. \left| \psi_{\rm s}(0) \right> =1$ initially. 

We next consider the equation of motion for the normalized state $\left| \psi_{\rm sa}^{\uparrow} (t) \right> \in \mathcal{H}_{\rm sa}$. Note that the logic is similar to the master equation, in that we want to derive a dynamical equation for a subpart of the full system-ancilla problem.
%i.e., we want to give the differential equation this state has to fulfill. 
For this we examine
\begin{align}
 i \partial_t \ket{\psi_{\rm sa}^{\uparrow} (t)} 
 &=
 i \partial_t \frac{\left| \uparrow \right> \otimes \left| \psi_{\rm s}(t) \right>}{\left<  \psi_{\rm s}(t) \right. \left| \psi_{\rm s}(t) \right>^{1/2}} \nonumber \\
 & =  i \partial_t \frac{\left| \uparrow \right> \otimes U_{\rm s}(t) \left| \psi_{\rm s}(0) \right>}{\left<  \psi_{\rm s}(0) \right| U_{\rm s}^\dag(t) U_{\rm s}(t) \left| \psi_{\rm s}(0) \right>^{1/2}} \nonumber \\
& = \left[ \mathbbm{1}_{\rm a} \otimes H_{\rm s}(t) \right] \left| \psi_{\rm sa}^{\uparrow} (t) \right>  - \frac{1}{2} \frac{\left< \psi_{\rm s}(t) \right| \left[ H_{\rm s}(t)- H_{\rm s}^\dag(t) \right] \left| \psi_{\rm s}(t) \right> }{\left<  \psi_{\rm s}(t) \right. \left| \psi_{\rm s}(t) \right>}  \left| \psi_{\rm sa}^{\uparrow} (t) \right> .
\label{eq:ancilla_eqm}    
\end{align}
As $ \left| \psi_{\rm s}(t) \right>$ enters in $ \left| \psi_{\rm sa}^{\uparrow}(t) \right>$, see Eq.~(\ref{eq:ancilla_proj}), this is a non-linear differential equation. This has to be contrasted to the Schr\"odinger equation, which is linear. Equation (\ref{eq:ancilla_eqm}) is equivalent to a norm-preserving equation of motion for a system with gain and loss which was written down mainly based on phenomenological reasoning in \cite{Gisin1981,Gisin1982,Gisin1983,Brody2012,Sergi2013}.\footnote{Note that Eq.~(2) of \cite{Brody2012} contains an additional term  $\sim i \left< \psi_{\rm s}(t) \right| \left[ H_{\rm s}(t) + H_{\rm s}^\dag(t) \right] \left| \psi_{\rm s}(t) \right>$. This is, however, irrelevant as it only changes the phase of the solution. Similarly, this holds for the equation of motion given in \cite{Gisin1981,Gisin1982,Gisin1983}.}

This rationale leads to a rather pragmatic reason why it is useful to solve the Schr\"odinger equation (\ref{eq:ancilla_SG}) with a non-Hermitian Hamiltonian $H_{\rm s}(t)$: It is linear and thus generically easier to solve than the non-linear equation~(\ref{eq:ancilla_eqm}). After the solution $ \left| \psi_{\rm s}(t) \right>$ of Eq.~(\ref{eq:ancilla_SG}) was found, one can compute the normalization Eq.~(\ref{eq:ancilla_norm}) and obtain $\left| \psi_{\rm sa}^{\uparrow}(t) \right>$ of Eq.~(\ref{eq:ancilla_proj}) \cite{Gal2022}. To put it reversely and, in particular, in more physical terms: Within the ancilla approach the non-unitary dynamics originating from the non-Hermitian $H_{\rm s}$ is effectively captured within the unitary evolution with $H_{\rm sa}$ and the ancilla spin-measurement. In fact, this idea was brought to life in an impressive experiment with a single nitrogen-vacancy center in diamond \cite{Wu2019}.

The remaining objective is to find a Hermitian Hamiltonian $H_{\rm sa}(t)$, an initial state $\left| \psi_{\rm sa}^{\rm sub}(0) \right> \in \mathcal{H}_{\rm sa}^{\rm sub}$, and an operator $g(t)$  such that $\left| \psi_{\rm sa}^{\rm sub}(t) \right>$ Eq.~(\ref{eq:ancilla_state_form}) fulfills the Schr\"odinger equation with $H_{\rm sa}(t)$. To determine a normalized $\left| \psi_{\rm sa}^{\rm sub}(0) \right>$ is straightforward. We simply take 
\begin{equation}
\left| \psi_{\rm sa}^{\rm sub}(0) \right> = K \left[ \left| \uparrow \right> \otimes \left| \psi_{\rm s}(0) \right> + \left| \downarrow \right> \otimes g(0) \left| \psi_{\rm s}(0) \right> \right],
\label{eq:ancilla_initialstate_form}
\end{equation}
with the given initial state $ \left| \psi_{\rm s}(0) \right> \in \mathcal{H}_{\rm s}$ and 
\begin{equation}
K = \left[\left< \psi_{\rm s}(0) \right. \left| \psi_{\rm s}(0) \right> +\left< \psi_{\rm s}(0) \right| g^\dag(0) g(0) \left| \psi_{\rm s}(0) \right>   \right]^{-1/2}  .   
\label{eq:K_determ}    
\end{equation}

\subsection{How to determine \texorpdfstring{$H_{\rm sa}(t)$}{Hsa(t)} and \texorpdfstring{$g(t)$}{g(t)}}
\label{subsec:constr_H_sa}

To construct the Hermitian Hamiltonian  $H_{\rm sa}(t)$ on $\mathcal{H}_{\rm sa}$ and the linear operator $g(t)$ on  $\mathcal{H}_{\rm s}$ for a given time-dependent $H_{\rm s} (t)$ we write the former in its most general form as
\begin{equation}
H_{\rm sa} = \sum_{\sigma,\sigma' = \uparrow,\downarrow} \left| \sigma \right> \left< \sigma' \right| \otimes H_{\rm sa}^{\sigma \sigma'} ,
\label{eq:ansatz_H_sa}    
\end{equation}
were we suppressed the time argument. The $H_{\rm sa}^{\sigma \sigma'}$ are linear operators acting on $\mathcal{H}_{\rm s}$ to be determined. For an obvious reason we refer to them as the components of $H_{\rm sa}$.  For $H_{\rm sa}$ to be Hermitian we have to require that 
\begin{equation}
\left( H_{\rm sa}^{\sigma \sigma} \right)^\dag = H_{\rm sa}^{\sigma \sigma} , \quad \left( H_{\rm sa}^{\sigma \bar \sigma} \right)^\dag = H_{\rm sa}^{\bar \sigma \sigma},
\label{eq:H_sa_Herm}    
\end{equation}
with $\bar \sigma$ being the complement of $\sigma$. Inserting the ansatz Eq.~(\ref{eq:ansatz_H_sa}) and the assumed form of the state Eq.~(\ref{eq:ancilla_state_form}) into the Schr\"odinger equation, replacing $i \partial_t \left| \psi_{\rm s}(t) \right>$ by $H_{\rm s}  \left| \psi_{\rm s}(t) \right>$, and taking the orthonormality of the ancilla basis states $\left| \uparrow \right>$ and $\left| \downarrow \right>$ into account we end up with the two coupled equations (with the time arguments suppressed) 
\begin{equation}
i \partial_t g + g H_{\rm s} =   H_{\rm sa}^{\uparrow \uparrow} g  +  H_{\rm sa}^{\uparrow \downarrow} , \quad
H_{\rm s} =  H_{\rm sa}^{\downarrow \downarrow} +  H_{\rm sa}^{\downarrow \uparrow} g.
\label{eq:H_sa_eta_equations}    
\end{equation}
These and the corresponding adjoint equations can be rewritten as
\begin{align}
 H_{\rm sa}^{\downarrow \uparrow} & = -i \partial_t g^\dag + H_{\rm s}^\dag g^\dag - g^\dag H_{\rm sa}^{\uparrow \uparrow}  \nonumber \\
 H_{\rm sa}^{\uparrow \downarrow} & = i \partial_t g + g H_{\rm s} - H_{\rm sa}^{\uparrow \uparrow} g \nonumber \\
H_{\rm sa}^{\downarrow \downarrow} & = H_{\rm s} - \left[  -i \partial_t g^\dag + H_{\rm s}^\dag g^\dag - g^\dag H_{\rm sa}^{\uparrow \uparrow} \right] g  
\label{eq:H_sa_eta_rewritten}    
\end{align}
and 
\begin{equation}
i \partial_t \left[ g^\dag g + \mathbbm{1} \right]  = H_{\rm s}^\dag \left[ g^\dag g + \mathbbm{1} \right] - \left[ g^\dag g + \mathbbm{1} \right]  H_{\rm s} .
\label{eq:eta_dag_eta}
\end{equation}
Note that the first two lines of Eq.~(\ref{eq:H_sa_eta_rewritten}) are consistent with Eq.~(\ref{eq:H_sa_Herm}). Employing Eq.~(\ref{eq:H_sa_eta_rewritten}) all components of $H_{\rm sa}$ can be computed from $H_{\rm sa}^{\uparrow \uparrow}$, $g$, and the given $H_{\rm s}$. Equation (\ref{eq:eta_dag_eta}) can be solved formally (we reintroduce the time arguments)
\begin{equation}
M(t) =   g^\dag(t) g(t) + \mathbbm{1}  = T \left( e^{-i \int_0^t d t' H_{\rm s}^\dag(t') } \right)
M(0)  
\tilde T \left( e^{i \int_0^t dt' H_{\rm s}(t') } \right)
\label{eq:M_solved}    
\end{equation}
with the anti-time-ordering symbol $\tilde T$, the initial value $M(0)$, and $M^\dag(t) = M(t)$. Let us assume that we can chose $M(0)$ such that $M(t) - \mathbbm{1}$ is a positive operator for all $0 \leq t \leq \tau$. In this case
\begin{equation}
g(t) = \mathcal{U}(t) \left[ M(t) - \mathbbm{1} \right]^{1/2} ,
\label{eq:g_solved}    
\end{equation}
with an arbitrary unitary operator $\mathcal{U}(t)$, exists for $0 \leq  t \leq \tau$.

Following this reasoning we have to chose an arbitrary Hermitian operator $H_{\rm sa}^{\uparrow \uparrow}(t)$, an arbitrary unitary operator $\mathcal{U}(t)$, and the initial Hermitian operator $M(0)$, the latter such that $M(t)-\mathbbm{1}$ is positive for all $0 \leq t \leq \tau$. From these operators $H_{\rm sa}(t)$ and $g(t)$ can be computed using Eqs.~(\ref{eq:g_solved}), (\ref{eq:M_solved}), and (\ref{eq:H_sa_eta_rewritten}). The supplementary material of \cite{Wu2019} contains an argument that a $M(0)$ fulfilling the above condition can always be found. However, it, in general, depends on the largest time $\tau$ up to which the system is time-evolved. In this case it might be impossible to take $\tau \to \infty$ and the time evolution with $H_{\rm sa}(t)$ is restricted to a finite time interval. In special cases a $\tau$-independent $M(0)$ can be found; see below. We here do not repeat the rationale of \cite{Wu2019} but explicitly give appropriate $M(0)$ for the cases of interest to us.  

We next discuss the situation of time-independent, pseudo-Hermitian Hamiltonians $H_{\rm s}$ with entirely real spectra. To illustrate this important case we employ the formalism to our two-level toy model Eq.~(\ref{eq:spin_1_2}) with $|z|<1$.

\subsection{\texorpdfstring{$\eta_{\rm r}$}{ETAr}-pseudo-Hermitian Hamiltonians} 
\label{subsec:PT_unbroken}

We now assume that $H_{\rm s}$ is time independent which simplifies Eq.~(\ref{eq:M_solved}) to 
\begin{equation}
M(t) =  e^{-i  H_{\rm s}^\dag t } M(0)  e^{i H_{\rm s} t } . 
\label{eq:M_solved_time_ind}    
\end{equation}
If  $H_{\rm s}$ is, in addition, $\eta_{\rm r}$-pseudo-Hermitian, e.g., $\mathcal{PT}$-symmetric and in the $\mathcal{PT}$ symmetry unbroken phase, we make the ansatz $M(0) = c \eta_{\rm r}$ with $\eta_{\rm r}$ of Eq.~(\ref{eq:ansatz_eta}) and $c \in \mathbbm{R}$. Exploiting the pseudo-Hermiticity relation Eq.~(\ref{eq:eta_doesit}) we obtain
\begin{equation}
M(t) =  e^{-i  H_{\rm s}^\dag t } c \eta_{\rm r}  
e^{i H_{\rm s} t } = c \eta_{\rm r} e^{-i  H_{\rm s} t }  e^{i H_{\rm s} t } =   c \eta_{\rm r} = M(0) 
\label{eq:exploit}    
\end{equation}
and the time dependence drops out. We finally have to determine a $c$ such that $c \eta_{\rm r} - \mathbbm{1}$ is positive. As $\eta_{\rm r}$ is a positive Hermitian operator it has positive eigenvalues $\lambda_\nu$. If we now take 
\begin{align}
c=\sum_{\nu} \frac{1}{\lambda_\nu} ,
\label{eq:c_def}
\end{align}
the operator $c \eta_{\rm r} - \mathbbm{1}$ is positive. For $\mathcal{U}(t)$ we take the unity operator and obtain a time-independent, Hermitian
\begin{equation}
g = \left[ c \eta_{\rm r} - \mathbbm{1} \right]^{1/2} = g^\dag .
\label{eq:g_for_sym_unbr}    
\end{equation}
This considerably simplifies Eq.~(\ref{eq:H_sa_eta_rewritten}). If we, in addition, exploit the freedom to chose a $H_{\rm sa}^{\uparrow \uparrow}$ and assume that $H_{\rm sa}^{\uparrow \uparrow} = H_{\rm sa}^{\downarrow \downarrow}$, $H_{\rm sa}$ is time-independent and can be written as
\begin{equation}
H_{\rm sa} = \mathbbm{1}_2 \otimes A + \sigma_y \otimes B 
\label{eq:H_sa_form_ashida}    
\end{equation}
with the Pauli matrix $\sigma_y$ and
\begin{align}
A= \left( g H_{\rm s} + H_{\rm s} g^{-1} \right) \left(g^{-1} + g\right)^{-1} , \quad B= i \left(H_{\rm s} + g H_{\rm s} g^{-1} \right) \left(g^{-1} + g\right)^{-1} , %VM: Typo korrigiert!!!
\label{eq:A_B_def}
\end{align}
acting on $\mathcal{H}_{\rm s}$. This is the form given in the supplementary material of \cite{Kawabata2017}. We, however, emphasize that the above reasoning shows that $H_{\rm sa}$ is by no means unique. E.g., choosing a different unitary operator $\mathcal{U}(t)$ or relaxing the condition $H_{\rm sa}^{\uparrow \uparrow} = H_{\rm sa}^{\downarrow \downarrow}$ would lead to a different form of $H_{\rm sa}$.

We note that in the present case the expression for the normalization constant $K$ Eq.~(\ref{eq:K_determ}) simplifies. As $g^\dag g=c \eta_r - \mathbbm{1}$ we obtain
\begin{equation}
K = \left[c \left< \psi_{\rm s}(0) \right| \eta_{\rm r} \left| \psi_{\rm s}(0) \right>   \right]^{-1/2}  .   
\label{eq:K_determ_1}    
\end{equation}

\subsection{The \texorpdfstring{$\mathcal{PT}$}{PT}-symmetric two-level problem in its unbroken phase}
\label{subsec:example_spin_1_2}

To illustrate the considerations of the last subsection we next apply them to our (toy) model Hamiltonian Eq.~(\ref{eq:spin_1_2}), with $|z|<1$, as the non-Hermitian but $\mathcal{PT}$-symmetric $H_{\rm s}$ acting on $\mathcal{H}_{\rm s} = \mathbbm{C}^2$. The matrix representation of $\eta_{\rm r}$ in the systems $\left| \uparrow \right>$ and $\left| \downarrow \right>$ basis is given in Eq.~(\ref{eq:spin_1_2_eta_a}). Its eigenvalues are $\lambda_{\pm} = (1\pm z)/\sqrt{1-z^2}$ which leads to $c=2/\sqrt{1-z^2}$. With this we can determine $g$ of Eq.~(\ref{eq:g_for_sym_unbr}) and obtain $g=\eta_{\rm r}$ \cite{Kawabata2017}. Furthermore, $g^{-1} + g = c \mathbbm{1}$, which simplifies the computation of the operators $A$ and $B$  of Eq.~(\ref{eq:A_B_def}). A straightforward computation gives
\begin{equation}
A \doteq \left(  \begin{array}{cc}   
r \cos{\theta} & s  (1-z^2) e^{i \phi} \\
 s  (1-z^2) e^{-i \phi} & r \cos{\theta} 
\end{array} \right) , \quad B \doteq \sqrt{1-z^2} \left(  \begin{array}{cc}   - r \sin{\theta} & 0  \\
 0 & r \sin{\theta} 
\end{array} \right) .
\label{eq:A_B_example}
\end{equation}
Inserting this in Eq.~(\ref{eq:H_sa_form_ashida}) leads to the matrix representation of $H_{\rm sa}$
\begin{equation}
 H_{\rm sa} \!\! \doteq  \!\!\!  \left(  \begin{array}{cccc}   
r \cos{\theta} & s  (1-z^2) e^{i \phi} & i  r \sqrt{1-z^2} \sin{\theta} & 0\\
s  (1-z^2) e^{-i \phi} & r \cos{\theta} & 0 & -i  r \sqrt{1-z^2} \sin{\theta} \\
-i  r \sqrt{1-z^2} \sin{\theta} & 0 &  r \cos{\theta} & s  (1-z^2) e^{i \phi}\\
0 & i  r \sqrt{1-z^2} \sin{\theta} &   s  (1-z^2) e^{-i \phi} &  r \cos{\theta} 
\end{array} \right)    
\label{eq:H_sa_example}    
\end{equation}
in the $\left\{ \left| \uparrow_{\rm a}\uparrow_{\rm s} \right>,  \left| \uparrow_{\rm a}\downarrow_{\rm s} \right>, \left| \downarrow_{\rm a} \uparrow_{\rm s} \right> , \left| \downarrow_{\rm a}\downarrow_{\rm s} \right> \right\} $ basis of $ \mathcal{H}_{\rm sa}$ (in self-explaining notation). This Hermitian matrix has the same real eigenvalues Eq.~(\ref{eq:spin_1_2_eigenvalues}) as $H_{\rm s} = H(\phi)$ of Eq.~(\ref{eq:spin_1_2}), both being doubly degenerate. 

\begin{figure}
    \centering
    \includegraphics{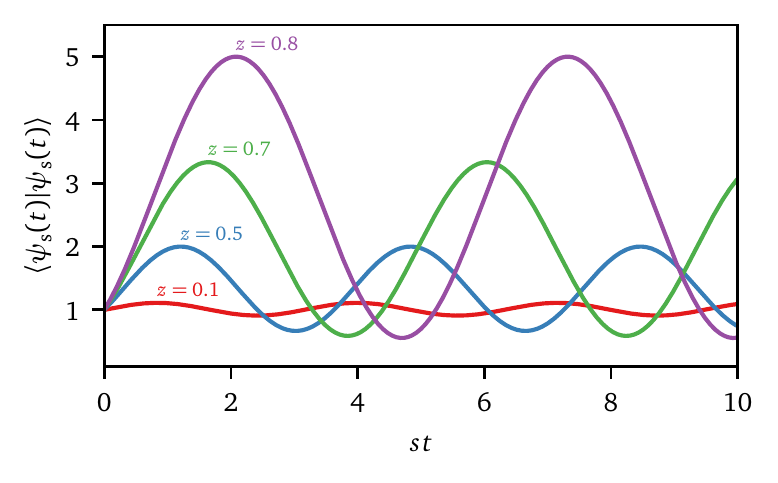}
    \caption{\textbf{Time dependence of the norm} \cref{eq:norm_psi_s} in the $\mathcal{PT}$-unbroken phase ($|z| < 1$). As $|z| \nearrow 1$ the period and amplitude of the oscillations increases.}
    \label{fig:norm_psi_s}
\end{figure}

We find it instructive to explicitly compute the states $\left| \psi_{\rm sa}^{\rm sub}(t) \right>$ and $\left| \psi_{\rm s}(t) \right>$ by applying the time evolution operator $e^{-i H_{\rm sa} t}$ and $e^{-i H_{\rm s} t}$ to initial states related by Eq.~(\ref{eq:ancilla_initialstate_form}) and show that the time-evolved ones are related according to the reasoning of Sect.~\ref{subsec:basic_idea}. Furthermore, we will reuse some of the results presented here when computing expectation values in Sect.~\ref{sec:obexp}. As our initial state in $\mathcal{H}_{\rm s}$ we take the normalized state $\left| \psi_{\rm s}(0) \right> = \left| \uparrow \right> $, that is, the particle initially occupies the left of the two lattice sites (or the spin-1/2 points upwards). Expressing $ \left| \uparrow \right> $ in terms of the right energy eigenstates Eq.~(\ref{eq:spin_1_2_eigenvectors_1_R}) it is straightforward to see that the non-unitarily time-evolved state is given by
\begin{align}
\left| \psi_{\rm s} (t) \right> =
e^{-i H_{\rm s} t} \left| \uparrow \right>  =
\frac{e^{-i \phi}}{2 \sqrt{1-z^2}} &
\left( e^{- i E_+ t} \left\{  1 + \left[ iz+\sqrt{1-z^2}  \right] \right\}^{1/2} \left|{\rm R}_+ \right>  \right. \nonumber \\
& - \left. e^{- i E_- t} \left\{  1 + \left[ iz-\sqrt{1-z^2}  \right] \right\}^{1/2} \left|{\rm R}_- \right> \right) 
\label{eq:time_evol_psi_s}    
\end{align}
or, in the systems $\left| \uparrow \right>$ and $\left| \downarrow \right>$ basis,
\begin{equation}
 \left| \psi_{\rm s} (t) \right>  \doteq \frac{1}{2 \sqrt{1-z^2}} \left(    \begin{array}{c}
e^{- i E_+ t} \left[ i z + \sqrt{1-z^2} \right] - e^{- i E_- t} \left[ i z - \sqrt{1-z^2} \right] \\[.1cm]
e^{- i \phi} \left[ e^{- i E_+ t} - e^{- i E_- t}  \right] 
\end{array} \right) .
\label{eq:time_evol_psi_s_basis}    
\end{equation}
From the last expression, the norm of $ \left| \psi_{\rm s} (t) \right>$ can be directly read off 
\begin{equation}
\left< \psi_{\rm s} (t) \right.\left| \psi_{\rm s} (t) \right> = 
\frac{1}{1-z^2} \left[ 1- z^2 \cos{\left(2 s \sqrt{1-z^2} t\right)} + z \sqrt{1-z^2} \sin{\left(2 s \sqrt{1-z^2} t\right)} \right].
\label{eq:norm_psi_s}
\end{equation}
This shows explicitly that, in the non-Hermitian case $z \neq 0$, the state $ \left| \psi_{\rm s} (t) \right>$ is no longer normalized to one for $t>0$ although it was so at $t=0$; the time evolution operator $e^{-i H_{\rm s} t}$ is non-unitary. The norm oscillates with the frequency given by the difference $E_+ - E_-$ of the (real) eigenenergies. It is shown in Fig.~\ref{fig:norm_psi_s} for different $z$. 

Next we compute the initial state of the combined system-ancilla setup associated to $\left| \psi_{\rm s}(0) \right> = \left| \uparrow \right>$. We consider
\begin{equation}
g   \left| \psi_{\rm s}(0) \right> \doteq    \frac{1}{\sqrt{1-z^2}} \left( 
\begin{array}{cc}
1 & -i z e^{i \phi} \\
i z e^{- i \phi} & 1  
\end{array} 
\right) \left(   \begin{array}{c} 1 \\ 0 \end{array} \right) = \frac{1}{\sqrt{1-z^2}}  \left(   \begin{array}{c} 1 \\ i z e^{-i \phi} \end{array} \right) .
\label{eq:ini_state_1}    
\end{equation}
Furthermore [see Eq.~(\ref{eq:K_determ_1})] 
\begin{equation}
K = \left[c \left< \psi_{\rm s}(0) \right| \eta_{\rm r} \left| \psi_{\rm s}(0) \right>   \right]^{-1/2}  = \left[ \frac{2}{\sqrt{1-z^2}} \, \frac{1}{\sqrt{1-z^2}} \right]^{-1/2} = \sqrt{\frac{1-z^2}{2}} .   
\label{eq:K_determ_2}    
\end{equation}
With Eq.~(\ref{eq:ancilla_initialstate_form}) and in the same basis as used in Eq.~(\ref{eq:H_sa_example}) this leads to
\begin{equation}
\left| \psi_{\rm sa}^{\rm sub} (0) \right>  \doteq \frac{1}{\sqrt{2}} \left(    \begin{array}{c}
\sqrt{1-z^2} \\[.1cm]
0 \\[.1cm]
1 \\[.1cm]
i z e^{-i \phi}
\end{array} \right) , \quad \left< \psi_{\rm sa}^{\rm sub} (0) \right. \left| \psi_{\rm sa}^{\rm sub} (0) \right> =  1 .
\label{eq:psi_sa_initial}    
\end{equation}

Determining the eigenvectors of the Hermitian $H_{\rm sa}$ Eq.~(\ref{eq:H_sa_example}) one finds that in each of the eigenspaces $\mathcal{H}_{\rm sa}^\pm \subset \mathcal{H}_{\rm sa}$ to the doubly degenerate eigenvalues $E_\pm$ one normalized eigenvector $\left| {\rm R}_{\pm}^{\rm sa} \right>$ can be constructed which is element of $\mathcal{H}_{\rm sa}^{\rm sub}$, i.e., which is of the form Eq.~(\ref{eq:ancilla_state_form}) (dropping time-dependence). We only need to consider those, as the initial state Eq.~(\ref{eq:psi_sa_initial}) is element of $\mathcal{H}_{\rm sa}^{\rm sub}$ and the time evolution leaves this subspace invariant. The states  $\left|{\rm R}_{\pm}^{\rm sa} \right>$ can most easily be found by inserting the two right eigenvectors Eq.~(\ref{eq:spin_1_2_eigenvectors_1_R}) for $\left| \psi_{\rm s} (0) \right>$ in Eq.~(\ref{eq:ancilla_initialstate_form}). 
Writing $\left| \psi_{\rm sa}^{\rm sub} (0) \right>$ as a linear combination of $\left|{\rm R}_{{\rm sa},\pm}^{\rm sub} \right>$, applying $e^{-i H_{\rm sa} t}$, and projecting with $\left| \uparrow \right> \left< \uparrow \right| \otimes \mathbbm{1}_{\rm s}$ gives the state Eq.~(\ref{eq:time_evol_psi_s_basis}) extended to a four dimensional vector by adding two rows containing $0$. 

For this example it is also straightforward to (at least numerically) show explicitly that if the state $\left| \psi_{\rm s}(t) \right> \in \mathcal{H}_{\rm s}$ fulfills the linear Schr\"odinger equation (\ref{eq:ancilla_SG}), the normalized state $\left| \psi^{\uparrow}_{\rm sa}(t) \right> \in \mathcal{H}_{\rm sa}$ fulfills the non-linear equation of motion (\ref{eq:ancilla_eqm}) and vice versa. All this is in full accordance with the formalism developed in Sect.~\ref{subsec:basic_idea}.  

We finally briefly investigate the case of a time-independent, $\mathcal{PT}$-symmetric Hamiltonian $H_{\rm s}$ in its symmetry broken phase, to contrast this to the unbroken one discussed in Sect.~\ref{subsec:PT_unbroken}.    

\subsection{\texorpdfstring{$\eta_{\rm cp}$}{ETACP}-pseudo-Hermitian Hamiltonians}
\label{subsec:ancilla_broken}

As in Sect.~\ref{subsec:PT_unbroken} we start out with Eq.~(\ref{eq:M_solved_time_ind}). Employing the ansatz $M(0) = c \eta_{\rm cp}$, with $\eta_{\rm cp}$ of Eq.~(\ref{eq:ansatz_eta_cp}) and the pseudo-Hermiticity relation Eq.~(\ref{eq:eta_doesit}) with $\eta_{\rm r} \to \eta_{\rm cp}$, in analogy to Eq.~(\ref{eq:exploit}), $M(t)$ becomes time-independent. Unfortunately, one cannot follow the steps of Sect.~\ref{subsec:PT_unbroken} any further as $\eta_{\rm cp}$ is not a positive definite operator. Thus, no $c$ can be found which renders $c \eta_{\rm cp} - \mathbbm{1}$ positive. This raises the question if, in the $\mathcal{PT}$ symmetry broken phase, one can find a $M(0)$ such that $M(t)$ in  Eq.~(\ref{eq:M_solved_time_ind}) becomes time-independent. We are not able to answer this question definitely but it appears unlikely that this is possible. 

As an alternative ansatz we employ $M(0) = m \mathbbm{1}_{\rm s}$ with $m \in \mathbb{R}$. Inserting this in  Eq.~(\ref{eq:M_solved_time_ind}) we obtain
\begin{equation}
M(t) = m \, e^{i ( H_{\rm s}-H^\dag_{\rm s} ) t} .    
\label{eq:sym_broken_M}    
\end{equation}
With this $M(t)$ is time-dependent and thus $g(t)$ Eq.~(\ref{eq:g_solved}), even if we still chose $\mathcal{U}(t) = \mathbbm{1}$. With Eqs.~(\ref{eq:ansatz_H_sa}) and (\ref{eq:H_sa_eta_rewritten}) also $H_{\rm sa}(t)$ becomes time-dependent even if $H_{\rm s}$ is time-independent. To ensure that $M(t) - \mathbbm{1}$ is positive for all $0 \leq t \leq \tau$ we furthermore have to chose a $m$ which depends on the largest time $\tau$ considered. To see this explicitly let us consider our toy model Hamiltonian Eq.~(\ref{eq:spin_1_2}) with $|z| >1$. For this we obtain
\begin{equation}
M(t) \doteq \left(  \begin{array}{cc}   
m \, e^{-2 r t \sin{\theta}} & 0\\
 0 & m \, e^{2 r t \sin{\theta}} 
\end{array} \right) 
\label{eq:sym_broken_M_expl}    
\end{equation}
Given this $M(t)$, $M(t) - \mathbbm{1}$ is positive in $0 \leq t \leq \tau$ if we take $m > e^{2 | r \sin{\theta}| \tau}$. As emphasized in Sect.~\ref{subsec:constr_H_sa}, $M(0)$ depends on the largest time $\tau$ considered and the limit $\tau \to \infty$ cannot be taken. 

This brief discussion hints at the difficulties one encounters when attempting to explicitly construct the Hermitian $H_{\rm sa}(t)$ given a time-independent, $\mathcal{PT}$-symmetric, non-Hermitian $H_{\rm s}$ in its symmetry broken phase. We, however, emphasize that with the rational of Sect.~\ref{subsec:constr_H_sa} we can be certain that such a $H_{\rm sa}(t)$ exists.

\subsection{Can a single spin-1/2 act as a complex environment?}

On first glance it appears to be very surprising, that adding just a single spin-1/2 degree of freedom and a post-selection measurement is sufficient to capture the non-Hermitian dynamics usually associated to open systems with much more complex environments \cite{Daley2014,Ashida2020-vp,Roccati2022,Bender2019}. However, one has to take a closer look.

The most microscopical view on justifying a non-unitary quantum dynamics associated with a non-Hermitian Hamiltonian of a system coupled to an environment we are aware of, is the master equation and quantum trajectory approach summarized in Sect.~\ref{subsec:master}. We use it here to illustrate that the environment and the system-environment coupling captured by the dynamics of a non-Hermitian Hamiltonian are both restricted. Already the starting equation (\ref{eq:lindblad_non_herm}) of the master equation approach is approximate in nature. It relies on the rotating wave, the Born, and the Markov approximations. This strongly restricts the nature of the environment and the system-environment coupling. Furthermore, the jump operators $a_l^{(\dag)}$ can frequently not been given based on microscopic considerations. This further restricts the universality of the approach. Finally, to end up with the von Neumann equation (\ref{eq:lindblad_non_herm}) with the last term dropped, one has to assume that up to time $t$ no quantum jump occurred---this is the post-selection inherent to this approach. If we do not want to limit the time too severely, it again restricts the nature of the environment and that of the system-environment coupling. Only after performing all these steps and this way giving up the idea of a rather general environment and system-environment coupling, the dynamics derived from the Master equation and quantum trajectory approach becomes equivalent to that of the ancilla approach. Under these assumption the environment in effect becomes equivalent to a single spin-1/2. Following this reasoning it might appear to be less surprising, that a single spin-1/2 degree of freedom is sufficient to emulate the non-unitary system dynamics. 

In this context we would like to again emphasize that the very interesting line of research of Bender et al. does not contain any derivation of a quantum dynamics described by $\mathcal{PT}$-symmetric non-Hermitian Hamlitonian from the unitary dynamics of a coupled system-environment setup with a Hermitian Hamlitonian. Concerning this, it relies on general considerations and intuitions.

\subsection{The dynamics using a biorthogonal inner product}
\label{ssec:biorthogonal_product}

Many readers will be familiar with or will, at least, have heard of the concept of a biorthogonal inner product \cite{Brody2013-br,Bender2019}. We here introduce this and discuss, in which sense it can be used to render the time evolution for a $\eta_{\rm r}$-pseudo-Hermitian Hamiltonian to be unitary. 

We assume that a general biorthonormal basis $\left\{\left|{\rm r}_\nu \right> , \left|{\rm l}_\nu \right>   \right\}$ is given; see Sect.~\ref{subsec:bio}. With the linear operator $\hat g$ Eq.~(\ref{eq:ansatz_hat_g}) being positive definite and given two arbitrary states $ \left| \psi_{\rm s}^{(i)}\right> \in \mathcal{H}_{\rm s}$, $i=1,2$,
\begin{equation}
\left< \psi_{\rm s}^{(1)} \right. \left| \psi_{\rm s}^{(2)}\right>_{\hat g} = \left< \psi_{\rm s}^{(1)} \right|\hat g \left| \psi_{\rm s}^{(2)}\right> 
\label{eq:alt_inner_prod}    
\end{equation}
fulfills all the defining axioms of an inner product.\footnote{On the right hand side of Eq.~(\ref{eq:alt_inner_prod}) we use the standard inner product in Dirac notation.} In particular, 
$\left< \psi_{\rm s}^{(i)} \right|\hat g \left| \psi_{\rm s}^{(i)} \right> \geq 0$.  We will refer to it as a biorthogonal inner product. The space of vectors from 
$\mathcal{H}_{\rm s}$ equipped with this inner product constitutes an alternative (biorthogonal) Hilbert space 
$\tilde{\mathcal{H}}_{\rm s}$. In this sense 
$\hat g$ can be viewed as a metric operator \cite{Pauli1943,Scholtz1992}. We here do not use the notation on the left hand side of Eq.~(\ref{eq:alt_inner_prod}) but rather always explicitly insert $\hat g$ in the standard inner product. 

To better understand the effect the insertion of $\hat g$ in the inner product has, let us consider the two arbitrary states 
\begin{equation}
\left| \psi_{\rm s}^{(i)} \right> = \sum_{\nu} c_\nu^{(i)} \left| {\rm r}_\nu \right>  , \quad i=1,2  ,
\label{eq:two_states}    
\end{equation}
with the right basis vectors $\left| {\rm r}_\nu \right>$, and $c_\nu^{(i)} \in \mathbbm{C}$. The standard inner product of these two states is
\begin{equation}
\left< \psi_{\rm s}^{(1)} \right. \left| \psi_{\rm s}^{(2)} \right> = \sum_{\nu,\mu} \left[ c_\nu^{(1)} \right]^\ast c_{\mu}^{(2)} \left< {\rm r}_\nu \right. \left| {\rm r}_\mu \right> .
\label{eq:standard_inner_p}    
\end{equation}
As the right basis states are not orthonormal, this expression does not simplify further. For the biorthogonal inner product one instead obtains 
\begin{equation}
\left< \psi_{\rm s}^{(1)} \right| \hat g \left| \psi_{\rm s}^{(2)} \right>  =  \sum_{\nu,\mu} \left[ c_\nu^{(1)} \right]^\ast c_{\mu}^{(2)} \left< {\rm r}_\nu \right| \hat g \left| {\rm r}_\mu \right> = 
\sum_{\nu} \left[  c_\nu^{(1)} \right]^\ast  c_{\nu}^{(2)} ,
\label{eq:alternative_inner_p}    
\end{equation}
were we used that $\hat g \left| {\rm r}_\mu \right> =  \left| {\rm l}_\mu \right> $, see Eq.~(\ref{eq:right_left_map}), and  $\left< \mbox{r}_\nu \right.  \left| \mbox{l}_\mu \right>  = \delta_{\nu,\mu}$, see Eq.~(\ref{eq:biorthogonal}). Equation (\ref{eq:alternative_inner_p}) is the familiar result for an inner product of Hermitian quantum mechanics. 

As there might be many different biorthonormal basis sets in a given vector space, these considerations define an entire family of biorthogonal inner products \cite{Brody2013-br}. However, there is only one biorthonormal basis set and thus one biorthogonal inner product which is used when studying the dynamics in biorthogonal and $\mathcal{PT}$-symmetric quantum mechanics. To introduce this, we consider the non-Hermitian, time-independent Hamiltonian $H_{\rm s}$ and its biorthonormal basis $\left\{\left|{\rm R}_\nu \right> , \left|{\rm L}_\nu \right>   \right\}$.  We time evolve the two arbitrary states Eq.~(\ref{eq:two_states}) by applying $e^{- i H_{\rm s} t}$. The standard inner product gives 
\begin{equation}
\left< \psi_{\rm s}^{(1)} (t) \right. \left| \psi_{\rm s}^{(2)} (t) \right> = \sum_{\nu,\mu} e^{-i(E_\mu - E_\nu^\ast) t} \left[  c_\nu^{(1)} \right]^\ast c_{\mu}^{(2)} \left< {\rm R}_\nu \right. \left| {\rm R}_\mu \right> ,
\label{eq:standard_inner_p_t}    
\end{equation}
which, for arbitrary states, cannot be simplified.  The time $t$ does not drop out which indicates that the dynamics is non-unitary. If we now assume that $H_{\rm s}$ is $\eta_{\rm r}$-pseudo-Hermitian, e.g., $\mathcal{PT}$-symmetric and in its unbroken phase, and use $\hat g \to \eta_{\rm r}$ as the metric operator in the biorthogonal inner product we obtain
\begin{align}
 \left< \psi_{\rm s}^{(1)} (t) \right| \eta_{\rm r} \left| \psi_{\rm s}^{(2)} (t) \right>  & =  \sum_{\nu,\mu} e^{-i(E_\mu - E_\nu^\ast) t} \left[ c_\nu^{(1)} \right]^\ast  c_{\mu}^{(2)} \left< {\rm R}_\nu \right| \eta_{\rm r} \left| {\rm R}_\mu \right>
= \sum_{\nu} \left[  c_\nu^{(1)} \right]^\ast c_{\nu}^{(2)} \nonumber \\  & = \left< \psi_{\rm s}^{(1)} (0) \right| \eta_{\rm r} \left| \psi_{\rm s}^{(2)} (0) \right>  
\label{eq:alternative_inner_p_t}    
\end{align}
Besides using Eqs.~(\ref{eq:biorthogonal}) and (\ref{eq:right_left_map}) we employed the reality of the eigenvalues $E_\nu$. The same result can be obtained without expressing the two states in terms of the right eigenvectors of $H_{\rm s}$ 
\begin{align}
 \left< \psi_{\rm s}^{(1)} (t) \right| \eta_{\rm r} \left| \psi_{\rm s}^{(2)} (t) \right> & =
  \left< \psi_{\rm s}^{(1)} (0) \right| e^{i H_{\rm s}^\dag t} \eta_{\rm r} e^{-i H_{\rm s} t} \left| \psi_{\rm s}^{(2)} (t) \right> \nonumber \\ & =  \left< \psi_{\rm s}^{(1)} (0) \right| \eta_{\rm r} e^{i H_{\rm s} t} e^{-i H_{\rm s} t} \left| \psi_{\rm s}^{(2)} (t) \right>  \nonumber \\
  & = \left< \psi_{\rm s}^{(1)} (0) \right| \eta_{\rm r} \left| \psi_{\rm s}^{(2)} (0) \right>  , 
\label{eq:alternative_inner_p_alternativ}    
\end{align}
were we used the pseudo-Hermiticity relation Eq.~(\ref{eq:eta_doesit}). In this sense the dynamics on $\tilde{\mathcal{H}}_{\rm s}$ is unitary. This, in particular, implies that a state $\left| \psi_{\rm s}(0) \right>$ normalized to one according to the biorthogonal inner product $\left< \psi_{\rm s}(0) \right| \eta_{\rm r}\left| \psi_{\rm s}(0) \right>=1$, remains normalized at all times, in accordance with what is known in standard Hermitian quantum mechanics.   

Note that this biorthogonal inner product, and thus $\tilde{\mathcal{H}}_{\rm s}$, depends on the Hamiltionian $H_{\rm s}$, as $\eta_{\rm r}$ Eq.~(\ref{eq:ansatz_eta}) depends on $H_{\rm s}$ (via its left eigenvectors). One first has to determine all the eigenvectors before the Hilbert space can be given \cite{Bender2019}. From the perspective of standard Hermitian quantum mechanics it appears odd that the Hamiltonian and the Hilbert space are tied together in this way.  Note also, that these considerations cannot be extended to the regime of complex eigenvalues, e.g., to a $\mathcal{PT}$-symmetric Hamiltonian in its symmetry broken phase. 
In other words, the mathematical option of introducing an inner product according to which the dynamics is unitary is tied to the particular physical properties of the dynamics in the phase of unbroken $\mathcal{PT}$ symmetry; for an example see Sect.~\ref{subsec:further_examples_2_2}.

The idea of an alternative inner product involving a metric operator can be extended to the case of explicitly time-dependent non-Hermitian Hamiltonians \cite{Mostafazadeh2018,Zhang2019}, where this approach leads to a time dependent Hilbert space. This is reviewed in \cite{Frith2020}.  

The observation that the time evolution for a non-Hermitian, $\mathcal{PT}$-symmetric Hamiltonian in its symmetry unbroken phase is unitary within the framework of biorthogonal and $\mathcal{PT}$-symmetric quantum mechanics is one of the reasons why these approaches are popular in mathematical physics \cite{Brody2013-br,Bender2019}. Other reasons will be given in Sects.~\ref{sec:obexp} and \ref{sec:genfun}. Note that all this is formalism (mathematics) not physics. Also in biorthogonal and $\mathcal{PT}$-symmetric quantum mechanics the initial state is time-evolved with the non-unitary time evolution operator $e^{-i H_{\rm s} t}$ and contains all the non-Hermitian physics. Having emphasized this, we here pose the question if, for an open quantum system with a non-Hermitian Hamiltonian, employing a formalism in which the dynamics becomes unitary is expedient. Regardless if the non-Hermitian Hamiltonian is derived from Feshbach projection \cite{Rotter2009-ar,Ashida2020-vp} or Lindblad master equations \cite{Daley2014,Ashida2020-vp} or embedded into a Hermitian one in the ancilla approach discussed in this section, a clear ``yes'' as an answer does not appear to be compelling. This holds even if all eigenvalues of $H_{\rm s}$ are real. Furthermore, employing a formalism in which the non-unitarity of the time evolution is hidden, might lead to misconceptions when it comes to the computation of expectation values of observables and correlation functions. In the next section we will elaborate on this.

\section{Observables and expectation values}
\label{sec:obexp}

The concept of observables lies at the heart of quantum mechanics. In this section we review the definition of observables and the notion of expectation values as it is introduced in $\mathcal{PT}$-symmetric as well as biorthogonal quantum mechanics and contrast this to what follows from a direct transfer of the ideas from Hermitian quantum mechanics to $\mathcal{PT}$-symmetric, non-Hermitian systems. The former two approaches postulate a modified definition of the notion of observables. They furthermore rely on a definition of ``expectation values'' evaluated in a modified scalar product---the biorthogonal inner product (see \cref{ssec:biorthogonal_product})---, which, while formally appealing, leads to physically questionable results. We illustrate this explicitly on the examples of the two-level model and the resonant level model. On the other hand, the direct application of the approach to observables and expectation values, as it is used in Hermitian quantum mechanics, does not suffer from these problems. Its validity for $\mathcal{PT}$-symmetric systems can be directly derived from the ancilla approach and leads to physically reasonable results in all cases considered. In addition, it is this Hermitian approach which is used for generic open quantum systems. We begin our argumentation with a short recap of standard Hermitian quantum mechanics. 

\subsection{Hermitian quantum mechanics}

In standard Hermitian quantum mechanics an observable is represented by a Hermitian operator $O$ \cite{Sakurai2017}. Hermiticity ensures that its eigenvalues $o_\nu$, with a multi-index $\nu$, are real. The $o_\nu$ are the possible outcomes of a measurement of the observable. Every state $\left| \psi \right>$ of the Hilbert space $\mathcal{H}$\footnote{We here drop the index s introduced in Sect.~\ref{sec:ancilla}.} can furthermore be expanded in the orthonormal (right) eigenstates $\left| {\rm R}_\nu^{O} \right>$ of $O$,\footnote{As $O$ is Hermitian, right and left eigenstates are identical. We still stick to the notation introduced earlier.}  which form a basis,
\begin{equation}
\left| \psi \right> = \sum_{\nu} c_\nu^{O} \left| {\rm R}_\nu^{O} \right> ,
\label{eq:ob_def_herm_exp}    
\end{equation}
with $c_\nu^{O} \in \mathbbm{C}$. Here $\left|c_\nu^{O}\right|^2$ is the relative probability to obtain $o_\nu$ when measuring $O$, if the quantum system is prepared in the state $\left| \psi \right>$ \cite{Sakurai2017}. Employing probability theory the (real) expectation value of a repeated measurement of $O$ in the state  $\left| \psi \right>$ is given by
\begin{equation}
\left< O \right>_{\left| \psi \right>} = \frac{\sum_{\nu}  o_\nu \left|c_\nu^{O}\right|^2 }{\sum_{\nu}\left|c_\nu^{O}\right|^2} .  
\label{eq:ex_value_def}    
\end{equation}
Using the spectral representation $O= \sum_\nu o_\nu \left| {\rm R}_\nu^{O} \right> \left< {\rm R}_\nu^{O} \right|$ of $O$ and the expansion Eq.~(\ref{eq:ob_def_herm_exp}) the probabilistic expression Eq.~(\ref{eq:ex_value_def}) for the expectation value can be rewritten as an expression of linear algebra
\begin{equation}
\left< O \right>_{\left| \psi \right>} = \frac{\left< \psi \right| O \left| \psi \right>}{\left< \psi \right. \left| \psi \right>} ,
\label{eq:ex_value_rewitten}    
\end{equation}
involving the matrix element of $O$ in the state $\left| \psi \right>$. If the latter is normalized to one, the denominator drops out. 

\subsection{\texorpdfstring{$\mathcal{PT}$}{PT}-symmetric quantum mechanics}
\label{subsec:ob_ex_PT}

In Sect.~\ref{ssec:biorthogonal_product} we discussed the biorthogonal inner product which is a defining element of $\mathcal{PT}$-symmetric quantum mechanics. We now present the concepts of an observable and that of the ``expectation value'' of an observable as they are introduced within this framework \cite{Bender2019}. 
%As already the biorthogonal inner product is strongly tied to the non-Hermitian Hamiltonian $H$, it does not come as a surprise that the same holds for these two concepts. However, we already now emphasizes, that this does not make these ties more reasonable when being interested in experimentally accessible emergent quantum many-body phenomena in open systems. 

Given a general biorthonormal basis $\left\{\left|{\rm r}_\nu \right> , \left|{\rm l}_\nu \right>   \right\}$, we can express any linear operator $O$ as
\begin{equation}
O = \mathbbm{1} O \mathbbm{1} = \sum_{\nu,\mu} \left|{\rm r}_\nu \right> \left< {\rm l}_\nu \right|O\left|{\rm r}_\mu \right>\left< {\rm l}_\mu \right| = \sum_{\nu,\mu} O_{\nu,\mu}  \left|{\rm r}_\nu \right> \left< {\rm l}_\mu \right| ,
\label{eq:op_bio_basis}    
\end{equation}
where, in the last step, we defined the matrix element $ O_{\nu,\mu}  = \left< {\rm l}_\nu \right|O\left|{\rm r}_\mu \right>$. Note that this left-right-matrix element provides a proper matrix representation of $O$ in the given biorthonormal basis in the following sense. If one has two linear operators $O^{(1)}$ and $O^{(2)}$ the matrix representation of the product of the two is given by the standard matrix multiplication $\sum_{\kappa} O^{(1)}_{\nu,\kappa} O^{(2)}_{\kappa,\mu}$. This would not be the case, if we would express $O$ in terms of the non-orthogonal basis $\left\{ \left| r_\nu \right> \right\}$ \cite{Brody2013-br}. 

If we take a $\eta_{\rm r}$-pseudo-Hermitian $H$, e.g. a $\mathcal{PT}$-symmetric $H$ in its symmetry unbroken phase, and the corresponding biorthonormal basis of eigenstates, using Eq.~(\ref{eq:right_left_map}), we can write for the left-right matrix element of a linear operator $O$ in the eigenstates of this $H$ 
\begin{equation}
O_{\nu,\mu}  = \left< {\rm L}_\nu \right|O\left|{\rm R}_\mu \right> = 
 \left< {\rm R}_\nu \right|\eta_{\rm r}  O\left|{\rm R}_\mu \right> . 
\label{eq:matrix_element_bio}    
\end{equation}
For this reason we also refer to $O_{\nu,\mu}$ as the biorthogonal matrix element; it involves the biorthogonal inner product.

Let us now assume that $O$ is $\eta_{\rm r}$-pseudo-Hermitian $\eta_{\rm r} O = O^\dag \eta_{\rm r}$, $\eta_{\rm r}=\eta_{\rm r}^\dag$, i.e., it is pseudo-Hermitian with the same Hermitian $\eta$-operator as $H$. Note that this is a severe constraint on the operator $O$, as it ties $O$ to $H$. With this we get
\begin{equation}
O_{\nu,\mu} = 
 \left< {\rm R}_\nu \right|\eta_{\rm r}  O\left|{\rm R}_\mu \right> 
 =  \left< {\rm R}_\nu \right| O^\dag \eta_{\rm r} \left|{\rm R}_\mu \right> =  \left< {\rm R}_\mu \right| \eta_{\rm r} O \left|{\rm R}_\nu \right>^\ast = O_{\mu,\nu}^\ast
\label{eq:matrix_element_bio_pseudo}    
\end{equation}
and $O_{\nu,\mu}$ is a Hermitian matrix. Taking into account that $\left\{ \left|{\rm R}_\nu \right> \right\}$ forms a basis this also holds the other way around: If $O_{\nu,\mu}$ is a Hermitian matrix in the biorthonormal basis of eigenstates of $H$, $O$ is $\eta_{\rm r}$-pseudo-Hermitian.

We are now in a position to introduce what is considered to be an observable within $\mathcal{PT}$-symmetric quantum mechanics ($\mathcal{PT}$-symmetric observable) \cite{Bender2019}. Given a $\eta_{\rm r}$-pseudo-Hermitian Hamiltonian $H$, a linear operator $O$ is denoted as an observable if it is $\eta_{\rm r}$-pseudo-Hermitian as well. This ensures that its eigenvalues $o_\nu$ are real and that its biorthogonal, diagonal matrix element $\left< \psi \right| \eta_{\rm r} O \left| \psi \right>$, in an arbitrary state $\left| \psi \right>$ is real \cite{Brody2013-br,Bender2019}. To show $o_\nu \in \mathbbm{R}$, consider the eigenvalue equation $ O  \left| {\rm R}_\nu^{O} \right> = o_\nu  \left| {\rm R}_\nu^{O} \right>$,\footnote{As in the last subsection the right eigenstates of a linear operator $O$ are denoted as $\left| {\rm R}_\nu^{O} \right>$. If $O$ is non-Hermitian the right and left eigenvectors are no longer the same. If $O=H$ we drop the superscript $O$ and recover our established notation for the right eigenvectors of $H$.} and apply $\left< {\rm R}_\nu^{O} \right| \eta_{\rm r}$ from the left
\begin{align}
  o_\nu = \left< {\rm R}_\nu^{O} \right| \eta_{\rm r} O \left| {\rm R}_\nu^{O} \right> =  \left< {\rm R}_\nu^{O} \right|  O^\dag \eta_{\rm r} \left| {\rm R}_\nu^{O} \right>
  =  \left< {\rm R}_\nu^{O} \right| \eta_{\rm r}  O\left| {\rm R}_\nu^{O} \right>^\ast = o_\nu^\ast ,
\label{eq:eva_real}    
\end{align}
where we used the $\eta_{\rm r}$-pseudo-Hermiticty of $O$ and, without loss of generality, assumed that $ \left| {\rm R}_\nu^{O} \right>$ is normalized to one according to the biorthogonal inner product $\left< {\rm R}_\nu^{O} \right| \eta_{\rm r}  \left| {\rm R}_\nu^{O} \right>=1$. To prove $\left< \psi \right| \eta_{\rm r} O \left| \psi \right> \in \mathbbm{R}$ for an arbitrary $ \left| \psi \right> $, we also employ the assumed $\eta_{\rm r}$-pseudo-Hermiticity of $O$
\begin{align}
\left< \psi \right| \eta_{\rm r} O \left| \psi \right> = \left< \psi \right| O^\dag \eta_{\rm r} \left| \psi \right> = \left< \psi \right| \eta_{\rm r} O \left| \psi \right>^\ast .
\label{eq:matrix_el_real}    
\end{align}
This suggests to postulate the $\mathcal{PT}$-symmetric ``expectation value'' of a $\mathcal{PT}$-symmetric observable in a state $\left| \psi \right>$ to be
\begin{align}
\left< O \right>_{\left| \psi \right>}^{\mathcal{PT}} = \frac{\left< \psi \right| \eta_{\rm r} O \left| \psi \right>}{\left< \psi \right| \eta_{\rm r}  \left| \psi \right>} \in  \mathbbm{R}.
\label{eq:PT_exp_def}    
\end{align}
It is (formally) appealing, that Eq.~(\ref{eq:PT_exp_def}) has the same form as the algebraic expression of an expectation value in Hermitian quantum mechanics Eq.~(\ref{eq:ex_value_rewitten}), with the standard inner product replaced by the biorthogonal one. In that sense Eq.~(\ref{eq:PT_exp_def}) is a matrix element within the biorthogonal Hilbert space $\tilde{\mathcal{H}}$.

%\LG{The expectation value of a $\mathcal{PT}$-symmetric observable $O$ is then defined as \cref{eq:PT_exp_def} [see also my overleaf-comment on this equation]}

This definition of an observable has obvious weaknesses if compared to the concept of an observable in standard Hermitian quantum mechanics. 
Firstly, in $\mathcal{PT}$-symmetric quantum mechanics what is considered to be an observable depends on $\eta_{\rm r}$ and thus on the Hamiltonian. On general grounds, one might wonder, if this is reasonable, in particular, if one thinks of experimentally accessible systems. As we will exemplify below, it certainly has consequences which appear to be odd. The concept of an observable is, secondly, restricted to the case of a $\eta_{\rm r}$-pseudo-Hermitian Hamiltonian, e.g., a $\mathcal{PT}$-symmetric $H$ in its symmetry unbroken phase. The $\mathcal{PT}$-symmetric observable ceases to exists if one crosses the $\mathcal{PT}$-transition, which can be achieved by varying model parameters, that might be directly accessible in experiments. This notion seems highly questionable. One might, thirdly, wonder, how this definition matches the concept of an observable as it is used for general open quantum systems with non-Hermitian Hamiltonians which are not $\mathcal{PT}$-symmetric. In this case an observable of the system would simply be a Hermitian operator which only contains system degrees of freedom \cite{Breuer2007,Weiss2012}. It seems odd to use an alternative definition of obervables for $\mathcal{PT}$-symmetric systems, which simply represent a subclass of open quantum systems. Let us emphasize that denoting a $\eta_{\rm r}$-pseudo-Hermitian operator an observable, is semantics not physics. The ultimate question is, if the concept of a $\mathcal{PT}$-symmetric  observable $O$ is of physical and not only of mathematical relevance. 
%An observable should, e.g., be measurable.  
An observable should, in particular, be measurable in experiments.

There is a reason, why we put the expression ``expectation value''  above Eq.~(\ref{eq:PT_exp_def})  into quotation marks. To reveal this, we ask if this equation has a proper probabilistic interpretation. Let us assume that the right eigenstates of $O$, $ \left\{ \left| {\rm R}_\nu^{O} \right> \right\}$ constitute a (non-orthonormal) basis. In this case Eq.~(\ref{eq:ob_def_herm_exp}) holds for any $\left| \psi \right>$. Using this and the definition of $\eta_{\rm r}$ Eq.~(\ref{eq:ansatz_eta}),  Eq.~(\ref{eq:PT_exp_def}) can be rewritten as
\begin{align}
\left< O \right>_{\left| \psi \right>}^{\mathcal{PT}} = \frac{ \sum_{\nu,\mu,\kappa} o_\mu \left( c_\nu^O \right)^\ast c_\mu^O \left< {\rm R}_\nu^O \right. \left| {\rm L}_\kappa \right> \left< {\rm L}_\kappa \right. \left| {\rm R}_\mu^O \right>}{\sum_{\nu,\mu,\kappa} \left( c_\nu^O \right)^\ast c_\mu^O \left< {\rm R}_\nu^O \right. \left| {\rm L}_\kappa \right> \left< {\rm L}_\kappa \right. \left| {\rm R}_\mu^O \right>} .
\label{eq:bio_exp_rewr}    
\end{align}
This reduces to Eq.~(\ref{eq:ex_value_def}) only if either the right eigenvectors of $O$ are identical to those of $H$, i.e., if a common system of right eigenvectors of $O$ and $H$ exists,---the trivial option would be that $O=H$---or if $\eta_{\rm r}=1$, i.e., if $H$ and $O$ are Hermitian. For all other cases, the algebraic definition of a $\mathcal{PT}$-symmetric ``expectation value'' does not have a straightforward probabilistic interpretation. Once again, one can dismiss this as semantics, but 
%On the one side, it is mathematically appealing, that Eq.~(\ref{eq:PT_exp_def}) has the form of an expectation value as computed in standard quantum mechanics, with the inner product replaced by the biorthogonal one. On the other 
one cannot deny the danger of confusion which might occur if an ``expectation value'' does not have an obvious probabilistic interpretation. The crucial question might, however, be, if the $\mathcal{PT}$-symmetric ``expectation value'' of the $\mathcal{PT}$-symmetric observable $O$ is the outcome of a repeated measurement of $O$ in the state $\left| \psi \right>$. 

We want to elaborate on the  $\mathcal{PT}$-symmetric energy ``expectation value'' in detail. Using the expansion of the arbitrary state $\left| \psi \right> = \sum_{\nu} c_\nu \left| {\rm R}_\nu \right>$ in the right eigenstates of $H$ and the biorthonormality of the $\left\{ \left| {\rm R}_\nu \right>, \left| {\rm L}_\nu \right> \right\}$ we obtain 
\begin{align}
\left< H \right>_{\left| \psi \right>}^{\mathcal{PT}} = \frac{\left< \psi \right| \eta_{\rm r} H \left| \psi \right>}{\left< \psi \right| \eta_{\rm r}  \left| \psi \right>} = \frac{\sum_{\nu} E_\nu \left|c_\nu \right|^2}{\sum_{\mu} \left|c_\mu \right|^2},
\label{eq:bio_exp_H}    
\end{align}
i.e., a probabilistic expectation value as in Eq.~(\ref{eq:ex_value_def}). For the $\mathcal{PT}$-symmetric energy expectation value we can thus drop the quotation marks. To put it differently, if one requires that a non-Hermitian, but pseudo-Hermitian Hamiltonian $H$ with an entirely real spectrum is an observable, then Eq.~(\ref{eq:bio_exp_H}) is a reasonable definition of its (real) expectation value in the arbitrary state $\left| \psi \right>$. We will return to this in Sect.~\ref{subsec:obs_right}.  

\subsection{The position and the occupancy of the two-level model}
\label{subsec:ob_ex_PT_ex}

The definition of a $\mathcal{PT}$-symmetric observable as an operator which is $\eta_{\rm r}$-pseudo-Hermitian has serious consequences. Let us illustrate this in our two-level toy model  Eq.~(\ref{eq:spin_1_2}). Within standard Hermitian quantum mechanics the position operator on the two-site lattice is given by the Hermitian Pauli matrix $\sigma_z$. The two possible outcomes of a position measurement are $+1$, meaning that the particle was found on the left site, and $-1$, for the particle being on the right. To check whether $\sigma_z$ is also a valid observable within $\mathcal{PT}$-symmetric quantum mechanics we compute 
\begin{equation}
 \sqrt{1-z^2} \eta_{\rm r} \sigma_z \doteq 
\left( \begin{array}{cc}
1 & -i z e^{i \phi} \\
i z e^{- i \phi} & 1  
\end{array} 
\right) \left( 
\begin{array}{cc}
1 &  0 \\
0 & -1  
\end{array} 
\right) = \left( 
\begin{array}{cc}
1 & i z e^{i \phi} \\
i z e^{- i \phi} & -1  
\end{array} 
\right)
\label{eq:eta_r_sigma_z}    
\end{equation}
and 
\begin{equation}
 \sqrt{1-z^2} \sigma_z^\dag \eta_{\rm r} \doteq 
\left( \begin{array}{cc}
1 &  0 \\
0 & -1  
\end{array} 
\right)  \left( 
\begin{array}{cc}
1 & -i z e^{i \phi} \\
i z e^{- i \phi} & 1  
\end{array} 
\right)  =  \left( 
\begin{array}{cc}
1 & - i z e^{i \phi} \\
- i z e^{- i \phi} & -1  
\end{array} 
\right) .
\label{eq:sigma_z_eta_r}    
\end{equation} 
As $\eta_{\rm r} \sigma_z \neq \sigma_z^\dag \eta_{\rm r}$, for $z \neq 0$, $\sigma_z$ is not $\eta_{\rm r}$-pseudo-Hermitian and thus not an observable in the sense of $\mathcal{PT}$-symmetric quantum mechanics; to put it differently, what one would like to call the position operator is not an observable. 

The same holds for the continuum model with Hamiltonian $H= (\hat p^2 + \hat x^2)/2 + ig \hat x^3$ [see \cref{eq:sci_post_ham}]. In this the position operator $\hat x$ is not an observable in the sense of $\mathcal{PT}$-symmetric quantum mechanics \cite{Bender2019,Grunwald2022}. Bender argues that there is no fundamental problem with the absence of a position operator in  $\mathcal{PT}$-symmetric quantum mechanics; see Sect.~3.4 of \cite{Bender2019}. Relating to the continuum model Eq.~(\ref{eq:sci_post_ham}) (and similar ones), an analogy to relativistic quantum field theory is drawn. In this the notion of a position operator is only meaningful in the non-relativistic limit which, in our context, would correspond to the Hermitian limit. However, we have severe difficulties to accept this as a reason to deny the existence of a position operator and, for that matter, a position measurement, in our toy model of a particle hopping between two lattice sites. 

Another Hermitian operator in our  two-level problem, which one would
% like to
intuitively associate to a measurement, is 
\begin{equation}
\hat d_{\rm L} = \left| \uparrow \right> \left< \uparrow \right| \doteq 
 \left( 
\begin{array}{cc}
1 & 0 \\
0 & 0 
\end{array} 
\right) .
\label{eq:left_lattice_site_occ}
\end{equation}
It represents the occupancy of the left lattice site and has eigenvalues 1 (occupied) and 0 (unoccupied). A simple calculation similar to that of Eqs.~(\ref{eq:eta_r_sigma_z}) and (\ref{eq:sigma_z_eta_r}) shows that also $\hat d_{\rm L}$ is not $\eta_{\rm r}$-pseudo-Hermitian. One would thus conclude that it cannot be measured. In Sect.~\ref{subsec:further_examples_2_2} we will discuss that the occupancy of the left lattice site, 
%\LG{evaluated as $\expval{\hat d\_{L}} = \mel{\psi}{\hat d\_{L}}{\psi} / \ip{\psi}{\psi}$} 
was, however, measured in experiments \cite{Wu2019,Dogra2021}. One cannot reconcile this within $\mathcal{PT}$-symmetric quantum mechanics. The application of the conventional ideas of an observable and expectation value from Hermitian quantum mechanics will provide a way out of this dilemma; see Sects.~\ref{subsec:obs_right} and \ref{subsec:further_examples_2_2}.  

As we are at a crucial stage of the development of a methodology to treat non-Hermitian and pseudo-Hermitian quantum many-body systems let us summarize what is accomplished within the formalism of $\mathcal{PT}$-symmetric quantum mechanics. The above postulates of a $\mathcal{PT}$-symmetric observable and of a $\mathcal{PT}$-symmetric ``expectation value'' fulfill the mandatory requirements that the eigenvalues of observables and the  ``expectation value'' are real. They have two more advantages. Firstly, in this framework the non-Hermitian but $\eta_{\rm r}$-pseudo-Hermitian Hamiltonian $H$ is a valid observable and the expression for the energy expectation value has a probabilistic interpretation. Secondly, for all observables a formal equivalence to the algebraic expression for the expectation value  known from Hermitian quantum mechanics is achieved, if the standard inner product is replaced by the biorthogonal one. This leads to a consistent formalism within the biorthogonal Hilbert space $\tilde{ \mathcal{H}}$. As we will discuss in Sect.~\ref{sec:genfun} $\mathcal{PT}$-symmetric quantum mechanics can be extended to quantum field theory allowing to use standard concepts such as generating functionals, path integrals, Green and vertex functions, diagrammatic perturbation theory etc. \cite{Bender2019,Mostafazadeh2007-xq,Rivers2011-fo,Grunwald2022}. For these reasons the above definitions are used in mathematical physics \cite{Brody2013-br,Bender2019} and non-Hermitian quantum field theory \cite{Bender2005_dual,Bender2019,Mostafazadeh2007-xq,Jones2007,Rivers2011-fo,Jones2009-hx}. They have also been adopted  by parts of the condensed matter community to study quantum many-body systems \cite{Ghatak2018,Yamamoto2019-pn,Yamamoto2022,Zhang2022,simQuantumMetric2023,Yamamoto2023,Yu2023,Kornich2022a,Kornich2023}  %VM: added Yamamoto2019-pn and Yamamoto2023 and Yu2023 %VM: added Kornich2023 and Kornich2022a
and non-Hermitian topological systems \cite{Kunst2018,Ghatak2019,Bergholtz2019-ss,Groenendijk2021}.\footnote{Another important work on non-Hermitian topological systems is \cite{Yao2018}. In this, the term non-Hermitian skin effect was coined and the non-Hermitian bulk-boundary correspondence was investigated.} %VM: Footnote with Yao et al added

However, the above postulates of a $\mathcal{PT}$-symmetric observable and of a $\mathcal{PT}$-symmetric ``expectation value''  have limiting weaknesses which cannot be ignored. On the one hand, this concerns general issues, such as the restriction to systems with $\eta_{\rm r}$-pseudo-Hermitian Hamiltonians, the linkage of the concept of an observable to the Hamiltonian, and the lack of a probabilistic interpretation of the algebraic expression which is considered to be the ``expectation value''. The first, e.g., implies that one cannot investigate the behavior of the  ``expectation value'' of an observable across a $\mathcal{PT}$ transition; it is simply not defined in the symmetry broken phase. There, furthermore, appears a discontinuity in the definition and evaluation of observables, when comparing general open system formalisms with $\mathcal{PT}$-symmetric quantum mechanics.
On the other, hand specific problems occur, such as in the above examples, in which operators which we intuitively consider to be the observable for the position of a quantum particle or the occupancy of a level turn out not to be an observable in the $\mathcal{PT}$-symmetric sense. 
%In Sect.~\ref{subsec:further_examples} we will give further examples which illustrate the problems of the postulates of a biorthogonal observable and the biorthogonal ``expectation value''. 

All of this are formal (and in parts even aesthetic) considerations. The ultimate question is, if $\mathcal{PT}$-symmetric observables and $\mathcal{PT}$-symmetric ``expectation values'' can be measured. This does not seem to be the case. We are not aware of a single work in which a $\mathcal{PT}$-symmetric observable (besides the energy) 
%\LG{more on this later})
was measured or in which a potential way how to measure such was described \cite{Yamamoto2023}. From the perspective of physics the %VM: added Yamamoto2023
formalism of $\mathcal{PT}$-symmetric quantum mechanics does not appear to be the first choice however beautiful it may seem mathematically. After extending our analysis to the slightly different definitions of biorthogonal observables and biorthogonal ``expectation values'' \cite{Brody2013-br} in the next subsection, in Sect.~\ref{subsec:obs_right} we discuss that applying the definitions of observables and expectation values of standard Hermitian quantum mechanics to non-Hermitian systems does not lead to any of the above problems and is further more consistent with the broader range of open system formalism.

\subsection{Biorthogonal quantum mechanics}
\label{subsec:ob_ex_bio}

Although $\mathcal{PT}$-symmetric quantum mechanics and biorthogonal quantum mechanics have a common ground \cite{Brody2013-br,Bender2019} the definition of the concepts of observables and  ``expectation values'' differ slightly. In biorthogonal quantum mechanics, instead of referring to the biorthonormal basis of eigenstates of a given $H$, one considers a general biorthonormal basis $\left\{\left|{\rm r}_\nu \right> , \left|{\rm l}_\nu \right>   \right\}$ and the corresponding metric operator $\hat g$ Eq.~(\ref{eq:ansatz_hat_g}). The biorthogonal matrix representation of an operator $O$ is then given by Eq.~(\ref{eq:op_bio_basis}). One now denotes an operator $O$ as a biorthogonal observable if $O_{\nu,\mu}$, represented in some biorthogonal basis, is Hermitian. This ensures that the biorthogonal  ``expectation values''
\begin{align}
\left< O \right>_{\left| \psi \right>}^{\rm bo} = \frac{\left< \psi \right| \hat g O \left| \psi \right>}{\left< \psi \right| \hat g  \left| \psi \right>}
\label{eq:bio_exp_def}    
\end{align}
is real for all states $\left| \psi \right>$. If  $\left| \psi \right>$  is an eigenstate of $O$ this also implies that the eigenvalues of $O$ are real. The minimal requirements for an observable and an ``expectation value'' are thus fulfilled. As in $\mathcal{PT}$-symmetric quantum mechanics this ``expectation value'' does not have a probabilistic interpretation (up to the exceptional case that the $\left\{\left|{\rm r}_\nu \right>  \right\}$ are eigenvectors of $O$ \cite{Brody2013-br}) such that the quotation marks are justified here as well.
 
In biorthogonal quantum mechanics the selected metric operator affects what is denoted as an observable. It might turn out that a certain operator is an observable for one fixed biorthogonal basis but not for another one. This is equally troubling as the dependence of the concept of an observable on the Hamiltonian in ${\mathcal PT}$-symmetric quantum mechanics. The formalism does not provide an answer which of the possibly many biorthogonal basis sets to select. On physical grounds, there appears to be only one which is distinguished. This is the set of eigenvectors of a given non-Hermitian Hamiltonian. With this one recovers the definitions of $\mathcal{PT}$-symmetric quantum mechanics (at least in the phase of real eigenvalues of $H$; see below) and all the criticism raised in the last subsection applies. It is, in addition, not obvious how the definition of biorthogonal quantum mechanics matches the one used for open systems with general non-Hermitian Hamiltonians (see the last subsection). The only advantage as compared to the $\mathcal{PT}$-symmetric observable is, that the definition is not bound to the phase in which the eigenvalues of a given Hamiltonian are entirely real. The biorthogonal basis exists in both phases such that one can follow an ``expectation value'' across the phase transition. 

To summarize, the definitions of a biorthogonal observable and a biorthogonal ``expectation value'' have their mathematical justification but their usefulness is restricted to the analysis of spectral properties.

\subsection{Non-Hermitian quantum systems}
\label{subsec:obs_right}

As the basis of another definition of the concepts of observables and expectation values for non-Hermitian quantum systems, three properties from Hermitian quantum mechanics can be adopted: 
\begin{enumerate}
\item An observable $O$ is independent of the Hamiltonian $H$ and the selected basis. 
%In particular, one must be able to define it for every Hamiltonian.  
\item The real eigenvalues $o_\nu$ of $O$ are the possible outcomes of a measurement of the observable $O$.
\item The expectation value of $O$ in an arbitrary state is real and has a probabilistic interpretation. 
\end{enumerate}
All this can be achieved if we simply take an observable $O$ to be a Hermitian operator. This  implies that its eigenvalues $o_\nu$ are real and that the right eigenstates $\left\{ \left|{\rm R}_\nu^O \right> \right\}$ form an orthonormal basis right away. An arbitrary state can be expanded in these; see Eq.~(\ref{eq:ob_def_herm_exp}). With this, the appropriate probabilistic expression for the real expectation value is Eq.~(\ref{eq:ex_value_def}), which can be rewritten in algebraic form as
\begin{equation}
\left< O \right>_{\left| \psi \right>} = \frac{\left< \psi \right| O \left| \psi \right>}{\left< \psi \right. \left| \psi \right>} .
\label{eq:ex_value_rewitten_nochmal}    
\end{equation}
We thus do not use the biorthogonal inner product to define the expectation value but rather the standard one. We refer to these definitions as Hermitian observables and Hermitian expectation values.

The definition of a Hermitian observable implies that the non-Hermitian Hamiltonian $H$ is not an observable, even if it has an entirely real spectrum. As $H$ describes the dynamics of an open quantum system, this does not seem to be utterly surprising. We believe that this restriction is less severe than the shortcomings of the postulates of $\mathcal{PT}$-symmetric and biorthogonal observables and ``expectation values'' discussed in the last two subsection. It can, furthermore, be overcome. If one considers a $\eta_{\rm r}$-pseudo-Hermitian Hamiltonian with real spectrum,   for the definition of an energy expectation value Eq.~(\ref{eq:bio_exp_H}) can be used, which has a probabilistic interpretation. A matrix element of this type appears in the computation of the ground state energy in perturbation theory. There, the leading order correction to the energy is given by the biorthogonal matrix element of the part of the Hamiltonian, that is considered as the perturbation  \cite{Budich2020}.\footnote{This was employed to show that the ground state energy in a non-Hermitian system can strongly depend on the boundary conditions \cite{Budich2020}. The authors suggested to use this effect in a highly sensitive sensor.} Equation (\ref{eq:bio_exp_H}) can also be employed for other non-Hermitian operators $O$ with entirely real spectra and a complete biorthonormal set  
$\left\{ \left| {\rm R}_\nu^O \right> ,  \left| {\rm L}_\nu^O \right>  \right\}$ of eigenvectors by replacing $H \to O$ and $\eta_{\rm r} \to \eta_{\rm r}^O = \sum_{\nu}  \left| {\rm L}_\nu^O \right>  \left< {\rm L}_\nu^O \right|$. Note that the use of the operator $\eta_{\rm r}^O$ specific to $O$ prevents the linking of $H$ and the observable $O$. If desired, this opens a path towards observables which are represented by non-Hermitian operators with real eigenvalues. However, in the following we focus on Hermitian observables.
% we here restrict our considerations to Hermitian ones. 

As emphasized above some authors from the quantum many-body community use the postulates for observables and ``expectation values'' of $\mathcal{PT}$-symmetric and biorthogonal quantum mechanics \cite{Ghatak2018,Yamamoto2019-pn,simQuantumMetric2023,Groenendijk2021,Yu2023,Kornich2022a}. However, others employ %VM: added Yamamoto2019-pn and Yu2023 %VM: added Kornich2022a
the alternative ones described in the present subsection \cite{doraKibbleZurekMechanism2019,Dora2020-oi,Moca2021,Dora2022,Sticlet2022,Dora2022a,Tetling2022,Turkeshi2023}. Some even present results for expectation values of observables obtained by both definitions in a single paper \cite{Herviou2019,Yamamoto2022,Zhang2022,Yamamoto2023,Kornich2023}. One of the %VM: added Yamamoto2023 %VM: added Kornich2023
central goals of this review is to contribute to a resolution of this conflict. We do so by showing that the  postulates directly adopted from Hermitian quantum mechanics are more reasonable on general grounds and give physically sensible results in all our examples. Furthermore, they are consistent with established open system formalism.

\subsection{Time dependence and the Heisenberg picture}
\label{subsec:exp_time_dep_state}

We next discuss the difference between time-dependent expectation values in $\mathcal{PT}$-symmetric quantum mechanics and employing the postulates adopted from Hermitian quantum mechanics. Note that in the former case the non-Hermitian Hamiltonian $H$ must be $\eta_{\rm r}$-pseudo-Hermitian while in the latter, $H$ can be arbitrary.  

We start out with the perspective of $\mathcal{PT}$-symmetric quantum mechanics. Let us assume that $O$ is a $\mathcal{PT}$-symmetric observable. The ``expectation value'' of $O$ in the state $\left| \psi(t) \right>$ time-evolved according the Schr\"odinger equation with a time-independent Hamiltonian is
\begin{align}
 \left< O \right>_{\left| \psi (t)\right>}^{\mathcal{PT}} & = \frac{\left< \psi (t) \right| \eta_{\rm r} O \left| \psi (t) \right>}{\left< \psi (t) \right| \eta_{\rm r}  \left| \psi (t) \right>} =    \frac{\left< \psi (0) \right| e^{i H^\dag t} \eta_{\rm r} O e^{-i H t} \left| \psi (0) \right>}{\left< \psi (0) \right| \eta_{\rm r}  \left| \psi (0) \right>} \nonumber \\
 & =  \frac{\left< \psi (0) \right| \eta_{\rm r} e^{i H t} O e^{-i H t} \left| \psi (0) \right>}{\left< \psi (0) \right| \eta_{\rm r}  \left| \psi (0) \right>}  \nonumber \\
 & =  \frac{\left< \psi (0) \right| \eta_{\rm r} O_{\rm H}(t) \left| \psi (0) \right>}{\left< \psi (0) \right| \eta_{\rm r}  \left| \psi (0) \right>} =  \left< O_{\rm H}(t) \right>_{\left| \psi (0)\right>}^{\mathcal{PT}} .
\label{eq:exp_val_time_evol_bio}    
\end{align}
In the first line, we employed the effective unitarity of the time evolution when using the biorthogonal inner product Eq.~(\ref{eq:alternative_inner_p_alternativ}) and in the second, the $\eta_{\rm r}$-pseudo-Hermiticity relation Eq.~(\ref{eq:eta_doesit}).
In the third line, the observable in the Heisenberg picture $O_{\rm H}(t)$ is introduced. It is defined as in Hermitian quantum mechanics
\begin{equation}
O_{\rm H}(t) =   e^{i H t} O e^{-i H t}   .
\label{eq:heisenberg_pict}    
\end{equation}
Note that the non-Hermiticity of $H$ is not apparent from \cref{eq:heisenberg_pict}. Without loss of generality we can assume that $\left| \psi(0) \right>$ is normalized to one according to the biorthogonal inner product. In this case the denominator in Eq.~(\ref{eq:exp_val_time_evol_bio}) can be dropped. That the Heisenberg picture can be employed is another formally appealing feature of the postulates of a $\mathcal{PT}$-symmetric  observable and of a $\mathcal{PT}$-symmetric  ``expectation value''. With this, the Heisenberg equation of motions for the observable and the Ehrenfest theorem for its ``expectation value'' hold as in Hermitian quantum mechanics \cite{Sakurai2017}. This further extends the mathematical analogy between  $\mathcal{PT}$-symmetric quantum mechanics for a $\eta_{\rm r}$-pseudo-Hermitian $H$ and the formalism of standard quantum mechanics.   

Within the framework taken from Hermitian quantum mechanics the expression for the time-dependent expectation value turns out to be mathematically less appealing. We obtain
\begin{equation}
 \left< O \right>_{\left| \psi (t)\right>} = \frac{\left< \psi (t) \right| O \left| \psi (t) \right>}{\left< \psi (t) \right.  \left| \psi (t) \right>} =    \frac{\left< \psi (0) \right| e^{i H^\dag t} O e^{-i H t} \left| \psi (0) \right>}{\left< \psi (0) \right|  e^{i H^\dag t}  e^{-i H t}  \left| \psi (0) \right>} .
 \label{eq:exp_val_time_evol_nonbio}    
\end{equation}
No conventional Heisenberg picture can be employed \cite{Dattoli1990}\footnote{Note that in \cite{Dattoli1990} the definition of the operator expectation value lacks the denominator of Eq.~(\ref{eq:exp_val_time_evol_nonbio}).} and, even worse, the denominator never drops out. Even if we assume that the initial state $\left| \psi (0) \right>$ is normalized to one according to the standard inner product, due to the non-unitarity of the time evolution, the denominator is a non-trivial function of $t$ and cannot be dropped. This will turn out to be one of the main challenges when it comes to the explicit computation of expectation values (and correlation functions) of quantum many-body systems. Needless to say, we cannot discard this methodology based on the lack of formal beauty or the abundance of technical challenges. What matters is the physical sensibility of an approach.   

\subsection{The ancilla approach}
\label{subsec:ob_anc}

In this subsection we show, that the concepts of Hermitian observables and Hermitian expectation values of Sect.~\ref{subsec:obs_right}, naturally fit into the framework of the ancilla approach which can be used to embed  non-Hermitian systems into Hermitian ones; see Sect.~\ref{sec:ancilla}. In fact, the ancilla approach enforces a definition of observables according to the Hermitian postulates.

To study the physics of a non-Hermitian system with Hamiltonian $H_{\rm s}$ we determine the Hermitian $H_{\rm sa}$ and the linear operator $g$ as described in  Sect.~\ref{sec:ancilla}. This might be difficult in practice, given a non-Hermitian (maybe even time-dependent) $H_{\rm s}$, but it is formally always possible. We start in the initial state Eq.~(\ref{eq:ancilla_initialstate_form}) and time evolve this up to time $t$, applying the unitary time evolution operator $U_{\rm sa}$ associated to $H_{\rm sa}$. At that time we perform the measurement on the ancilla spin. Right after the measurement we are in the state $\left| \psi_{\rm sa}^\uparrow (t) \right>$. This is, by construction, normalized to one with respect to the standard inner product on $\mathcal{H}_{\rm sa}$. We now perform a measurement of the observable associated to the operator $\mathbbm{1}_{\rm a} \otimes O$ acting on $\mathcal{H}_{\rm sa}$ in the state $\left| \psi_{\rm sa}^\uparrow (t) \right>$. This operator is Hermitian as long as $O$ is a Hermitian operator on $\mathcal{H}_{\rm s}$, which we assume to be the case. As the Hamiltonian $H_{\rm sa}$ is Hermitian by construction we are not tempted to employ any other formalism than the one of standard quantum mechanics to compute the operator expectation value
\begin{align}
\left< \mathbbm{1}_{\rm a} \otimes O \right>_{\left| \psi_{\rm sa}^\uparrow (t) \right>} & = \left< \psi_{\rm sa}^\uparrow (t) \right| \mathbbm{1}_{\rm a} \otimes O \left| \psi_{\rm sa}^\uparrow (t) \right>  \nonumber \\
& = \frac{\left< \psi_{\rm s} (t) \right| O \left| \psi_{\rm s} (t) \right> }{\left< \psi_{\rm s} (t) \right. \left| \psi_{\rm s} (t) \right>} = \left< O \right>_{\left| \psi_{\rm s} (t) \right>} ,
\label{eq:exp_value_ancilla}
\end{align}
where we used Eqs.~(\ref{eq:ancilla_proj}) and (\ref{eq:ancilla_norm}). Note that the inner products in the second line are standard ones on $\mathcal{H}_s$. The second line is directly obtained if one does not extend the non-Hermitian system to a Hermitian one and uses the formalism put forward in Sect.~\ref{subsec:obs_right}. The corresponding state is $\left| \psi_{\rm s} (t) \right> $, time-evolved out of the initial one  $\left| \psi_{\rm s} (0) \right> $ employing the non-unitary time evolution operator $U_{\rm s}$, associated to $H_{\rm s}$, and the observable is represented by the Hermitian operator $O$. 

\subsection{Employing the isospectral Hamiltonian}
\label{subsec:isospectral}

Theorem $\mbox{T}^{\mathcal{PT}}_{2}$ states that for every ${\mathcal{PT}}$-symmetric Hamiltonian $H$ with unbroken ${\mathcal{PT}}$ symmetry, i.e., if $H$ has an entirely real spectrum, there exists a Hermitian Hamiltonian $h$, which is isospectral to $H$ and is related to this by a similarity transformation: $h=S H S^{-1}$. We have clarified that $S=\eta_{\rm r}^{1/2}$. In the special cases, in which $h$ and $\eta_{\rm r}^{1/2}$ are known explicitly, see, e.g., Sect.~\ref{subsec:examples_2_2}, one can also use this theorem to compute expectation values. Depending on which set of postulates for the concept of an observable and that of the expectation value is used, this leads to different expressions.

Let us start out with the Hermitian postulates of Sect.~\ref{subsec:obs_right}. In this case $O$ is Hermitian and we obtain
\begin{align}
 \left< O \right>_{\left| \psi (t)\right>} &  = \frac{\left< \psi (t) \right| O \left| \psi (t) \right>}{\left< \psi (t) \right.  \left| \psi (t) \right>} =    \frac{\left< \psi (0) \right| e^{i H^\dag t} O e^{-i H t} \left| \psi (0) \right>}{\left< \psi (0) \right|  e^{i H^\dag t}  e^{-i H t}  \left| \psi (0) \right>} \\* \nonumber
 & = \frac{\left< \psi (0) \right| \eta_{\rm r}^{1/2} e^{i h t} \eta_{\rm r}^{-1/2} O   \eta_{\rm r}^{-1/2} e^{-i h t} \eta_{\rm r}^{1/2} \left| \psi (0) \right>}{\left< \psi (0) \right| \eta_{\rm r}^{1/2}  e^{i h t} \eta_{\rm r}^{-1} e^{-i h t} \eta_{\rm r}^{1/2} \left| \psi (0) \right>} .
\end{align}
This shows that using the isospectral, Hermitian Hamiltonian does neither simplify the formal structure of the expectation value of an observable in a time-dependent state nor can one expect that it simplifies its computation. 

If $O$, in contrast, is an observable in the sense of $\mathcal{PT}$-symmetric quantum mechanics and we use the corresponding ``expectation value'' (see Sect.~\ref{subsec:ob_ex_PT}) we obtain

\begin{align}
\left< O \right>_{\left| \psi (t)\right>}^{\mathcal{PT}} &= \frac{\left< \psi (t) \right| \eta_{\rm r} O \left| \psi (t) \right>}{\left< \psi (t) \right| \eta_{\rm r}  \left| \psi (t) \right>} =    \frac{\left< \psi (0) \right| e^{i H^\dag t} \eta_{\rm r} O e^{-i H t} \left| \psi (0) \right>}{\left< \psi (0) \right| \eta_{\rm r}  \left| \psi (0) \right>} \notag \\
& =  \frac{\left< \psi (0) \right| \eta_{\rm r}^{1/2}  e^{i h t} \eta_{\rm r}^{1/2} O \eta_{\rm r}^{-1/2} e^{-i h t} \eta_{\rm r}^{1/2} \left| \psi (0) \right>}{\left< \psi (0) \right| \eta_{\rm r}  \left| \psi (0) \right>} .
\label{eq:obersable_time_evolution_pt}    
\end{align}
With $O$ being $\eta_{\rm r}$-pseudo-Hermitian, $ \tilde{O} = \eta_{\rm r}^{1/2} O \eta_{\rm r}^{-1/2} $, appearing in the numerator, is a Hermitian operator and thus an observable in the Hermitian sense
\begin{equation}
\left[ \eta_{\rm r}^{1/2} O \eta_{\rm r}^{-1/2}\right]^\dag \!\! = \!  \eta_{\rm r}^{-1/2} O^\dag  \eta_{\rm r}^{1/2} =  \! \eta_{\rm r}^{-1/2} O^\dag  \eta_{\rm r} \eta_{\rm r}^{-1/2} = \! \eta_{\rm r}^{-1/2} \eta_{\rm r} O  \eta_{\rm r}^{-1/2} = \! \eta_{\rm r}^{1/2} O \eta_{\rm r}^{-1/2} ,
\label{eq:eta_O_eta}    
\end{equation}
where we used the pseudo-Hermiticity relation Eq.~(\ref{eq:eta_doesit}) for $O$. Written in terms of the Hermitian operator $\tilde O$, Eq.~(\ref{eq:obersable_time_evolution_pt}) simplifies to 
\begin{align}
 \left< O \right>_{\left| \psi (t)\right>}^{\mathcal{PT}} 
 & =  \frac{\left< \psi (0) \right| \eta_{\rm r}^{1/2}  e^{i h t}  \tilde O e^{-i h t} \eta_{\rm r}^{1/2} \left| \psi (0) \right>}{\left< \psi (0) \right| \eta_{\rm r}  \left| \psi (0) \right>} \nonumber \\
 & =  \frac{\left< \varphi(0) \right|  e^{i h t}  \tilde O e^{-i h t} \left| \varphi (0) \right>}{\left< \varphi (0) \right. \left| \varphi (0) \right>} ,  
\end{align}
with $ \left| \varphi (0) \right> = \eta_{\rm r}^{1/2}  \left| \psi (0) \right>$. All the operators appearing on the right hand side are Hermitian and the time evolution is unitary. The formal structure is simple.  

In fact, given a  $\mathcal{PT}$-symmetric observable $O_{\mathcal{PT}}$ a corresponding Hermitian one $O_{\rm herm}$ can always be defined as
\begin{equation}
O_{\rm herm} = \eta_{\rm r}^{1/2} O_{\mathcal{PT}} \eta_{\rm r}^{-1/2} 
\label{eq:op_pair}    
\end{equation}
and vice versa
\begin{equation}
O_{\mathcal{PT}} = \eta_{\rm r}^{-1/2} O_{\rm herm} \eta_{\rm r}^{1/2} . 
\label{eq:op_pair_vice}    
\end{equation}
Consider for example our $2 \times 2$-matrix Hamiltonian Eq.~(\ref{eq:spin_1_2}). As discussed in Sect.~\ref{subsec:ob_ex_PT_ex}, we intuitively would like to interpret the Hermitian operator $\sigma_z$ as the observable for the position. However, it is not $\eta_{\rm r}$-pseudo-Hermitian and thus not an observable in the biorthogonal sense. Within $\mathcal{PT}$-symmetric quantum mechanics one could now take $\eta_{\rm r}^{-1/2} \sigma_z \eta_{\rm r}^{1/2}$ instead, which is $\eta_{\rm r}$-pseudo-Hermitian and thus a valid observable in this framework. However, the physical meaning of this operator is far from obvious. A reasoning similar to this plays an important role when setting up functional integrals for $\mathcal{PT}$-symmetric and pseudo-Hermitian systems \cite{Mostafazadeh2007-xq,Jones2007,Rivers2011-fo,Jones2009-hx,Grunwald2022}. We will return to this in Sect.~\ref{sec:genfun}.

\subsection{The occupancy in the non-Hermitian two-level problem}
\label{subsec:further_examples_2_2}

We now return to our non-Hermitian toy problem of a two-site lattice system occupied by a single particle. It was introduced in Sect.~\ref{subsec:examples_2_2}. As discussed in Sect.~\ref{subsec:ob_ex_PT_ex}, on physical grounds we consider $\hat d_{\rm L} = \left| \uparrow \right> \left< \uparrow \right|$ to be the Hermitian operator which represents the occupancy of the left lattice site. Initially the particle is assumed to be localized on this site, associated to the eigenvalue $+1$ of $\hat d_{\rm L}$. We thus start in the state $\left| \psi (0) \right> = \left| \uparrow \right> $. 

\begin{figure}
    \centering
    \includegraphics{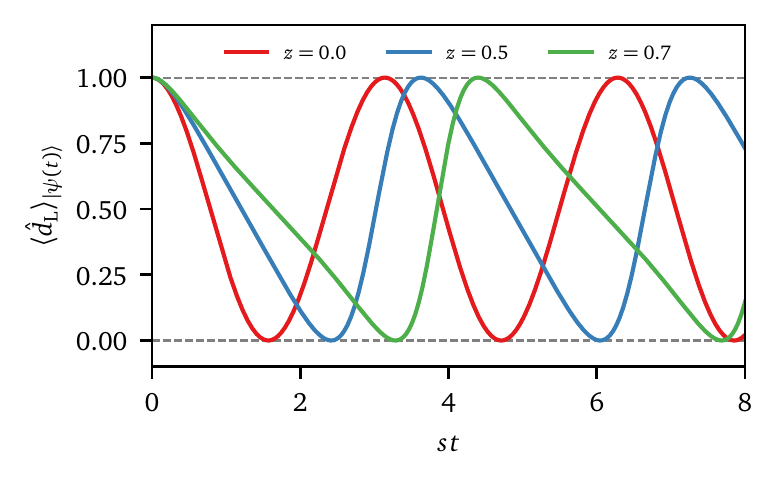}
    \caption{\textbf{Occupancy of the left lattice site \cref{eq:ex_eval_spin_1_2} in the $\bm{\mathcal{PT}}$-unbroken phase} of the two-level toy model  for various $|z| < 1$. Starting from Rabi oscillations at $z = 0$ [see \cref{eq:ex_eval_spin_1_2_z_0}] the non-Hermiticity deforms the time evolution with increasing $z$, leading to a distinct non-sinusoidal time dependence.}
    \label{fig:complex_rabi}
\end{figure}

We first consider the regime of unbroken $\mathcal{PT}$ symmetry with $|z|<1$. The time-evolved state $\left| \psi(t) \right>$ under the action of the non-unitary time evolution operator $e^{-i H t}$ was computed in Sect.~\ref{subsec:example_spin_1_2} and is given in Eq.~(\ref{eq:time_evol_psi_s}). To obtain the Hermitian expectation value 
\begin{equation}
\left<  \hat d_{\rm L}  \right>_{\left|\psi(t) \right>} = 
\frac{ \ip{\psi(t)}{ \uparrow} \ip{\uparrow}{\psi(t)}}{\left< \psi(t) \right. \left| \psi(t) \right>}
\label{eq:herm_exp_spin_1_2}    
\end{equation}
we still need to compute the numerator. The denominator is given in Eq.~(\ref{eq:norm_psi_s}). In a straightforward computation using the results of Sects.~\ref{subsec:examples_2_2}  and  \ref{subsec:example_spin_1_2} we obtain
\begin{equation}
  \left<  \hat d_{\rm L} \right>_{\left|\psi(t) \right>} = 
\frac{\frac{1}{2} \left[ 1+ \cos{\left(\omega_z s t \right)} \right]
-  z^2 \cos{\left(\omega_z s t\right)}
+ z \sqrt{1-z^2}  \sin{\left(\omega_z s t\right)}  }{1- z^2 \cos{\left(\omega_z s t\right)} + z \sqrt{1-z^2} \sin{\left(\omega_z s t\right)}} .
\label{eq:ex_eval_spin_1_2}    
\end{equation}
This is a periodic function with a (dimensionless) $z$-dependent frequency set by the difference of the two eigenvalues $\omega_z = (E_+-E_-)/s = 2 \sqrt{1-z^2}$. In the Hermitian case $z=0$, this simplifies to
\begin{equation}
\left<   \hat d_{\rm L} \right>_{\left|\psi(t) \right>}  \stackrel{z=0}{=}
\frac{1}{2} \left[ 1+ \cos{\left(2 s t\right)} \right] ,
\label{eq:ex_eval_spin_1_2_z_0}    
\end{equation}
which is a manifestation of Rabi oscillations for two energetically degenerate, coupled levels. The particle is oscillating back and  forth between the left and the right lattice site in a simple, sinusoidal form.  Figure \ref{fig:complex_rabi} shows $\left<  \hat d_{\rm L}\right>_{\left|\psi(t) \right>}$  Eq.~(\ref{eq:ex_eval_spin_1_2}) as a function of $t$ for different $z$, including the case $z=0$. For increasing $z \in [0,1)$ the oscillation becomes increasingly non-sinusoidal. This complex line shape originates from the non-Hermiticity of the Hamiltonian, or, to put it in physical terms, from the coupling of the hopping particle to an environment. However, the coupling is special as it leads to an open system Hamiltonian which is $\mathcal{PT}$-symmetric. As a consequence the oscillation is undamped. Achieving this symmetry, and thus the undamped but non-sinusoidal oscillation requires fine-tuning of the coupling which manifests in the special form of the Hamiltonian Eq.~(\ref{eq:spin_1_2}). This is often phrased as a balance of gain and loss \cite{Bender2005-py,Bender2007-tf,Bender2019,Ashida2020-vp}.

\begin{figure}
    \centering
    \includegraphics{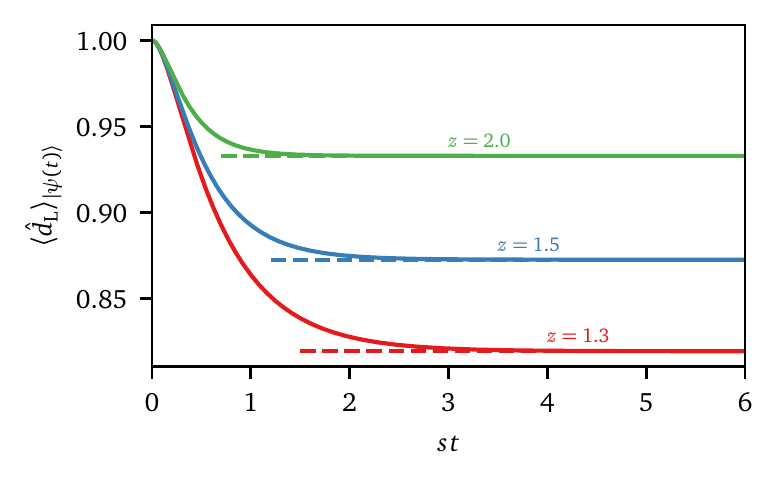}
    \caption{\textbf{Occupancy of the left lattice site \cref{eq:ex_eval_spin_1_2_broken} in the $\bm{\mathcal{PT}}$-broken phase} of the two-level toy model for various $|z| > 1$. Color matched dashed lines indicate the asymptotic limit $t \to \infty$ [see \cref{eq:ex_eval_spin_1_2_broken_t_inf}], which always remains different from 1/2 due to the balanced loss and gain of the $\mathcal{PT}$-symmetric system.}
    \label{fig:decay_finite}
\end{figure}

As we are using the concepts of observables and expectation values from Hermitian quantum mechanics we can also compute the expectation value $\left<   \hat d_{\rm L} \right>_{\left|\psi(t) \right>}$ in the $\mathcal{PT}$ symmetry broken phase $|z|>1$. We again start in the state with the particle located on the left lattice site. Using the results for the eigenvalues and eigenvectors presented in Sect.~\ref{subsec:examples_2_2} the calculation is straightforward and we do not give any details here. We obtain
\begin{equation}
\left<  \hat d_{\rm L} \right>_{\left|\psi(t) \right>}  =
\frac{e^{\gamma_z st} \left( z^2 + z \sqrt{z^2-1} -\frac{1}{2} \right)  +  e^{- \gamma_z st} \left( z^2 - z \sqrt{z^2-1} - \frac{1}{2} \right) -  1}{e^{\gamma_z st} \left( z^2 + z \sqrt{z^2-1} \right) +  e^{-\gamma_z st} \left( z^2 - z \sqrt{z^2-1}\right)  - 2} .
\label{eq:ex_eval_spin_1_2_broken}    
\end{equation}
The expectation value decays monotonically from 1 to the asymptotic limit 
\begin{equation}
 \left<  \hat d_{\rm L}  \right>_{\left|\psi(t) \right>} \stackrel{t \to \infty}{\longrightarrow}
\frac{z^2 + z \sqrt{z^2-1} -\frac{1}{2}}{ z^2+ z \sqrt{z^2-1}} .
\label{eq:ex_eval_spin_1_2_broken_t_inf}    
\end{equation}
The (dimensionless) rate is given by $\gamma_z = 2 \sqrt{z^2-1}$. The time-dependence is shown in Fig.~\ref{fig:decay_finite} for different $z$. Remarkably, and in contrast to the behavior of a open quantum system with generic decoherence, the expectation value does not decay to 1/2 (for $|z|>1$). In this sense the $\mathcal{PT}$ symmetry also manifests in the dynamics of the symmetry broken phase. 

\begin{figure}
    \centering
    \includegraphics[width = 13cm]{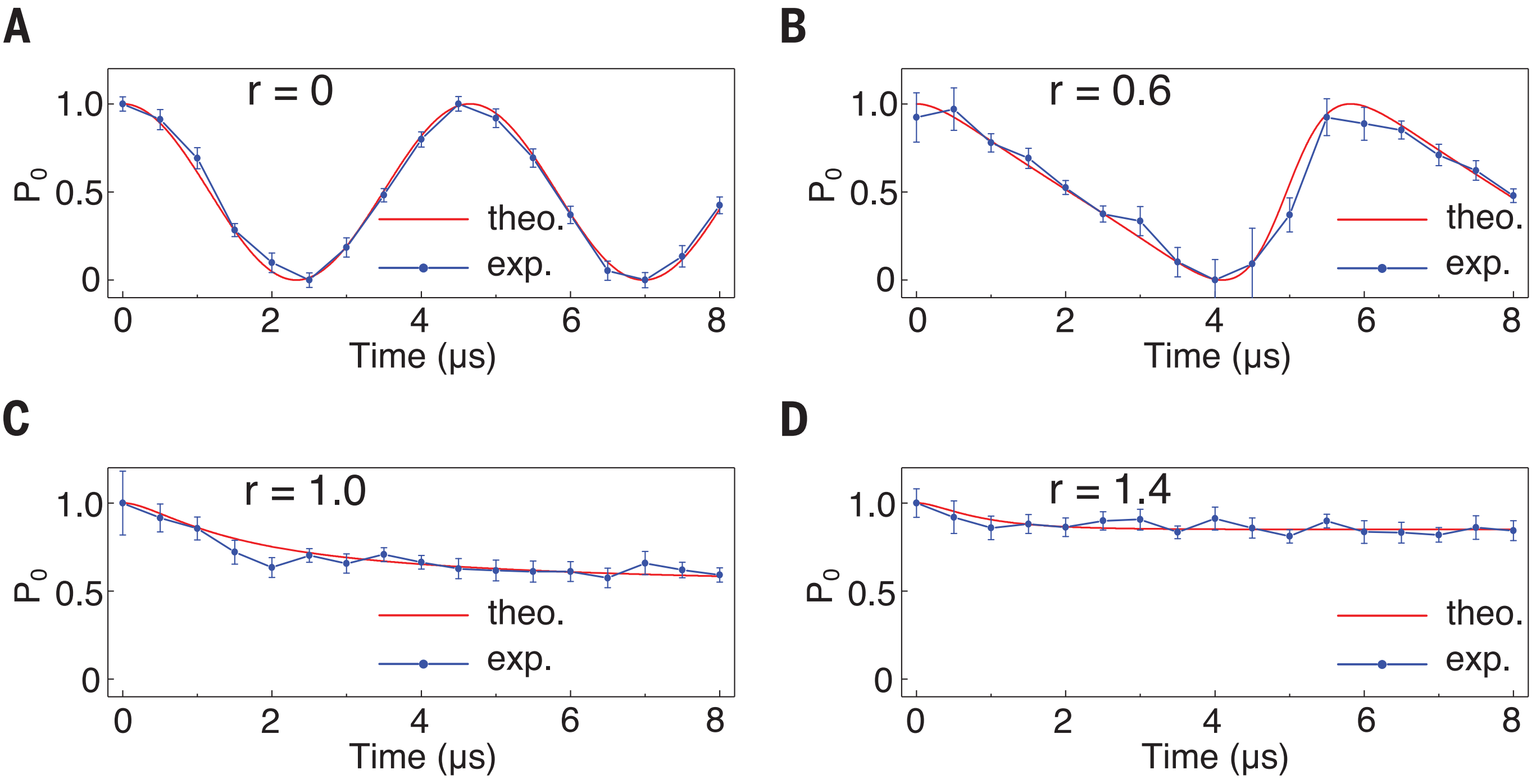}
    \caption{\textbf{Time evolution of the occupancy of a level in a single nitrogen-vacancy center in diamond}, modeled by the $\mathcal{PT}$-symmetric two-level model with Hamiltonian \cref{eq:spin_1_2} for $\phi = 0, \theta = \pi / 2, s = 1$, implying $z=r$. Shown are measurements and theoretical results for $P_0 =  \left<  \hat d_{\rm L} \right>_{\left|\psi(t) \right>} $ [see \cref{eq:ex_eval_spin_1_2,eq:ex_eval_spin_1_2_broken}]
    in the regime of unbroken (panels A, B) and broken (panels C, D) $\mathcal{PT}$-symmetry. Note the excellent agreement between the theoretical prediction and the measurement in both regimes.
    From \cite{Wu2019}. Reprinted with permission from AAAS.}
    \label{fig:experiment}
\end{figure}

The Hamiltonian Eq.~(\ref{eq:spin_1_2}) was realized in an experiment on a single nitrogen-vacancy center in diamond \cite{Wu2019}. The left and right lattice sites correspond to two energy levels of the nitrogen-vacancy center. The setup is tunable such that it was possible to drive the system through the $\mathcal{PT}$ transition. The occupancy of the energy level corresponding to $\left| \uparrow \right>$ was measured as a function of time. The data were compared to the theoretical results Eqs.~(\ref{eq:ex_eval_spin_1_2}) and (\ref{eq:ex_eval_spin_1_2_broken}). As shown in Fig.~\ref{fig:experiment} excellent quantitative agreement was reached. In addition, this Hamiltonian was implemented on a superconducting quantum processor \cite{Dogra2021}. 

Two important lessons can be learned from the preceding discussion:
\begin{enumerate}
\item The physics of non-Hermitian, $\mathcal{PT}$-symmetric systems is  unusual and interesting, even in very simple setups such as our two-level toy problem.
\item Hermitian observables and expectation values, evaluated using \cref{eq:ex_value_rewitten_nochmal}, can be directly measured in non-Hermitian systems. It is not evident how the concepts of an observable and its ``expectation value'' from $\mathcal{PT}$-symmetric or biorthogonal quantum mechanics might be useful in the present context. 
\end{enumerate}

\subsection{The dot occupancy in the \texorpdfstring{$\mathcal{PT}$}{PT}-symmetric resonant level model}
\label{subsec:further_examples_RL}

\subsubsection{The single-particle case}

In a first step we compute the occupancy of the dot site of the resonant level model in the single-particle scattering and bound eigenstates discussed in Sect.~\ref{subsec:examples_RL} \cite{Yoshimura2020}. Avoiding second quantization,  we can write the corresponding operator as $\hat d = \left| j=0 \right> \left< j=0 \right| = \left| 0 \right> \left< 0 \right|$, with the Wannier state $\left| j \right> $. It is obviously Hermitian and thus an observable in the conventional sense. 
However, $\hat d$ is also an observable in the $\mathcal{PT}$-symmetric as well as in the biorthogonal sense, hence providing the opportunity to directly compare the expectation values obtained within the different approaches. To show that $\hat d$ is an observable in the biorthogonal sense, we have to verify that there exists a matrix representation of $\hat d$ in some biorthonormal basis that is Hermitian  (see \cref{subsec:ob_ex_bio}), while for $\hat d$ to be an observable in the $\mathcal{PT}$-symmetric sense we need the matrix representation of $\hat d$ in the biorthonormal eigenbasis of the Hamiltonian to be Hermitian (see \cref{subsec:ob_ex_PT}). Remember that the entire concept of a $\mathcal{PT}$-symmetric observable is only defined if the spectrum is entirely real. In the $\mathcal{PT}$-symmetric resonant level model we are thus restricted to the regime of unbroken $\mathcal{PT}$ symmetry $\gamma_{\rm i} < \gamma_{\rm r}$. In the present case it will turn out that the biorthogonal matrix representation of $\hat d$ is always Hermitian in the eigenstates of the Hamiltonian, also in the symmetry broken phase.

To investigate the Hermiticity of the matrix representation of $\hat d$ we have to compute all matrix elements of $\hat d$ in the single-particle states determined in Sect.~\ref{subsec:examples_RL}. As $\hat d$ is a projector on the dot site, all matrix elements involving the odd scattering solution $\left| {\rm R}_k^- \right>$ vanish; see Eq.~(\ref{eq:RL_ss_odd0}). Employing the results of  Sect.~\ref{subsec:examples_RL} it is straight forward to show that $\left<{\rm L}_k^+ \right| \hat d \left| {\rm R}_{k'}^+ \right>$ is real and symmetric ($k \leftrightarrow k'$). This proves that the biorthogonal matrix representation of $\hat d$ in these states is Hermitian. Considering the single-particle bound states with real energy one can furthermore show that 
\begin{align}
\left<{\rm L}_\lambda^{\rm r} \right| \hat d \left| {\rm R}_{k}^+ \right> & =
\left<{\rm L}_k^+ \right| \hat d \left| {\rm R}_{\lambda}^{\rm r} \right>  \in {\mathbbm R} , \nonumber \\
\left<{\rm L}_\lambda^{\rm r} \right| \hat d \left| {\rm R}_{\lambda'}^{\rm r} \right> & =
\left<{\rm L}_{\lambda'}^{\rm r} \right| \hat d \left| {\rm R}_{\lambda}^{\rm r} \right>  \in {\mathbbm R} . 
\label{eq:biorth_matrix}
\end{align}
The same holds for the bound states with imaginary energy, i.e., for ${\rm r} \to {\rm i}$. This finalizes our argumentation showing that the matrix representation of $\hat d$ in the biorthonormal basis of eigenstates is Hermitian and $\hat d$ is a
\begin{enumerate}
\item $\mathcal{PT}$-symmetric observable for $\gamma_{\rm i} < \gamma_{\rm r}$,
\item a biorthogonal observable for all parameter regimes, and
\item a Hermitian observable for all parameter regimes. 
\end{enumerate}
In the symmetry unbroken regime the $\mathcal{PT}$-symmetric and the biorthogonal ``expectation values'' of course agree. It is meaningful to compute expectation values of $\hat d$ both within Hermitian and biorthogonal quantum mechanics, a situation we did not encounter in the two-level model. 

We thus compute the expectation values of $\hat d$ in one of the scattering eigenstates $\left| {\rm R}_{k}^+ \right>$  according to Eqs.~(\ref{eq:ex_value_rewitten_nochmal}) and (\ref{eq:bio_exp_def}), respectively. We start out with the latter
%
% \left< {\rm L}_k^+ \right| \left. 0^{\phantom{+}} \!\!\! \right> \left<  0^{\phantom{+}} \!\!\!\right| \left. {\rm R}_k^+ \right>
\begin{align}
\left< \hat d \, \right>_{\left| {\rm R}_{k}^+ \right>}^{\rm bo}  
= \frac{\left< {\rm L}_k^+ \right| \hat d \left|{\rm R}_k^+ \right>}{\left< {\rm L}_k^+ \right. \left|{\rm R}_k^+ \right>} 
=  \frac{2 \tilde \Gamma^2 \cos^2{\left(k\left[ 1- \delta_+\right]  \right)}}{E_k^2  \left[ \frac{1}{2} + \frac{N}{4} + \frac{2 \tilde \Gamma (J^2 - \tilde \Gamma)}{\Delta_{+}(k)} \right] } ,
\label{eq:scat_exp_L_R}
\end{align}
where we used the biorthonormalization of the scattering states. Note that this ``expectation value'' vanishes as $1/N$ in the thermodynamic limit. This is reasonable as, in this limit, a single scattering state does have a vanishing weight on every lattice site including the dot site. Using the results of Sect.~\ref{subsec:examples_RL} we obtain for the Hermitian expectation value
\begin{align}
\left< \hat d \, \right>_{\left| {\rm R}_{k}^+ \right>}  = 
\frac{\left< {\rm R}_k^+ \right| \hat d \left|{\rm R}_k^+ \right>}{\left< {\rm R}_k^+ \right. \left|{\rm R}_k^+ \right>} = \frac{\left< {\rm L}_k^+ \right| \hat d \left|{\rm R}_k^+ \right>}{\left< {\rm R}_k^+ \right. \left|{\rm R}_k^+ \right>} .
\label{eq:scat_exp_R_R}
\end{align}
The numerator agrees with Eq.~(\ref{eq:scat_exp_L_R}), while in the non-Hermitian case $\gamma_{\rm i} \neq 0$ the denominator is not unity so that the two results Eqs.~(\ref{eq:scat_exp_L_R}) and (\ref{eq:scat_exp_R_R}) differ. The expression for the denominator is lengthy and not of particular interest, but simplifies in the thermodynamic limit to 
% We here only give it in the thermodynamic limit $N \to \infty$
%
\begin{equation}
\left< {\rm R}_k^+ \right. \left|{\rm R}_k^+ \right>   \stackrel{N \to \infty}{\longrightarrow} \frac{\left| \gamma \right|^2}{\tilde \Gamma}  \stackrel{\gamma_{\rm i}=0}{=} 1  .
\label{eq:scat_den_R_R}    
\end{equation}
We also find differences in the two expressions for the expectation value of $\hat d$ in the bound states. 

In the case of a single particle in the system and based on the above results for the dot occupancy in a scattering state we find it difficult to argue that the Hermitian expectation value is physically more reasonable than the biorthogonal  ``expectation value'', in particular, as both vanish in the thermodynamic limit. However, this changes if we consider the many-body case with $M$ fermions occupying the, in total, $N+1$ lattice sites \cite{Yoshimura2020}. That is the first time in this review that we explicitly consider a quantum many-body problem. 

\subsubsection{The many-body case}

We tackle the many-body problem numerically, since an analytic evaluation of expectation values of the form $\left< \psi \right| \hat d \left| \psi \right>$ is difficult (see \cref{sec:critical_num_corr} and \cite{Yoshimura2020}). Here $\ket{\psi}$ denotes a many-body eigenstate of the Hamiltonian, that is constructed as a $M$ body Slater-determinant of the single-particle eigenstates of the problem, just as in Hermitian quantum many-body theory \cite{Bruus2003}. We use exact diagonalization to determine all single-particle states for a fixed number of lattice sites $N+1$ (see \cref{sec:critical_num_corr}). As it is common in quantum many-body theory we consider large system sizes, ideally the thermodynamic limit $N \to \infty$, with the filling  $M/(N+1)$  being fixed. We focus on half-filling (in the thermodynamic limit) of the entire system. Parametrizing the tunnel coupling as $\gamma = |\gamma| e^{i \phi}$, we can vary the angle $\phi \in [0, \pi)$ to analyze the system both in the broken and unbroken phase. For $\phi < \pi/4$ all eigenvalues are real, while for $\phi > \pi / 4$, a pair of bound states with purely imaginary energy Eq.~(\ref{eq:RL_bi_spec}) emerges. 

\begin{figure}[t]
    \centering
    \includegraphics{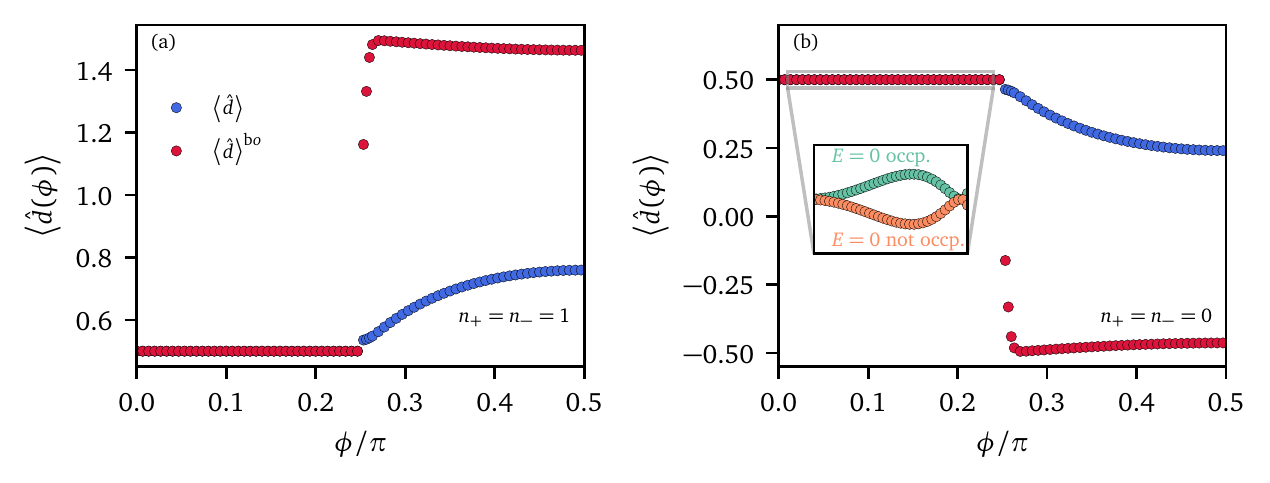}
    \caption{\textbf{The dot occupancy of the resonant level model in a many-body eigenstate} of the Hamiltonian Eq.~(\ref{eq:RL_ham}) for $\abs{\gamma} = 0.2$ and $N = 2018$ as a function of $\phi$ with $\gamma = |\gamma| e^{i \phi}$. For $\phi>\pi/4$ two single-particle bound states with complex energy appear. The red and blue data in the main panels are obtained taking the biorthogonal or conventional Hermitian expectation values, respectively. (a) The occupancy of the dot if both complex energy bound states are occupied $n_+ = n_- = 1$ and 
    (b) unoccupied $n_+ = n_- = 0$. The inset shows the expectation value $\expval*{\hat d}$ evaluated with the energy level at vanishing energy occupied and unoccupied. For more on this, see the main text. The main panels show the average of these two curves (speed-up of convergence to the thermodynamic limit).}
    \label{fig:rlm_occupation}
    \label[fig_a]{fig:rlm_occupation_a}
    \label[fig_b]{fig:rlm_occupation_b}
\end{figure}

More specifically, we would like to evaluate the occupancy of the dot level in a many-body state which one can reasonably regard as the ground state of the system. In the case that all eigenvalues are real, i.e., for $\phi < \pi/4$, it is clear how to construct this state \cite{Yoshimura2020}. We simply fill up all the scattering and real-energy bound states (if they appear) with negative eigenenergies. The corresponding single-particle right eigenstates are the entries of the Slater-determinant. Details of the numerics are presented in \cref{sec:critical_num_corr}. For a large but finite system with $N=2018$ sites this leads to the lower branch shown in the inset of Fig.~\cref{fig:rlm_occupation_b} for the dot occupancy $\left< \hat d \right>$. The deviation of this curve from $1/2$ decreases for increasing $N$ (finite size corrections; not shown). We observe numerically that the system hosts a scattering state with energy $E_\nu = 0$. If this state is, in addition, occupied in the Slater-determinant we obtain the upper branch of the inset of  \cref{fig:rlm_occupation_b}.\footnote{Note that in Fig.~10 of \cite{Yoshimura2020} only the lower branch is shown.} In the thermodynamic limit, it must be irrelevant whether this zero energy state is occupied or not, since it only has a contribution of measure 0. To speed up the convergence to the thermodynamic limit, we hence average %$\expval{d} = 1/2 ( \expval{d}_{E=0 \text{occp.}} + \expval{d}_{E=0 \text{not occp.}})$. 
the two dot occupancies. This procedure follows naturally when starting from a grand canonical ensemble and considering the case $\beta \to \infty$ for chemical potential $\mu = 0$, which would lead to an occupancy of $1/2$ for a single-particle state with vanishing energy. This averaging is performed in the main panels of \cref{fig:rlm_occupation}. As expected, in the thermodynamic limit the dot occupancy in the many-body ground state with half-filling is $1/2$ (for $\phi < \pi/4$). This holds for both ways to compute the dot occupancy; the expectation value according to conventional Hermitian quantum theory Eq.~(\ref{eq:ex_value_rewitten_nochmal}) (blue in \cref{fig:rlm_occupation} and for $\phi<\pi/4$ hidden by the red data) as well as that of biorthogonal quantum mechanics  Eq.~(\ref{eq:bio_exp_def}) (red in \cref{fig:rlm_occupation}). For $\phi < \pi / 4$ we therefore do not obtain any indication to which approach is physically more reasonable.

In the $\mathcal{PT}$-symmetry broken phase $\phi > \pi/4$, where complex bound state energies appear (even purely imaginary in the present case), it is not evident how to construct the many-body ground state \cite{Herviou2019}. We do not have a guiding principle telling us which complex energy states to fill and which to leave empty. This holds, in particular, for the case of half filling of the entire system that we are interested in \cite{Yoshimura2020}. To circumvent this problem we proceed as follows. To construct the many-body eigenstate the single-particle scattering states are filled as for $\phi < \pi/4$. On top of that, we either fill none of the bound states with purely imaginary energy Eq.~(\ref{eq:RL_bi_spec}), or both. We also analyzed the case where only one of the complex bound states is occupied, but we focus on the former two cases, as these illustrate the breakdown of the biorthogonal ``expectation value'' very nicely.
%On top we either fill none of the bound states with purely imaginary energy Eq.~(\ref{eq:RL_bi_spec}), the one with either negative or positive imaginary part of the energy, or both. We computed the expectation value of $\hat d$ in both ways (Hermitian and biorthogonal) for all four configurations.
In  \cref{fig:rlm_occupation_a} the data for both bound states being filled ($n_+ = n_- =1$) are shown. In  \cref{fig:rlm_occupation_b} both imaginary energy bound states are empty  ($n_+ = n_- =0$). Regardless if one considers one of the two cases as the ground state, they are certainly many-body eigenstates and the expectation value of $\hat d$ should thus show a physically sensible behavior. In contrast to this, for $\phi> \pi / 4$, $\expval*{\hat d}\^{bo}$ suddenly yields values larger than one, see \cref{fig:rlm_occupation_a}, or even smaller than zero, \cref{fig:rlm_occupation_b}. In both cases, a physically sensible interpretation is hard to find, since according to the Pauli-exclusion principle, the occupancy of a fermionic lattice site should always be between zero and one \cite{Sakurai2017}. This is indeed fulfilled when considering $\expval*{\hat d}$, which shows the same universal trend as $\expval*{\hat d}\^{bo}$, i.e. $\expval*{\hat d} > 1/2$ in \cref{fig:rlm_occupation_a} and $\expval*{\hat d} < 1/2$ in \cref{fig:rlm_occupation_b}, but which always remains in a physically sensible range. It is hence directly amenable to a physical interpretation. 

This observation calls into question the physical significance of the biorthogonal observables and their ``expectation values''. This is to be expected, since the definition of the non-Hermitian observables can be rigorously derived from the ancilla approach, see \cref{subsec:ob_anc}, while the definition of the biorthogonal observable remains a postulate. The main allure of the latter approach are convenient mathematical properties, but at the cost of loosing a direct physical interpretation, as explicitly shown in this section.

\section{Statistical ensembles and the density matrix}
\label{sec:statens}

Theoretical studies of quantum many-body systems are not restricted to states but routinely extended to statistical ensembles \cite{Bruus2003}. The density matrix $\rho(t)$ for a  non-Hermitian Hamiltonian obeys the generalized  von Neumann equation
\begin{equation}
i \partial_t \rho(t)= H \rho(t) -  \rho(t) H^\dag , \quad \rho(t=0) = \rho_0 ,
\label{eq:von_Neuman}    
\end{equation}
with the initial density matrix $\rho_0$. The right hand side cannot be rewritten as a commutator as $H^\dag \neq H$. This equation was already presented in Sect.~\ref{subsec:master} when discussing the derivation of non-Hermitian Hamiltonians in the master equation approach, with the recycling term being dropped; compare Eq.~(\ref{eq:lindblad_non_herm}). The latter is justified if no quantum jumps occurred, which has to be monitored by a continuous measurement (and is getting more and more unlikely with increasing time). In \cref{subsec:eqm_ancilla} we, in addition, explicitly derive Eq.~(\ref{eq:von_Neuman}) from the ancilla approach. The formal solution of Eq.~(\ref{eq:von_Neuman}) for a time-independent Hamiltonian $H$ is
\begin{equation}
\rho(t)= e^{-i H t} \rho_0 e^{i H^\dag t}.
\label{eq:sol_von_Neuman}    
\end{equation}
In this section we will discuss the time evolution of density matrices of non-Hermitian systems, but also investigate equilibrium ensembles for which the density matrix becomes stationary: $\rho(t) = \rho_0$. In Hermitian quantum statistical mechanics these are central to derive an equilibrium thermodynamics \cite{Huang1987}.   

To investigate statistical ensembles and density matrices in non-Hermitian and pseudo-Hermitian ($\mathcal{PT}$-symmetric) systems in detail, we proceed in several steps. We start out by deriving the time-evolution equation of the density matrix employing the ancilla approach. This shows that the embedding of the non-Hermitian system in a larger, Hermitian one, provides a transparent approach to the (sub-) systems dynamics, in complete analogy to states (see Sect.~\ref{sec:ancilla}),

\subsection{The equation of motion from the ancilla approach}
\label{subsec:eqm_ancilla}

In accordance with the notation of Sect.~\ref{sec:ancilla} we reintroduce the labels s and a, for system and ancilla, and follow the steps of Sect.~\ref{subsec:basic_idea}. 
%We assume that the non-Hermitian system Hamiltonian $H_{\rm s}$ is time-independent, has an entirely real spectrum, and is $\eta_{\rm r}$-pseudo-Hermitian. With this, we can use the results for the time-independent, Hermitian system-ancilla Hamiltonian $H_{\rm sa}$ Eq.~(\ref{eq:H_sa_form_ashida}), the time-independent Hermitian operator $g$ Eq.~(\ref{eq:g_for_sym_unbr}), and the normalization constant $K$ Eq.~(\ref{eq:K_determ_1}) derived in Sect.~\ref{subsec:PT_unbroken}.
Instead of a single normalized state from the subspace $\mathcal{H}_{\rm sa}^{\rm sub}$ of the system-ancilla Hilbert space $\mathcal{H}_{\rm sa}$ of the form Eq.~(\ref{eq:ancilla_initialstate_form}), we consider several of these (index $n$)
\begin{equation}
\left| \psi_{\rm sa}^{{\rm sub},(n)}(0) \right> = K_n \left[ \left| \uparrow \right> \otimes \left| \psi_{\rm s}^{(n)}(0) \right> + \left| \downarrow \right> \otimes g \left| \psi_{\rm s}^{(n)}(0) \right> \right],
\label{eq:ancilla_state_ensemble}
\end{equation}
where the $ \left| \psi_{\rm s}^{(n)}(0) \right>$ are taken from $\mathcal{H}_{\rm s}$. As usual in the construction of an ensemble in (standard) quantum statistical mechanics we assign a probability $p_n$, with $\sum_{n} p_n =1$, to each state $\left| \psi_{\rm sa}^{{\rm sub},(n)}(0) \right>$.
%\footnote{Think, e.g., of the thermal canonical ensemble for a Hermitian Hamiltonian in which $p_n=e^{-\beta E_n}/\sum_{m} e^{-\beta E_m}$, with the eigenenergies $E_n$ and the states are the corresponding orthonormal energie eigenstates.} 
The density matrix of the ensemble on $\mathcal{H}_{\rm sa}$ is then written as a weighted sum of the state projectors 
\begin{equation}
\rho_{\rm sa}(0) = \sum_{n} p_n \left| \psi_{\rm sa}^{{\rm sub},(n)}(0) \right> \left< \psi_{\rm sa}^{{\rm sub},(n)}(0) \right| .
\label{eq:density_matrix_sa}    
\end{equation}
As the states $\left| \psi_{\rm sa}^{{\rm sub},(n)}(0) \right>$ are normalized to one according to the standard inner product on $\mathcal{H}_{\rm sa}$, also $\rho_{\rm sa}(0)$ is normalized 
\begin{align}
\Tr_{\mathcal{H}_{\rm sa}} \left[ \rho_{\rm sa}(0) \right] 
&=
\sum_{j} 
\mel{e_j^{\rm sa}}{\rho_{\rm sa}}{e_j^{\rm sa}}
= \sum_{n,j} p_n  
\ip{e_j^{\rm sa}}{\psi_{\rm sa}^{{\rm sub},(n)}(0)}
\ip{\psi_{\rm sa}^{{\rm sub},(n)}(0)}{e_j^{\rm sa}}
\nonumber \\
&=
\sum_{n,j} p_n  
\ip{\psi_{\rm sa}^{{\rm sub},(n)}(0)}{e_j\^{sa}}
\ip{e_j^{\rm sa}}{\psi_{\rm sa}^{{\rm sub},(n)}(0)}
\nonumber \\ \nonumber
&= 
\sum_{n} p_n  \left< \psi_{\rm sa}^{{\rm sub},(n)}(0) \right.  \left| \psi_{\rm sa}^{{\rm sub},(n)}(0) \right> \\ 
&= \sum_{n} p_n = 1.
\label{eq:density_matrix_ancilla_norm}
\end{align}
Here $\left\{\ket*{e_j^{\rm sa}}\right\}$ denotes an orthonormal basis in  $\mathcal{H}_{\rm sa}$ and $\Tr_{\mathcal{H}_{\rm sa}}$ the corresponding trace. 

With the Hermitian Hamiltonian $H_{\rm sa}$ a density matrix on $\mathcal{H}_{\rm sa}$ obeys the von Neumann equation 
\begin{equation}
i \partial_t \rho_{\rm sa}(t)= \left[ H_{\rm sa}, \rho_{\rm sa} (t) \right] .  
\label{eq:von_Neuman_standard}     
\end{equation}
To determine the density matrix $ \rho_{\rm sa}(t)$, which evolves out of the initial one Eq.~(\ref{eq:density_matrix_sa}), we thus have to time evolve the states $\left| \psi_{\rm sa}^{{\rm sub},(n)}(0) \right> $, forming the ensemble, with the unitary time-evolution operator $U_{\rm sa }(t)$. Using the results of Sect.~\ref{subsec:basic_idea} we obtain
\begin{equation}
\left| \psi_{\rm sa}^{{\rm sub},(n)}(t) \right> = U_{\rm sa }(t)  \left| \psi_{\rm sa}^{{\rm sub},(n)}(0) \right> =  K_n \left[ \left| \uparrow \right> \otimes \left| \psi_{\rm s}^{(n)}(t) \right> + \left| \downarrow \right> \otimes g \left| \psi_{\rm s}^{(n)}(t) \right> \right] ,
\label{eq:single_state_time_evol}    
\end{equation}
with $ \left| \psi_{\rm s}^{(n)}(t) \right> =  U_{\rm s}(t) \left| \psi_{\rm s}^{(n)}(0) \right>$ from $\mathcal{H}_{\rm s}$. The
density matrix at time $t$ is given by Eq.~(\ref{eq:density_matrix_sa}) with $0$ replace by $t$
\begin{equation}
\rho_{\rm sa}(t) = \sum_{n} p_n \left| \psi_{\rm sa}^{{\rm sub},(n)}(t) \right> \left< \psi_{\rm sa}^{{\rm sub},(n)}(t) \right| .
\label{eq:density_matrix_sa_time_evol}    
\end{equation}
As the time evolution is unitary we have  $\Tr_{\mathcal{H}_{\rm sa}} \left[ \rho_{\rm sa}(t) \right]=1$.

Following the steps of Sect.~\ref{subsec:basic_idea} one now performs a measurement of the ancilla spin at time $t$ and post-selects for incidents with up-spin. According to the formalism of standard quantum statistical mechanics \cite{Huang1987} the density matrix right after the measurement with outcome ``up'' is given by 
% \begin{equation}
% \rho_{\rm sa}^\uparrow(t) = \frac{\left( \left| \uparrow \right> \left< \uparrow \right| \otimes \mathbbm{1}_{\rm s} \right) \rho_{\rm sa}(t) \left( \left| \uparrow \right> \left< \uparrow \right| \otimes \mathbbm{1}_{\rm s} \right) }{\Tr_{\mathcal{H}_{\rm sa}} \left[
% \left( \left| \uparrow \right> \left< \uparrow \right| \otimes \mathbbm{1}_{\rm s} \right)
% \rho_{\rm sa}(t) \left( \left| \uparrow \right> \left< \uparrow \right| \otimes \mathbbm{1}_{\rm s} \right) \right] },  
% \label{eq:dens_mat_measured}    
% \end{equation}
\begin{equation}
\rho_{\rm sa}^\uparrow(t) = \frac{\big( \left| \uparrow \right> \left< \uparrow \right| \otimes \mathbbm{1}_{\rm s} \big) \rho_{\rm sa}(t) \big( \left| \uparrow \right> \left< \uparrow \right| \otimes \mathbbm{1}_{\rm s} \big) }{\Tr_{\mathcal{H}_{\rm sa}} \left[
\big( \left| \uparrow \right> \left< \uparrow \right| \otimes \mathbbm{1}_{\rm s} \big)
\rho_{\rm sa}(t) \big( \left| \uparrow \right> \left< \uparrow \right| \otimes \mathbbm{1}_{\rm s} \big) \right] },  
\label{eq:dens_mat_measured}    
\end{equation}
with  $\Tr_{\mathcal{H}_{\rm sa}} \left[ \rho_{\rm sa}^\uparrow(t) \right]=1$. Inserting Eqs.~(\ref{eq:density_matrix_sa_time_evol}) and (\ref{eq:single_state_time_evol}) we end up with 
\begin{equation}
\rho_{\rm sa}^\uparrow(t) = \frac{ \sum_{n} p_n K_n^2 \left[ \left| \uparrow \right> \otimes \left| \psi_{\rm s}^{(n)}(t) \right>  \right] \left[ \left< \uparrow \right| \otimes \left< \psi_{\rm s}^{(n)}(t) \right| \right] }{\Tr_{\mathcal{H}_{\rm sa}} \left\{ \sum_{n} p_n K_n^2 \left[ \left| \uparrow \right> \otimes \left| \psi_{\rm s}^{(n)}(t) \right>  \right] \left[ \left< \uparrow \right| \otimes \left< \psi_{\rm s}^{(n)}(t) \right| \right] \right\}}.  
\label{eq:dens_mat_measured_expl}    
\end{equation}
Taking the time derivative of this expression and employing the Schr\"odinger equation (on $\mathcal{H}_{\rm s}$, i.e., with $H_{\rm s}$) we obtain the equation of motion of $\rho_{\rm sa}^\uparrow(t)$ 
\begin{align}
i \partial_t \rho_{\rm sa}^\uparrow(t) = & \left[ \mathbbm{1}_{\rm a} \otimes H_{\rm s}\right]    \rho_{\rm sa}^\uparrow(t) -  \rho_{\rm sa}^\uparrow(t) \left[\mathbbm{1}_{\rm a} \otimes H_{\rm s}^\dag \right]   \nonumber \\ & - \rho_{\rm sa}^\uparrow(t) \; \Tr_{\mathcal{H}_{\rm sa}} \left\{ \rho_{\rm sa}^\uparrow(t) \left[\mathbbm{1}_{\rm a} \otimes \left( H_{\rm s} - H_{\rm s }^\dag \right) \right]  \right\} .
\label{eq:von_Neuman_sa_measured}    
\end{align}
This is the analogue of Eq.~(\ref{eq:ancilla_eqm}) obtained for states.  In \cite{Brody2012} the result Eq.~(\ref{eq:von_Neuman_sa_measured}) was written down for mixed states (apparently) based on purely phenomenological reasoning.  For pure states it was derived in \cite{Sergi2013,Herviou2019}.  Its formal solution is 
\begin{equation}
  \rho_{\rm sa}^\uparrow (t) = \left| \uparrow \right> \left< \uparrow \right| \otimes \frac{\rho_{\rm s}(t)}{\Tr_{\mathcal{H}_{\rm s}} \left[\rho_{\rm s}(t) \right]} , \quad 
\rho_{\rm s}(t) =  U_{\rm s}(t) \rho_{\rm s}(0) U_{\rm s}^\dag (t) ,
\label{eq:solution_von_Neuman_sa_measured}    
\end{equation}
The density matrix $\rho_{\rm s}(t)$ on $\mathcal{H}_{\rm s}$  obeys the von Neumann equation (\ref{eq:von_Neuman}) with the non-Hermitian Hamiltonian $H_{\rm s}$. We thus derived this equation within the ancilla approach. The trace in the denominator is now only taken in the system Hilbert space $\mathcal{H}_{\rm s}$. It is carried out with respect to an orthonormal basis $\left\{ \left| e_j^{\rm s} \right> \right\}$ in $\mathcal{H}_{\rm s}$
\begin{equation}
\Tr_{\mathcal{H}_{\rm s}} \left[ \ldots  \right]  = \sum_{j}   \left< e_j^{\rm s} \right| \ldots   \left| e_j^{\rm s} \right> .
\label{eq:trace_def}    
\end{equation}
For an arbitrary linear operator $A$ on $\mathcal{H}_{\rm s}$ the trace can, however, be rewritten as
\begin{align}
 \Tr_{\mathcal{H}_{\rm s}} \left[ A  \right]  & = \sum_{j}   \left< e_j^{\rm s} \right| A  \left| e_j^{\rm s} \right>  = \sum_{j}   \left< e_j^{\rm s} \right| \mathbbm{1} A  \left| e_j^{\rm s} \right> = \sum_{j,\nu} 
 \left< e_j^{\rm s} \right. \left| {\rm R}_\nu \right> \left< {\rm L}_\nu \right| A  \left| e_j^{\rm s} \right> \nonumber \\  & =    \sum_{j,\nu} 
  \left< {\rm L}_\nu \right| A  \left| e_j^{\rm s} \right> \left< e_j^{\rm s} \right. \left| {\rm R}_\nu \right> = \sum_{\nu}  \left< {\rm L}_\nu \right| A  \left| {\rm R}_\nu \right> , 
\label{eq:trace_rewritten}    
\end{align}
where we used the completeness relation Eq.~(\ref{eq:complete}) for the biorthonormal basis of eigenstates of $H_{\rm s}$. The trace in $\mathcal{H}_{\rm s}$ can thus be taken in the biorthonormal basis of energy eigenstates as well (and in any other biorthonormal basis of $\mathcal{H}_{\rm s}$ for that matter). Note, however, that it cannot be taken in the non-orthogonal basis $\left\{ \left| {\rm R}_\nu \right>  \right\}$ of right energy eigenstates. 

In analogy to the ancilla approach for states, the differential (operator) equation for the density matrix $\rho_{\rm sa}^\uparrow (t)$ on $\mathcal{H}_{\rm sa}$ is non-linear, while the equation of motion for $\rho_{\rm s}(t)$ on $\mathcal{H}_{\rm s}$ is linear. In general, the latter will be easier to solve. The density matrix  $\rho_{\rm sa}^\uparrow (t)$ after the measurement is that of a Hermitian system and thus a Hermitian operator. With this also $\rho_{\rm s}(t)$ Eq.~(\ref{eq:solution_von_Neuman_sa_measured}) of the non-Hermitian (sub-) system has to be a Hermitian operator, which is ensured if the initial one $\rho_{\rm s}(0)$ is Hermitian. This is fully consistent with (reduced) density matrices of general open systems which have to be Hermitian \cite{Breuer2007}; see also Eqs.~(\ref{eq:lindblad}) and (\ref{eq:lindblad_non_herm}).

To understand the non-unitary dynamics of the (sub-) system's statistical ensemble one can now follow the reasoning of the ancilla approach developed for states in Sect.~\ref{subsec:basic_idea}. Instead of solving the non-linear operator equation (\ref{eq:von_Neuman_sa_measured}) for the density matrix, one solves the linear von Neumann equation (\ref{eq:von_Neuman}) with a non-Hermitian $H_{\rm s}$ and normalizes afterwards by taking the trace of the solution. 

\subsection{Equilibrium ensembles on \texorpdfstring{$\mathcal{H}_{\rm s}$}{Hs}}
\label{subsec:eq_ensembles}

The most commonly used equilibrium ensembles for Hermitian systems are the canonical and the grand canonical ones. They are constructed from the projectors to the eigenstates of the Hamiltonian which are weighted by the Boltzmann factor $e^{- \beta E_\nu}$, with the inverse temperature $\beta$. 

For a time-independent, non-Hermitian Hamiltonian $H_{\rm s}$ the canonical density matrix is often assumed to be \cite{Brody2012,Li2015,Yamamoto2019-pn,Zhang2022,Yamamoto2022,Groenendijk2021,Chen2018} %VM: added Yamamoto2019-pn
\begin{equation}
\rho_{\rm s}^{\rm can} = \frac{1}{Z_{\rm s}^{\rm can}} \; e^{- \beta H_{\rm s}} = \frac{1}{Z_{\rm s}^{\rm can}} \, \sum_{\nu}  e^{- \beta E_\nu} \left| {\rm R}_\nu\right> \left< {\rm L}_\nu \right| ,
\label{eq:rho_can}    
\end{equation}
with the partition function
\begin{equation}
Z_{\rm s}^{\rm can}= \Tr_{\mathcal{H}_{\rm s}} \left[ e^{- \beta H_{\rm s}} \right] = \sum_{\nu} e^{- \beta E_\nu} ,
\label{eq:rho_can_part}
\end{equation}
ensuring that $\rho_{\rm s}^{\rm can}$ is normalized to one: $\Tr_{\mathcal{H}_{\rm s}} \left[ \rho_{\rm s}^{\rm can}  \right] =1$. In the second step of Eq.~(\ref{eq:rho_can}) we used the spectral representation Eq.~(\ref{eq:specrepr}) and expressed  $\rho_{\rm s}^{\rm can}$ in terms of the biorthonormal basis of right and left eigenstates of the non-Hermitian Hamiltonian. 

If $\rho_{\rm s}^{\rm can}$ is used as the initial density matrix in Eq.~(\ref{eq:sol_von_Neuman}), it is evidently not stationary, unless $\comm*{H\_s}{H\_s^\dagger} = 0$.\footnote{The canonical density matrix  $\rho_{\rm s}^{\rm can}$ is stationary with respect to the usual von Neumann equation $i \partial_t \rho_{\rm s}(t) = \left[ H_{\rm s},  \rho_{\rm s}(t) \right]$ even for a non-Hermitian $H_{\rm s}$ \cite{Herviou2019}. However, thinking of the the non-Hermitian system as an open one, this equation of motion does not appear to be a proper choice. We have shown this using the Master equation as well as the ancilla approach.} The same holds for the grand canonical density matrix. This is in striking contrast to the case of a Hermitian Hamiltonian. Note that this problem occurs regardless if the spectrum is entirely real or not. Therefore, $\rho_{\rm s}^{\rm can}$ does not correspond to an equilibrium ensemble. 

With a non-Hermitian $H_{\rm s}$, $\rho_{\rm s}^{\rm can}$ Eq.~(\ref{eq:rho_can}) is furthermore not Hermitian, but density matrices of open systems should still be Hermitian \cite{Breuer2007}, as we also emphasized in Sect.~\ref{subsec:master}. In fact, this is fully consistent with the two constructive ways towards non-Hermitian open systems, the Master equation approach of Sect.~\ref{subsec:master}  as well as the ancilla approach of Sect.~\ref{subsec:eqm_ancilla}. Both lead to Hermitian (sub-) system density matrices. 

Based on these insights we conclude that $\rho_{\rm s}^{\rm can}$ of Eq.~(\ref{eq:rho_can}) is not a proper density matrix for a non-Hermitian, $\mathcal{PT}$-symmetric (or pseudo-Hermitian) Hamiltonian even if the symmetry is unbroken. It does certainly not represent an equilibrium ensemble. It is thus surprising, that this (grand) canonical density matrix was still used by many authors to compute thermodynamic expectation values of non-Hermitian quantum (many-body) systems \cite{Brody2012,Li2015,Yamamoto2019-pn,Zhang2022,Yamamoto2022,Groenendijk2021,Chen2018}.   %VM: added Yamamoto2019-pn

For a time-independent $H_{\rm s}$ which is $\eta_{\rm r}$-pseudo-Hermitian (real eigenvalues), it is, however, easy to construct  stationary density matrices on $\mathcal{H}_{\rm s}$, and thus equilibrium ensembles, which are Hermitian operators. As an ansatz we take the normalized density matrix 
\begin{equation}
 \rho_{\rm s}^{\rm eq} = \frac{\sum_{\nu} \pi_\nu \left| {\rm R}_\nu\right> \left< {\rm R}_\nu \right| }{\Tr_{\mathcal{H}_{\rm s}} \left[ \sum_{\nu} \pi_\nu \left| {\rm R}_\nu\right> \left< {\rm R}_\nu \right|\right]  } , \quad  \Tr_{\mathcal{H}_{\rm s}} \left[  \rho_{\rm s}^{\rm eq} \right] =1 ,
\label{eq:rho_eq_ansatz}    
\end{equation} 
with the right eigenstates $\left| {\rm R}_\nu\right>$ of the non-Hermitian $H_{\rm s}$ and $\pi_\nu \in \mathbbm{R}$. With this Eq.~(\ref{eq:sol_von_Neuman}) gives 
\begin{align}
\rho_{\rm s}^{\rm eq}(t) & = e^{-i H_{\rm s} t}  \frac{\sum_{\nu} \pi_\nu \left| {\rm R}_\nu\right> \left< {\rm R}_\nu \right| }{\Tr_{\mathcal{H}_{\rm s}} \left[ \sum_{\nu} \pi_\nu \left| {\rm R}_\nu\right> \left< {\rm R}_\nu \right|\right]  } e^{i H_{\rm s}^\dag t} \nonumber \\ & =    \frac{\sum_{\nu} \pi_\nu e^{- i E_\nu t} \left| {\rm R}_\nu\right> \left< {\rm R}_\nu \right| e^{i E_\nu t}}{\Tr_{\mathcal{H}_{\rm s}} \left[ \sum_{\nu} \pi_\nu \left| {\rm R}_\nu\right> \left< {\rm R}_\nu \right|\right]} \nonumber \\ & = \rho_{\rm s}^{\rm eq},   
\label{eq:rho_eq_ansatz_time_evol}    
\end{align}
where in the second row we used $E_\nu \in \mathbb{R}$.
We intentionally denoted the amplitude with which the eigenstate $\left| {\rm R}_\nu \right>$ contributes to the equilibrium ensemble by $\pi_\nu$ and not by $p_\nu$. With the $\left| {\rm R}_\nu\right>$ not being orthonormal one cannot directly interpret this weight as a probability  (see below). 
%Regardless of this, one can of course take $\pi_\nu = e^{-\beta E_{\nu}}$, but even for this choice $\rho_{\rm s}^{\rm eq} \neq \rho_{\rm s}^{\rm can}$.    

We now construct the density matrix $\rho_{\rm sa}^{\rm{eq}}$ on $\mathcal{H}_{\rm sa}$ which leads to $\rho_{\rm s}^{\rm eq}$  Eq.~(\ref{eq:rho_eq_ansatz}) when following the steps of the ancilla approach of the last subsection. For this we first have to go back to the foundations of the ancilla approach. We consider a time-independent state $ \left| {\rm R}_\nu^{\rm sa} \right>$ of the form Eq.~(\ref{eq:ancilla_initialstate_form}) with $\left| \psi_{\rm s} \right> \to \left| {\rm R}_\nu \right>$. According to the action of $H_{\rm sa}$ Eq.~(\ref{eq:H_sa_form_ashida}) on states from $\mathcal{H}_{\rm sa}^{\rm sub}$, $ \left| {\rm R}_\nu^{\rm sa} \right>$ is an eigenstate of this system-ancilla Hamiltonian\footnote{Given a $H_{\rm s}$, the system-ancilla Hamiltonian $H_{\rm sa}$ is, in fact, constructed just so that this is fulfilled. See also the supplementary material of \cite{Kawabata2017}.}
\begin{align}
H_{\rm sa} \left| {\rm R}_\nu^{\rm sa} \right> & = 
H_{\rm sa} K_\nu \left[ \left| \uparrow \right> \otimes \left| {\rm R}_\nu  \right> +   \left| \downarrow \right> \otimes g \left| {\rm R}_\nu  \right> \right] \nonumber \\*
& = K_\nu \left[ \left| \uparrow \right> \otimes H_{\rm s} \left| {\rm R}_\nu  \right> +   \left| \downarrow \right> \otimes g H_{\rm s} \left| {\rm R}_\nu  \right> \right] \nonumber \\* 
& = E_\nu  \left| {\rm R}_\nu^{\rm sa} \right> . 
\label{eq:system_ancilla_es}    
\end{align} 
As the left and right eigenstates of $H_{\rm s}$ form a biorthonormal basis the normalization constant $K_\nu$ is, in fact, $\nu$-independent
\begin{align}
K_\nu = \left[c \left< {\rm R}_\nu \right| \eta_{\rm r} \left| {\rm R}_\nu\right>  \right]^{-1/2} = \left[c \left< {\rm L}_\nu \right. \left| {\rm R}_\nu\right>  \right]^{-1/2} = c^{-1/2} ,
\label{eq:K_nu_for_R_nu}    
\end{align}
where we used Eq.~(\ref{eq:K_determ_1}) and $c$ is given in Eq.~(\ref{eq:c_def}). It is crucial to keep in mind that states of the form  $ \left| {\rm R}_\nu^{\rm sa} \right>$ only constitute a subset of the eigenstates of $H_{\rm sa}$. This was exemplified for our two-level toy Hamiltonian in Sect.~\ref{subsec:example_spin_1_2}. We take the ensemble of states $\left\{   \left| {\rm R}_\nu^{\rm sa} \right>\right\}$ with weights $\left\{ p_\nu \right\}$, $\sum_\nu p_\nu=1$,
\begin{equation}
\rho_{\rm sa}^{\rm eq}= \sum_{\nu} p_\nu \left| {\rm R}_\nu^{\rm sa} \right> \left< {\rm R}_\nu^{\rm sa} \right| .
\label{eq:system_ancilla_embed}    
\end{equation}
When following the steps of the last subsection, it leads to an equilibrium density matrix on $\mathcal{H}_{\rm s}$ of the form Eq.~(\ref{eq:rho_eq_ansatz}), with $\pi_\nu = p_\nu/c$.  

The density matrix of the combined system-ancilla setup $\rho_{\rm sa}^{\rm eq}$ Eq.~(\ref{eq:system_ancilla_embed}), leading to the density matrix $\rho_{\rm s}^{\rm eq}$ of the system, is stationary under the time evolution with $\exp{ - i H_{\rm sa} t }$. It thus describes an equilibrium ensemble. 
%This allows us to speak of an equilibrium ensemble of the open system and not merely a steady state.  
Note, however, that even for $p_\nu=e^{-\beta E_\nu}/\sum_{\mu} e^{-\beta E_\mu}$ Eq.~(\ref{eq:system_ancilla_embed}) is not the conventional canonical density matrix of the system-ancilla setup with respect to the Hermitian Hamiltonian $H_{\rm sa}$ as only the energy eigenstates from $\mathcal{H}_{\rm sa}^{\rm sub}$ contribute to the ensemble. 

We will introduce the systems equilibrium density matrix which we consider to represent the proper extension of the canonical ensemble to non-Hermitian, $\eta_{\rm r}$-pseudo-Hermitian Hamiltonians when studying ensemble expectation values in Sect.~\ref{subsec:ens_exp_st}.

% \LG{Might be interesting to explicitly state that
% %
% \begin{align}
%     \Tr(\rho\^{can}(t) O) \neq \tr(\rho\^{can}O(t)\_H),
% \end{align}
% %
% i.e. that it matters if we evolve the density matrix or the operator via the Heisenberg picture. On the other hand, such a relation of course holds when evaluating expectation values with the Hermitian approach.}

\subsection{Stationary density matrices for broken \texorpdfstring{$\mathcal{PT}$}{PT}-symmetry}
\label{subsec:ensemble_broken}

We now consider the situation in which the system Hamiltonian $H_{\rm s}$ is in its $\mathcal{PT}$ symmetry broken phase. We still assume, that $H_{\rm s}$ is time independent. This appears to be necessary when aiming at a stationary density matrix on $\mathcal{H}_{\rm s}$. As the symmetry is broken at least one pair of complex conjugate eigenvalues occurs. As discussed in Sect.~\ref{subsec:ancilla_broken} the ancilla approach can be used in this case as well. We can thus follow all the steps of Sect.~\ref{subsec:eqm_ancilla} up to Eq.~(\ref{eq:solution_von_Neuman_sa_measured}). As the system-ancilla Hamiltonian becomes time dependent, the time evolution operator $\exp{- i H_{\rm sa} t}$ must be replaced by the general one $U_{\rm sa}(t)$. The equation of motion for $\rho_{\rm sa}^{\uparrow} (t)$ remains of the same form Eq.~(\ref{eq:von_Neuman_sa_measured}) as the operator $g(t)$ (see Sect.~\ref{subsec:basic_idea}) drops out when the ancilla spin is measured and the instances with spin down are disregarded. 

We now pose the question, if we can find an equilibrium, mixed-state ensemble such that the time dependence in Eq.~(\ref{eq:solution_von_Neuman_sa_measured}) drops out or more precisely the time dependence in the system part $\rho_{\rm s}(t)/\Tr_{\mathcal{H}_{\rm s}} \left[ \rho_{\rm s}(t) \right]$, with $\rho_{\rm s}(t) =e^{- i H_{\rm s} t} \rho_{\rm s}(0) e^{i H_{\rm s}^\dag t}$ drops out. As in the $\mathcal{PT}$ symmetry unbroken case an obvious ansatz for an equilibrium density matrix would be Eq.~(\ref{eq:rho_eq_ansatz}) made out of right eigenstates of $H_{\rm s}$. Considering Eq.~(\ref{eq:rho_eq_ansatz_time_evol}), the time dependence only drops out, if the weights $\pi_\nu$ for the eigenstates with complex energies vanish. If not, for $t \to \infty$ the contributing state with the largest ${\rm Im} E_\nu$ survives and the density matrix becomes the one of the corresponding pure state. One could avoid this by replacing the ket-state $\left< {\rm R}_\nu \right|$ in  Eq.~(\ref{eq:rho_eq_ansatz}) by the ``partner state'' with conjugate complex eigenvalue, as discussed in context of Eq.~(\ref{eq:ansatz_eta_cp}). In this case, the density matrix would, however, lose its established form as a weighted sum over a product of the same ket and bra states. This goes even beyond the generalization of Eq.~(\ref{eq:rho_can}). 

We do not investigate this any further here and simply note that it is not obvious how to obtain a non-trivial, mixed-state stationary density matrix on $\mathcal{H}_s$ and thus an equilibrium ensemble if the $\mathcal{PT}$ symmetry is broken.

\subsection{Ensemble expectation values for stationary density matrices}
\label{subsec:ens_exp_st}

Given a stationary, statistical ensemble of system states for a time-independent Hamiltonian we now study ensemble expectation values of observables. As we will not make any further explicit contact with the ancilla approach we drop the corresponding indices s and a to lighten the notation.

We start out by considering the definitions of observables and ``expectation values'' of $\mathcal{PT}$-symmetric and biorthogonal quantum mechanics introduced in Sects.~\ref{subsec:ob_ex_PT} and \ref{subsec:ob_ex_bio} for states. We aim to generalize them to ensembles. In Sect.~\ref{sec:obexp} we already argued that employing these definitions is not compelling from the perspective of physics. Still we want to show that using these and the canonical density matrix $\rho_{\rm s}^{\rm can}$ Eq.~(\ref{eq:rho_can}) leads to a mathematically consistent formalism that was used in \cite{Brody2012,Li2015,Yamamoto2019-pn,Zhang2022,Yamamoto2022,Groenendijk2021,Chen2018}. We already %VM: added Yamamoto2019-pn
emphasized that $\rho_{\rm s}^{\rm can}$ is not stationary and the reader might wonder why we discuss this situation in the present subsection whose title contains the word ``stationary''. The reasons are, that the computations formally fit into the present subsection and that, in any case, the non-stationarity was ignored in \cite{Brody2012,Li2015,Yamamoto2019-pn,Zhang2022,Yamamoto2022,Groenendijk2021,Chen2018}. As the %VM: Yamamoto2019-pn
biorthonormal basis appearing in Eq.~(\ref{eq:rho_can}) is the one made out of right and left eigenstates of $H$, it is reasonable to use this in the metric operator $\hat g$ and the $\mathcal{PT}$-symmetric and biorthogonal definitions of an observable become equal (see Sects.~\ref{subsec:ob_ex_PT} and \ref{subsec:ob_ex_bio}). We thus start out with the ``expectation value'' of an observable $O$ ($\eta_{\rm r}$-pseudo-Hermitian operator) in a right energy eigenstate. According to Eq.~(\ref{eq:PT_exp_def}) this is given by 
\begin{align}
\left< O \right>_{\left| {\rm R}_\nu \right>}^{\mathcal{PT}} = \frac{\left< {\rm R}_\nu \right| \eta_{\rm r} O \left| {\rm R}_\nu \right>}{\left< {\rm R}_\nu  \right| \eta_{\rm r}  \left| {\rm R}_\nu \right>} = \left< {\rm L}_\nu \right| O \left| {\rm R}_\nu \right> ,
\label{eq:PT_exp_R}    
\end{align}
where we used Eqs.~(\ref{eq:right_left_map}) and (\ref{eq:biorthogonal}). We now assign the Boltzmann weight $e^{- \beta E_\nu}/Z^{\rm can}$ to each pair of right and left eigenstates and sum over $\nu$ to obtain
\begin{align}
\sum_{\nu} p_\nu \left< O \right>_{\left| {\rm R}_\nu \right>}^{\mathcal{PT}} = \frac{1}{Z^{\rm can}} \sum_{\nu} e^{- \beta E_\nu}  \left< {\rm L}_\nu \right| O \left| {\rm R}_\nu \right> = \Tr \left[ \rho^{\rm can} O \right] = \left< O \right>_{\rho^{\rm can}}^{\mathcal PT} ,
\label{eq:PT_exp_stat}    
\end{align}
where we used Eq.~(\ref{eq:rho_can}) and in the last step defined the $\mathcal{PT}$-symmetric ``expectation value'' with respect to the canonical density matrix. 
%In  \cite{Brody2012,Li2015,Yoshimura2020,Zhang2022,Yamamoto2022} it was used to obtain ``expectation values'' of observables and correlation functions of $\mathcal{PT}$-symmetric systems in thermal equilibrium. This ignores that $\rho^{\rm can}$ is not stationary with respect to the von Neuman equation.

When discussing generating functionals and correlation functions in Sect.~\ref{sec:genfun}, we will be interested in the $\beta \to \infty$ limit (zero temperature) of Eq.~(\ref{eq:PT_exp_stat}). In this limit only the addend with the smallest (real) energy $E_0$ survives, the Boltzmann weight drops out, and one ends up with the ground state ``expectation value'' of $\mathcal{PT}$-symmetric quantum mechanics as defined in Eq.~(\ref{eq:PT_exp_def})
\begin{equation}
  \left< O \right>_{\rho^{\rm can}}^{\mathcal PT} \stackrel{\beta \to \infty}{\longrightarrow} \left< {\rm L}_0 \right| O \left| {\rm R}_0 \right> = \left< {\rm R}_0 \right| \eta_{\rm r} O \left| {\rm R}_0 \right>  =  \left< O \right>_{\left|{\rm R}_0\right>}^{\mathcal PT} .
\label{eq:PT_exp_stat_0}    
\end{equation}
The non-Hermiticity of $\rho^{\rm can}$ and the lack of stationarity do not play a role in this limit. 

We next use the concepts of an observable and the expectation value of Hermitian quantum mechanics  as favoured by us on physical grounds; see Sect.~\ref{subsec:obs_right}. In this case we take an arbitrary set $\left\{ \left| \psi_n \right> \right\} $ of (un-normalized) states to construct an ensemble. The expectation value of a Hermitian observable with spectral representation 
\begin{equation}
O = \sum_{\nu} o_\nu \left| {\rm R}^O_\nu \right>  \left< {\rm R}^O_\nu \right|
\label{eq:spec_O}
\end{equation}
in a state is given in Eq.~(\ref{eq:ex_value_rewitten}). We assign a weight $p_n$ to every state leading to the ensemble average
\begin{align}
\sum_{n} p_n \left< O \right>_{\left| \psi_n \right>} & = \sum_{n} p_n \frac{\left< \psi_n \right| O \left| \psi_n \right>}{\left< \psi_n \right. \left| \psi_n \right>} = \sum_{n} p_n \frac{ \sum_{\nu} o_\nu \left< \psi_n \phantom{{\rm R}^O_\nu} \!\!\!\! \! \!\!\!\! \! \right. \left| {\rm R}^O_\nu \right>  \left< {\rm R}^O_\nu \right|  \left. \phantom{{\rm R}^O_\nu} \!\!\!\! \! \!\!\!\! \!  \psi_n \right>}{\left< \psi_n \right.\!\! \left| \psi_n \right>} 
\nonumber \\
& = \sum_{\nu}  \left< {\rm R}^O_\nu \right|  
 \underbrace{ \left[ \sum_{n}\frac{p_n}{\left< \psi_n \right.\!\! \left| \psi_n \right>} 
\left| \psi_n \right> \left< \psi_n \right| \right]}_{=\rho} O \left| {\rm R}^O_\nu \right> \nonumber \\  
& = \Tr \left[\rho  O \right] = \left< O \right>_{\rho} .
\label{eq:herm_exp_stat}    
\end{align}
Using $\sum_{n} p_n =1$, it is straightforward to show that $\Tr[ \rho ] =1$ and the Hermitian operator $\rho$ is a valid density matrix as known from standard quantum statistical mechanics \cite{Huang1987}.  Note that the formal expression $\Tr[\rho O]$ is exactly the same as that of Eq.~(\ref{eq:PT_exp_stat}). As shown in Eq.~(\ref{eq:trace_rewritten}), the trace can either be taken in an orthonormal or in a biorthonormal basis of $\mathcal{H}_{\rm s}$. If the states $\left\{ \left| \psi_n \right> \right\} $ contributing to the ensemble are taken as a subset of the set of right eigenstates of $H$,  $\left\{ \left| {\rm R}_\nu \right> \right\} $, $\rho$ in Eq.~(\ref{eq:herm_exp_stat}) is stationary.

The (normalized) stationary density matrix for a system with a non-Hermitian Hamiltonian and right eigenstates $\left\{  \left| {\rm R}_\nu \right>  \right\}$, in which the Boltzmann weight is assigned to each eigenstate, is given by
\begin{equation}
\rho_{\rm nH}= \frac{1}{Z^{\rm can}} \sum_{\nu} \frac{e^{- \beta E_\nu}}{\left< {\rm R}_\nu \right.\!\! \left| {\rm R}_\nu \right>}  \left| {\rm R}_\nu \right> \left< {\rm R}_\nu \right| =\frac{1}{Z^{\rm can}} \,  e^{- \beta H} \sum_{\nu} \frac{\left| {\rm R}_\nu \right> \left< {\rm R}_\nu \right|}{\left< {\rm R}_\nu \right.\!\! \left| {\rm R}_\nu \right>}    ,
\label{eq:nH_can}
\end{equation}
with $Z^{\rm can}$ Eq.~(\ref{eq:rho_can_part}). We consider this as the proper generalization of the canonical ensemble to a pseudo-Hermitian Hamiltonian with real eigenvalues. It obviously differs from Eq.~(\ref{eq:rho_can}) by the factor involving the sum over $\nu$. In contrast to $\rho^{\rm can}$, $\rho\_{nH}$ is a stationary as well as a Hermitian operator and thus suitable to set up an equilibrium thermodynamics.

For the $\beta \to \infty$ (zero temperature) limit we obtain from Eqs.~(\ref{eq:herm_exp_stat}) and (\ref{eq:nH_can})
\begin{equation}
  \left< O \right>_{\rho_{\rm nH}} \stackrel{\beta \to \infty}{\longrightarrow} \frac{\left< {\rm R}_0 \right| O \left| {\rm R}_0 \right>}{\left<  {\rm R}_0 \! \right.  \left| {\rm R}_0 \right>} =  \left< O \right>_{\left|{\rm R}_0\right>} ,
\label{eq:herm_exp_stat_can_0}    
\end{equation}
that is, the ground state expectation value as in Hermitian quantum mechanics. This appears to be necessary if we want to speak about a generalization of the canonical ensemble. In the last step we used the definition Eq.~(\ref{eq:ex_value_rewitten_nochmal}). 

In Hermitian quantum statistical mechanics the inverse temperature in the canonical density matrix plays the role of the Lagrange multiplier which can be adjusted to fix the energy expectation value at the desired value. We now show that (formally) the same holds for the $\beta$ appearing in $\rho_{\rm nH}$. Ignoring that, within the formalism we prefer on physical grounds, a non-Hermitian $H$ is not an observable (see Sect.~\ref{subsec:obs_right}), we still compute
\begin{equation}
\Tr [\rho_{\rm nH} H] = 
 \sum_{\nu,\mu} \frac{e^{- \beta E_\mu}}{Z^{\rm can}} \frac{ \left< {\rm L}_\nu \right. \left| {\rm R}_\mu \right> \left< {\rm R}_\mu \right| H \left| {\rm R}_\nu \right> }{\left< {\rm R}_\mu \right. \left| {\rm R}_\mu \right>} = \frac{1}{Z^{\rm can}} \sum_{\nu} E_\nu  e^{- \beta E_\nu}.
\label{eq:langrange}    
\end{equation}
This is the statistical expectation value of the (real) eigenenergies $E_\nu$ under the probability distribution given by the Boltzmann weight $p_\nu =  e^{- \beta E_\nu}/Z^{\rm can}$ just as in the Hermitian case employing the canonical density matrix. The right hand side defines a unique relation between this expectation value and the inverse temperature $\beta$. This completes the reasoning that Eq.~(\ref{eq:nH_can}) is indeed the analogue of the canonical density matrix for a pseudo-Hermitian Hamiltonian with real eigenvalues.     

The computation of Eq.~(\ref{eq:langrange}) leads us to a curiosity we would like to emphasize. If $O$ is taken as an arbitrary function $f(H)$, with the Hamiltonian $H$ (in the simplest case $O=H$), using $\rho_{\rm nH}$ Eq.~(\ref{eq:nH_can}) and $\rho^{\rm can}$ Eq.~(\ref{eq:rho_can}) give the same result for the quantum statistical  ``expectation value''
\begin{align}
\Tr [\rho_{\rm nH} f(H)] &  = 
\sum_{\nu} \left< {\rm L}_\nu \right| \rho_{\rm nH} f(H) \left| {\rm R}_\nu \right> = \sum_{\nu,\mu} \frac{e^{- \beta E_\mu}}{Z^{\rm can}} \frac{ \left< {\rm L}_\nu \right. \left| {\rm R}_\mu \right> \left< {\rm R}_\mu \right| f(H)\left| {\rm R}_\nu \right> }{\left< {\rm R}_\mu \right. \left| {\rm R}_\mu \right>} \nonumber \\
& = \sum_{\nu} \frac{e^{- \beta E_\nu}}{Z^{\rm can}}  f(E_\nu)= \sum_{\nu} \left< {\rm L}_\nu \right| \frac{e^{- \beta H}}{Z^{\rm can}} f(H)  \left| {\rm R}_\nu \right> \nonumber \\
&  =  \Tr [\rho^{\rm can} f(H)].
 \label{eq:the_same}
\end{align} 
Note, however, that the quotation marks around the expression  ``expectation value'' are this time advisable in both cases. When using $\rho^{\rm can}$ [last line of Eq.~(\ref{eq:the_same})] the quantum part of the averaging lacks a straightforward probabilistic interpretation (see Sect.~\ref{subsec:ob_ex_PT}). Employing the definition of an observable from Hermitian quantum mechanics [left hand side of Eq.~(\ref{eq:the_same})] $f(H)$ is not an observable  as it is not Hermitian; see Sect.~\ref{subsec:obs_right}. Note, that the equality of the two results does not extend to other observables beyond this rather peculiar case.   

\subsection{Linear response theory}
\label{subsec:lin_resp}

Using Eqs.~(\ref{eq:solution_von_Neuman_sa_measured}) and (\ref{eq:herm_exp_stat}) the expectation value of a Hermitian observable $O$, for a time-dependent density matrix, is given by 
\begin{equation}
\left< O \right>_{\rho(t)} = \frac{\Tr\left[ \rho(t) O \right]}{\Tr\left[ \rho(t)\right]}     
\label{eq:exp_time_dep_rho}    
\end{equation}
within the Hermitian scheme (see \cref{subsec:obs_right}). This general expression can be used to set up a linear response theory for a time-independent, Hermitian or non-Hermitian (but $\eta_0$-pseudo-Hermitian) Hamiltonian $H_0$ in the presence of a time-dependent perturbation $V_t = f(t) D$.\footnote{A biorthogonal linear response theory, which essentially boils down to the expressions known for Hermitian Hamiltonians, was set up and used in \cite{Groenendijk2021}.} The real-valued function $f$ vanishes for $t<0$, i.e., the perturbation is switched on at time $t=0$ and in the most general case the perturbing operator $D$, might (also) be non-Hermitian. The crucial assumptions of linear response theory are, that the problem with $H_0$ is exactly solvable and that the perturbation $V_t$ is small. Linear response theory is central in quantum many-body physics as it allows to make contact with a variety of experimental techniques to probe a system \cite{Bruus2003}. It leads to the concept of response functions. Different variants of the general setup were investigated in the literature. In two papers only the perturbation was assumed to be non-Hermitian \cite{Pan2020,Geier2022}. Here we are not interested in this case, as it is the nature of $H_0$ which leads to major differences in the formalism. The most general case of non-Hermitian $H_0$ and $D$ was comprehensively introduced and employed in \cite{Sticlet2022}, while in \cite{Tetling2022} the focus was on non-Hermitian $H_0$ but Hermitian perturbations. Here we mainly consider the latter case and refer the interested reader to \cite{Sticlet2022} for the most general one. 

To start out we rewrite Eq.~(\ref{eq:exp_time_dep_rho}) for the general case of a time-dependent non-Hermitian Hamiltonian as
\begin{equation}
 \left< O \right>_{\rho(t)} = \frac{\Tr\left[ \rho(t) O \right]}{\Tr\left[ \rho(t)\right]}    =
\frac{\Tr\left[ U(t) \rho  U^\dag(t) O \right]}{\Tr\left[ U(t) \rho U^\dag(t) \right]} = \frac{\Tr\left[ \rho  U^\dag(t) O U(t) \right]}{\Tr\left[ \rho U^\dag(t) U(t) \right]} ,
\label{eq:exp_time_dep_rho_lr}    
\end{equation}
with the time-evolution operator $U(t)$ and the initial density matrix $\rho$. In the last step we used the invariance of the trace under cyclic permutations. Note that in \cite{Tetling2022} the denominator is missing. As this will give a contribution to linear order in $V_t$ \cite{Sticlet2022}, the analysis of  \cite{Tetling2022} is incomplete. Writing 
\begin{equation}
U(t) = e^{-i H_0 t} S(t) , 
\label{eq:ansatz_U}    
\end{equation}
the operator $S(t)$ fulfills the differential equation
\begin{equation}
i \partial_t S(t) = e^{i H_0 t} V_t e^{-i H_0 t} S(t) ,
\label{eq:diff_eq_S}    
\end{equation}
with the solution
\begin{equation}
 S(t) = \mathbbm{1} - i \int_{0}^{t} dt' e^{i H_0 t'} V_{t'} e^{- i H_0 t'} + \ldots =   \mathbbm{1} - i \int_{0}^{t} dt' f(t') D_{\rm D}(t') + \ldots   , 
\label{eq:sol_S}    
\end{equation}
where the dots denote terms $\mathcal{O}\left( V_{t}^2 \right)$. We introduced the operator $D_{\rm D}(t)$ as known from the Dirac (or interaction) picture of standard quantum mechanics\footnote{Note that in the Dirac picture no $H_0^\dag$ appears even though $H_0$ is non-Hermitian.}
\begin{equation}
D_{\rm D}(t) = e^{i H_0 t} D e^{- i H_0 t} .
\label{eq:Dirac_pic}
\end{equation}
This holds regardless if $D$ is Hermitian or not. We next consider the numerator and the denominator in Eq.~(\ref{eq:exp_time_dep_rho_lr}) separately. 

The denominator can be written as
\begin{align}
\label{eq:den_interaction_pic}
\Tr \left[ \rho U^\dag(t) U(t) \right]  &= \Tr \left[ \rho S^\dag(t) e^{i H_0^\dag t} e^{-i H_0 t} S(t) \right] \\* \nonumber
& = \Tr \left[ \rho e^{i H_0^\dag t} e^{-i H_0 t} \!\! + \!\! i \rho  e^{i H_0^\dag t} \!\!\! \int_0^t  \!\! dt' f(t') \left\{ D_{\rm D}^\dag(t'-t) -  D_{\rm D}(t'-t)   \right\} e^{-i H_0 t} \right] + \ldots .
\end{align}
As usual in linear response theory, we assume that the initial density matrix is stationary with respect to the unperturbed time evolution \cite{Bruus2003} and that it is normalized $\Tr[\rho] = 1$. Taking into account the considerations of Sects.~\ref{subsec:eq_ensembles} and \ref{subsec:ensemble_broken} we are essentially restricted to Hamiltonians $H_0$ with entirely real spectra ($\mathcal{PT}$ symmetry unbroken phase). Using this and the cyclic invariance of the trace we end up with
\begin{align}
\Tr\left[ \rho U^\dag(t) U(t) \right]   =
1 + \Tr\left[ \rho \int_0^t dt' f(t')  \left\{ D_{\rm D}^\dag(t'-t) -  D_{\rm D}(t'-t)   \right\} \right] + \ldots .
\label{eq:den_interaction_pic_2}
\end{align}
Without specifying $H_0$ and $V_t$ this cannot be simplified any further. Crucially, the denominator has a contribution of order $V_t$. Up to now we did not use that $D$ is Hermitian. 

For the computation of the numerator of Eq.~(\ref{eq:exp_time_dep_rho_lr}) we study
\begin{align}
 U^\dag(t) O U(t) & = e^{i H_0^\dag t} O e^{-i H_0 t} + i \int_{0}^{t} dt' f(t') e^{i H_0^\dag t'} D^\dag  e^{-i H_0^\dag (t'-t)} O  e^{-i H_0 t} \nonumber \\
& -i  e^{i H_0^\dag t} O \int_{0}^{t} dt' f(t') e^{i H_0 (t'-t)} D  e^{-i H_0 t'} + \ldots .
\label{eq:num_1}
\end{align}
If we now use $D^\dag = D$, introduce the two operators $\Omega = \eta_0^{-1} O$ and $\Delta = \eta_0^{-1} D$, with $\eta_0$ corresponding to $H_0$, and employ the pseudo-Hermiticity relation Eq.~(\ref{eq:eta_doesit}) we can rewrite this expression as
\begin{align}
 U^\dag(t) O U(t) & = \eta_0 \Omega_{\rm D}(t) + i \eta_0 \int_{0}^{t} dt' f(t') e^{i H_0 t'} \Delta \eta_0  \Omega_{\rm D}(t-t') e^{-i H_0 t'} \nonumber \\*
& -i  \eta_0 \int_{0}^{t} dt' f(t') e^{i H_0 t'} \Omega_{\rm D}(t-t')  \eta_0 \Delta  e^{-i H_0 t'}  + \ldots .
\label{eq:num_2}
\end{align}
Multiplying by the initial density matrix $\rho$, again assumed to be stationary under the time-evolution generated by $H_0$, and using the cyclic invariance of the trace we end up with
\begin{align}
\Tr \left[ \rho  U^\dag(t) O U(t) \right]  = 
\Tr \left[ \rho \eta_0 \Omega_{\rm D}(t) \right]  -  i  \int_{0}^{t} dt' f(t')
\Tr \left\{  \rho  \left[\eta_0 \Omega_{\rm D}(t-t') , \eta_0 \Delta_{\rm D}(0) \right] \right\}  + \ldots .
\label{eq:num_3}
\end{align}
In analogy to Hermitian quantum many-body theory \cite{Bruus2003} one is now tempted to define a response function of the unperturbed systems as
\begin{equation}
\chi_{O,D}(t) = - i \Theta(t) \Tr\left\{  \rho  \left[\eta_0 \Omega_{\rm D}(t) , \eta_0 \Delta_{\rm D}(0) \right] \right\}      
\label{eq:def_response}    
\end{equation}
such that the denominator is given by
\begin{align}
 \Tr\left[ \rho  U^\dag(t) O U(t) \right] =
\Tr\left[ \rho \eta_0 \Omega_{\rm D}(t) \right] + 
\int_{-\infty}^{\infty} \!\! dt' f(t') \chi_{O,D}(t-t') + \ldots .
\label{eq:num_4}
\end{align}
The response function is computed with respect to the time-evolution of the unperturbed system with non-Hermitian Hamiltonian $H_0$ and a density matrix which is stationary with respect to the time evolution induced by $H_0$, e.g., the analogue of the canonical density matrix Eq.~(\ref{eq:nH_can}). This is the extension  of the result given in \cite{Tetling2022} for states\footnote{Note that in \cite{Tetling2022} it is not mentioned that the state considered must be a right eigenstate of $H_0$. This corresponds to our requirement of the stationarity of the initial density matrix.} to density matrices.

It is formally appealing that by introducing the response function we obtained an expression similar to that of Hermitian quantum theory. However, for practical calculations this formulation of linear response theory for a non-Hermitian (but $\eta_0$-pseudo-Hermitian) $H_0$ is by far less useful than in the Hermitian case. There are several reasons for this. Firstly, the entire response to the perturbation $V_t$ switched on at $t=0$ to linear order is given by
\begin{align}
\label{eq:response_full} 
\delta \left< O \right>_{\rho(t)} & =  
\int_{-\infty}^{\infty}  dt' f(t') \chi_{O,D}(t-t') \\ 
& \hspace{1cm}
- \Tr\left[ \rho \eta_0 \Omega_{\rm D}(t) \right] \Tr\left[ \rho \int_0^t dt' f(t')  \left\{ D_{\rm D}^\dag(t'-t) -  D_{\rm D}(t'-t)   \right\} \right] .
\nonumber
\end{align}
The second line, which is absent in the Hermitian case, cannot be expressed in terms of a response function of the unperturbed system. It results out of the required normalization of the density matrix which, for a non-unitary time evolution, is not preserved. Already in Sect.~\ref{subsec:exp_time_dep_state} we emphasized that it is this normalization denominator which spoils much of the many-body formalism established for Hermitian systems. The present case is an example for this. More on this will be discussed in the next section. Secondly, the operators appearing in the response function are not the observable $O$ and the Hermitian system operator $D$ to which the perturbation couples but rather operators transformed by multiplying $\eta_0^{-1}$ from the left. As emphasized above, for most non-Hermitian Hamiltonians $H_0$ of interest a corresponding $\eta_0$ will not be known explicitly. Thirdly, the approach is limited to the case of an unbroken $\mathcal{PT}$-symmetry of $H_0$ and cannot be extended to the symmetry broken case (due to the lack of a stationary density matrix with respect to $H_0$).    

\section{Correlation functions and functional integrals}
\label{sec:genfun}

\subsection{Correlation functions}
\label{subsec:corr_fu}

To investigate the physical properties of  quantum many-body systems, besides studying expectation values of observables directly, one often computes correlation functions. They contain crucial information about spatial and temporal correlations of a system, are central to make contact with experiments, and can be used to extract expectation values of various observables \cite{Bruus2003}. The response functions appearing in linear response theory introduced in Sect.~\ref{subsec:lin_resp} [see Eq.~(\ref{eq:def_response})] are special types of correlation functions. The general structure of a correlation function is $\left< A_1 A_2 A_3 \ldots \right>$ with generic operators $A_i$ and an averaging $ \left< \ldots \right>$. The operators might depend on real or imaginary time. The different $A_i$ could, e.g., be the same operator at different times or at different spatial locations but the same time. Neither the individual operator nor their product needs to be an observable. For now we do not specify of what type the averaging (expectation value) is. It could be a state expectation value or a quantum statistical one (involving a density matrix). It could also be an expectation value of Hermitian quantum mechanics or one involving a metric operator $\hat g$ (of $\mathcal{PT}$-symmetric or biorthogonal quantum mechanics).

To be concrete, let us consider two examples. The first concerns the $\mathcal{PT}$-symmetric harmonic oscillator with imaginary, cubic anharmonicity Eq.~(\ref{eq:sci_post_ham}). For this model one might be interested in the ground state or canonical $\mathcal{PT}$-symmetric ``expectation value'' of the product of two position operators at different times in the imaginary time Heisenberg picture \cite{Grunwald2022}. Remember that a conventional Heisenberg picture exists only when using this ``expectation value''. The imaginary time evolution [for real times, see Eq.~(\ref{eq:heisenberg_pict})] is given by \cite{Bruus2003,NegeleOrland_QTCM} 
\begin{equation}
\hat x_{\rm H}(\tau) =   e^{H \tau} \hat x e^{- H \tau}   .
\label{eq:heisenberg_pict_im_time}    
\end{equation}
The correlation function is defined as ($\tau \geq 0$) 
\begin{equation}
\left< \hat x_{\rm H}(\tau) \hat x_{\rm H}(0) \right>_{\rho^{\rm can}}^{\mathcal{PT}}  = \Tr \left[\rho^{\rm can}  \hat x_{\rm H}(\tau) \hat x_{\rm H}(0)  \right] .
\label{eq:correl_x_x}    
\end{equation}
For $\beta \to \infty$ this yields the ground state ``expectation value''; see Eq.~(\ref{eq:PT_exp_stat_0}).  
%Note that $\hat x_{\rm H}(\tau) \hat x_{\rm H}(0)$ is not an observable in the sense of $\mathcal{PT}$-symmetric quantum mechanics. It is not $\eta_{\rm r}$-pseudo Hermitian. 
Equation (\ref{eq:correl_x_x}) is an auto-correlation function. For $\tau=0$ it enters the expression for the square of the fluctuations of the particles ``position''\footnote{Remember that $\hat x$ is not an observable in the sense of $\mathcal{PT}$-symmetric quantum mechanics; see Sect.~\ref{subsec:ob_ex_PT_ex}. In fact, its ground state expectation value for the present model is purely imaginary; for more see \cite{Grunwald2022}.} around its mean value $\left< \hat x^2_{\rm H}(0) \right>_{\rho^{\rm can}}^{\mathcal{PT}}  - \left( \left< \hat x_{\rm H}(0) \right>_{\rho^{\rm can}}^{\mathcal{PT}} \right)^2$ \cite{Grunwald2022}. 

In the course of this review we already collected several arguments why  $\left< \hat x_{\rm H}(\tau) \hat x_{\rm H}(0) \right>_{\rho^{\rm can}}^{\mathcal{PT}}$ evaluated at finite or infinite $\beta$ will very likely be of mathematical interest but not of direct physical relevance. Its computation fits into the elegant formalism of $\mathcal{PT}$-symmetric or biorthogonal quantum mechanics (see below) but defining it, several concepts are used, which turned out to be questionable from the physics perspective. This, in particular, concerns the Heisenberg picture (see Sect.~\ref{subsec:exp_time_dep_state}) as well as the averaging with the non-Hermitian and non-stationary $\rho^{\rm can}$ at finite $\beta$ (see Sect.~\ref{subsec:eq_ensembles}) or the left-right state ``expectation value'' [see Eq.~(\ref{eq:PT_exp_stat_0}) and Sect.~\ref{subsec:ob_ex_PT}] for $\beta \to \infty$. For the  introduction of  functional integrals it turns out that considering this correlation function is still very instructive. 

The second example is a spatial many-body correlation function of our staggered tight-binding chain with complex hopping Eq.~(\ref{eq:SC_ham}). To investigate the model's quantum critical behavior for $g = \delta$, in Sect.~\ref{subsec:critical_stag}, we explicitly compute temperature $T=0$ correlation functions within $\mathcal{PT}$-symmetric quantum mechanics
\begin{equation}
G_l^{\mathcal PT}(j) = \left< c^\dag_{l+j} c_{l} \right>_{\left| {\rm R}_0 \right>}^{\mathcal PT} = \left< {\rm L}_0 \right| c^\dag_{l+j} c_{j} \left| {\rm R}_0 \right> 
\label{eq:correl_qc_PT}    
\end{equation}
and Hermitian quantum mechanics 
\begin{equation}
G_l(j) = \left< c^\dag_{l+j} c_{l} \right>_{\left| {\rm R}_0 \right>} = \frac{\left< {\rm R}_0 \right| c^\dag_{l+j} c_{j} \left| {\rm R}_0 \right>}{\left< {\rm R}_0 \! \right. \left| {\rm R}_0 \right>}  
\label{eq:correl_qc}    
\end{equation}
in the many-body ground state $\left| {\rm R}_0 \right>$. The latter was earlier studied in  \cite{Dora2022a}. The operators entering these expressions are not observables; neither in the framework of  $\mathcal{PT}$-symmetric quantum mechanics nor in the Hermitian sense. But they contain information about criticality and symmetry breaking; more on this later. These type of single-particle correlation functions are often also denoted as Green functions, which explains our notation.  

\subsection{Functional integrals and generating functionals}
\label{subsec:FI}

To compute correlation functions in quantum many-body theory one often employs functional integrals \cite{NegeleOrland_QTCM}. For $\mathcal{PT}$-symmetric, non-Hermitian systems with Hamiltonian $H$, functional (or path) integrals were first used for specific models to construct the isospectral Hermitian operator $h$ and the corresponding similarity transformation $S$ according to theorem $\mbox{T}^{\mathcal{PT}}_{2}$ \cite{Bender2006,Jones2006a}. However, they are also crucial for an extension of $\mathcal{PT}$-symmetric quantum mechanics to $\mathcal{PT}$-symmetric quantum field theory \cite{Bender2004,Rivers2011-fo}. This is reviewed in chapter 5 of \cite{Bender2019}.

To introduce the use of functional integrals for non-Hermitian, $\mathcal{PT}$-symmetric systems let us rewrite the canonical partition function Eq.~(\ref{eq:rho_can_part}) of the Hamiltonian Eq.~(\ref{eq:sci_post_ham}) as 
\begin{equation}
Z^{\rm can} \! = \! \Tr \left[ e^{-\beta H} \right] \! = \sum_{\nu} \left< {\rm L}_\nu\right|  e^{-\beta H} \left| {\rm R}_{\nu} \right>
      = \!\! \int \! d x  \left< x \right| e^{-\beta H} \left| x  \right> = \!\! \int_{x(0)=x(\beta)} \!\!\!\!\! \mathcal{D} x  \; e^{-S[x]},
\label{eq:Z_FI}
\end{equation}
where for the second equal sign we used Eq.~(\ref{eq:complete}) as well as the (over-) completeness of the basis of position eigenstates $\left\{ \left| x \right> \right\}$. In the last step we employed the standard slicing derivation to obtain a functional integral \cite{NegeleOrland_QTCM}.  The imaginary time (Euclidean) action $S[x]$ for this model reads
\begin{equation}
    S[x] = \int_0^{\beta} d \tau \;
    \left\{ \frac{ \left[ \partial_{\tau}x (\tau) \right]^2}{2} 
    + \frac{x^2(\tau)}{2} 
    + \frac{ik}{3!}x^3(\tau)
    \right\}.
\label{eq:action_cubic_os}    
\end{equation}
After transforming  from the operator $\hat x_{\rm H}(\tau)$ in the imaginary time Heisenberg picture to the function $x(\tau)$ in the functional integral we, as usual, suppress the index H. Note that $Z^{\rm can}$ is an integral part of the unphysical (non-stationary, non-Hermitian) $\rho^{\rm can}$ Eq.~(\ref{eq:rho_can}) but also of the proper extension of the canonical density matrix to non-Hermitian systems $\rho_{\rm nH}$ Eq.~(\ref{eq:nH_can}). Since the functional integral with a cubic term cannot be solved exactly, Eqs.~(\ref{eq:Z_FI}) and (\ref{eq:action_cubic_os}) can be used as the starting point to derive expressions for $Z^{\rm can}$ perturbative in $k$  involving Feynman diagrams \cite{NegeleOrland_QTCM,Bruus2003}. Also more elaborate methods such as mean-field theory or a functional renormalization group approach can be employed starting from the functional integral presentation \cite{Zambelli2017,Bender2018-og,Bender2021-ib,Grunwald2022,Liegeois2022}. For $\beta \to \infty$ one can extract the ground state energy (eigenvalue) $E_0$ from  $Z^{\rm can}$ \cite{Grunwald2022}. 

By adding a source term 
\begin{equation}
S_{\rm s}[x,j] = \int_{0}^{\beta} d \tau \; x(\tau) \, j(\tau)      
\label{eq:action_source}    
\end{equation}
to the action,  the functional integral becomes an even more versatile tool for Hermitian systems \cite{NegeleOrland_QTCM}. Formally (and on first glance) this appears to be the case for non-Hermitian Hamiltonians as well. One can use the resulting functional 
\begin{equation}
\mathcal{W}[j] = \int_{x(0)=x(\beta)}  \mathcal{D} x  \; e^{-S[x] + S_{\rm s}[x,j]} 
\label{eq:generating_functional}    
\end{equation}
to generate correlation functions by applying functional derivatives with respect to $j(\tau_i)$ \cite{NegeleOrland_QTCM}. For $\tau >0$ we obtain
\begin{equation}
\left< \hat x_{\rm H}(\tau) \hat x_{\rm H}(0) \right>_{\rho^{\rm can}}^{\mathcal{PT}} = \frac{1}{Z^{\rm can}} \left. \frac{\delta^2 \mathcal{W}[J]}{\delta j(\tau) \delta j(0)} \right|_{j=0}  .   
\label{eq:funder_correl}    
\end{equation}
To derive this, one can follow the same steps as in the Hermitian case to relate operator averages (left hand side) to functional integral ones (right hand side) \cite{NegeleOrland_QTCM}. This exemplifies that within the framework of $\mathcal{PT}$-symmetric (and, for that matter, also in that of biorthogonal) quantum mechanics, functional integrals can be introduced and used just as established for Hermitian Hamiltonians. However, all this is rather formal as one has to ignore that $\rho^{\rm can}$ is not a proper stationary density matrix, and the ``expectation value'' is thus not taken for an equilibrium ensemble. This appears to be odd when thinking of physics and not merely of elegant mathematics. Furthermore, the operator $\hat x$ appearing in the correlation function is not the observable of the particles position (see Sect.~\ref{subsec:ob_ex_PT_ex}). 
%For the action \cref{eq:action_cubic_os} it is even purely imaginary. 
We conclude that for finite $\beta$ the correlation function $\left< \hat x_{\rm H}(\tau) \hat x_{\rm H}(0) \right>_{\rho^{\rm can}}^{\mathcal{PT}}$ is not a physically sensible object. The usefulness of its elegant functional integral representation appears to be questionable. 
% Thus its elegant function integral representation is useless. 

If one takes the zero temperature limit the missing stationarity and also the lack of Hermiticity of $\rho^{\rm can}$ are no longer an issue. In this case one ends up with the ground state ``expectation value'' of $\mathcal{PT}$-symmetric quantum mechanics; see Eq.~(\ref{eq:PT_exp_stat_0}). However, we already argued on physical grounds that the methodology of $\mathcal{PT}$-symmetric quantum mechanics does not appear to be the proper one, even for states. At least, employing a Lehmann representation \cite{Bruus2003,NegeleOrland_QTCM}, one can still extract energy eigenvalues of excited states \cite{Grunwald2022}. Even for $\beta \to \infty$, $\left< \hat x_{\rm H}(\tau) \hat x_{\rm H}(0) \right>_{\rho^{\rm can}}^{\mathcal{PT}}$ and its functional integral representation are thus only of limited usefulness.

The question that immediately arises is, if a functional integral can also be used to compute the expectation value of a product of time evolved position operators involving the proper generalization of the canonical density matrix $\rho_{\rm nH}$ Eq.~(\ref{eq:nH_can}). As emphasized in Sect.~\ref{subsec:ens_exp_st} $\rho_{\rm nH}$ and $\rho^{\rm can}$ differ by the factor $\sum_{\nu} \left| {\rm R}_\nu \right> \left< {\rm R}_\nu \right| / \left< {\rm R}_\nu \right.\!\! \left| {\rm R}_\nu \right>$. To investigate this, let us start with an even simpler expectation value, namely
\begin{align}
\left< \hat x \right>_{\rho_{\rm nH}} & = {\rm Tr} \left[  \rho_{\rm nH} \hat x \right] = {\rm Tr} \left[ \frac{1}{Z^{\rm can}} e^{- \beta H} \sum_{\nu} \frac{\left| {\rm R}_\nu \right> \left< {\rm R}_\nu \right|}{\left< {\rm R}_\nu \! \right.  \left| {\rm R}_\nu \right> }     \hat x \right] \nonumber \\
& = \frac{1}{Z^{\rm can}} \int dx \, dx' \; x \left<x\right|   e^{- \beta H} \left| x' \right> \sum_{\nu} \frac{\left< x' \! \right. \left| {\rm R}_\nu \right> \left< {\rm R}_\nu \! \right.  \left| x \right>}{\left< {\rm R}_\nu \! \right. \left| {\rm R}_\nu \right>} ,
\label{eq:x_exp_FI}    
\end{align}
where we took the trace in the (overcomplete) basis of position states $\left\{ \left| x \right> \right\}$. Now the problem becomes apparent. While the first factor in the integrand can be treated in the slicing derivation of functional integrals \cite{NegeleOrland_QTCM} this does not hold for the second one involving the sum. The (right) eigenfunctions of the Hamiltonian $\left| {\rm R}_\nu \right>$ do not drop out but cannot be part of a standard functional integral. We thus conclude that a standard functional integral cannot be used to compute more complex products of time evolved position operators.\footnote{This leaves open which type of time evolution one should use; the Heisenberg picture one of Eq.~(\ref{eq:heisenberg_pict}) or that of Eq.~(\ref{eq:exp_val_time_evol_nonbio}). Staying consistently within the Hermitian approach, \cref{eq:exp_val_time_evol_nonbio} appears to be the correct choice.} 

This insight has severe consequences for the application of a variety of established quantum many-body methods which are based on the functional integral approach to non-Hermitian systems. We emphasize that, as in linear response theory Eq.~(\ref{eq:response_full}) and the (Hermitian) expectation value of an observable Eq.~(\ref{eq:ex_value_rewitten_nochmal}), it is again the presence of  states which spoils the use of methods for non-Hermitian Hamiltonians which are at the heart of Hermitian quantum many-body theory. 

For completeness we mention the following. In the $\beta \to \infty$ limit Eq.~(\ref{eq:funder_correl}) becomes the ground state ``expectation value'' of $\hat x_{\rm H}(\tau) \hat x_{\rm H}(0)$ according to the definitions of $\mathcal{PT}$-symmetric quantum mechanics. While the latter explicitly contains the metric operator $\hat g = \eta_{\rm r}$ it does not appear in the functional integral expression. This surprising observation led to an extensive discussion in the mathematical physics community \cite{Jones2007,Mostafazadeh2007-xq,Jones2009-hx}. In this Mostafazadeh \cite{Mostafazadeh2007-xq} emphasized that the ``field'' $x(\tau)$, the source $j(\tau)$ is coupled to in Eq.~(\ref{eq:action_source}), is not an observable in the sense of $\mathcal{PT}$-symmetric quantum mechanics; see Sect.~\ref{subsec:ob_ex_PT_ex}. He argued that the external source should be coupled to an observable which he suggested to be the $\eta_{\rm r}$-pseudo-Hermitian operator $\hat x_{\mathcal PT} = \eta_{\rm r}^{-1/2} \hat x \eta_{\rm r}^{1/2}$; see Sect.~\ref{subsec:isospectral}. If using this, the metric operator would only drop out in $\mathcal{W}[j=0]=Z^{\rm can}$ but not for finite sources.

\subsection{Quantum critical behavior of the staggered tight-binding chain}
\label{subsec:critical_stag}

Correlation functions are routinely used to analyze the critical behavior of quantum many-body systems. Quantum (temperature $T = 0$) or thermal ($T \neq 0$) criticality, is generally associated with the emergence of long-range correlations, that manifest as spacial or temporal power-law decaying Green functions whose exponents provide insight into the universality class of the transition. In the case of quantum critical systems, such a critical behavior is  associated with the closing of an excitation gap \cite{Sachdev2011}.

These ideas have recently been extensively investigated in the context of $\mathcal{PT}$-symmetric quantum mechanics, unveiling new kinds of criticality, universality, and symmetry breaking \cite{Dora2022a,Ashida2016-ll,Li2015,Kawabata2017,Starkov2022,Li2014,Lenke2021,Xiao2019,doraKibbleZurekMechanism2019,xueNonHermitianKibbleZurek2021,Syed2022,Lourenco2018-kn,Nakagawa2018-yu,Kattel2023,Yamamoto2019-pn,Yamamoto2023,Roccati2021}. %VM: added Yamamoto2023 and Roccati2021
At this point we refrain from diving into this advanced topic and focus on the methodological aspects. Underlying many established methods for analyzing criticality, such as the renormalization group or loop-wise expansions \cite{zinn-justinQuantumField2002}, is the functional integral. But, as outlined in the last section (\cref{subsec:FI}), the conventional construction of the functional integral fundamentally relies on the non-stationary density matrix $\rho\^{can}$ and on  $\mathcal{PT}$-symmetric correlation functions. As we have argued throughout this review, these ``expectation values'' and the corresponding correlation functions do not seem to be of physical significance. 

One might, however, ask if aspects of criticality and symmetry breaking can be captured in the behavior of the $\mathcal{PT}$-symmetric correlation functions? Below we show by example, that this is not the case. The quantum critical behavior of a $\mathcal{PT}$-symmetric system can not be captured by analyzing the $\mathcal{PT}$-symmetric Green function. This complements our previous studies and strengthens the argument, that only the Hermitian correlation functions capture the complete physical picture.

The example we will consider is the staggered tight-binding chain with complex hopping \cref{eq:SC_ham}, that we introduced in \cref{subsec:examples_stag}. There we discussed, that the system shows a gap closing for $\delta = g$ and hence becomes quantum critical. Here we want to analyze the behavior of the Hermitian \cref{eq:correl_qc} and $\mathcal{PT}$-symmetric \cref{eq:correl_qc_PT} correlation functions at the quantum critical point, for a half-filled system in the thermodynamic limit. The Hermitian Green function was previously analyzed in \cite{Dora2022a}.

Due to the translational invariance of the Hamiltonian \cref{eq:SC_ham} we have $G_l^{(\mathcal PT)}(j) = G_{l + 2\mathbb{N}}^{(\mathcal PT)}(j)$ (for both, the Hermitian and the $\mathcal{PT}$-symmetric case), so that, without loss of generality, we fix $l = 1$ and drop the index in the following. Linearizing the model around $k = \pi / 2$, and mapping it onto the field theory Eq.~(\ref{eq:SC_ham_field}), analytical results for the scaling behavior of the Hermitian Green function have been obtained in \cite{Dora2022a}. In the thermodynamic limit $N \to \infty$ one finds at half-filling
\begin{align}
    \label{eq:crit_GF_scaling}
    \mathcal{S}(m) = \abs{G(2m + 1)} &\sim
    \begin{cases}
        m^{-1} &\text{for} \; 1 \ll m \ll J/\delta ,\\ 
        m^{-3} &\text{for} \; J/\delta \ll m
    \end{cases}, \\
    \mathcal{F}(m) = \abs{G(2m)} &\sim
    \begin{cases}
        \log(\abs{m})  &\text{for} \; 1 \ll m \ll J/\delta ,\\ 
        m^{-2}  &\text{for} \; J/\delta \ll m
    \end{cases},
\end{align}
i.e., exactly the kind of power-law scaling we would expect for a quantum critical model. The distinction between even and odd lattice sites can be intuitively understood, due to the staggered nature of the model [momenta $k$ and $k - \pi$ couple in the Hamiltonian; see \cref{eq:SC_ham_k}]. The spectrum of the linearized theory \cref{eq:SC_ham_k_exp} is given by $\epsilon(k) = \pm J \abs*{k-\pi/2}$ and therefore equivalent to that of a free Fermi gas. This in particular implies that the canonical partition function $Z\^{can}$ does not show signs of criticality \cite{Dora2022a}. Despite this, the Hermitian Green function, which in a free Fermi gas would decay as $G(m) \sim m^{-1}$, are suppressed at large separations. This feature can be explained by a measurement induced Fano effect \cite{Dora2022a}.

\begin{figure}
    \centering
    \includegraphics{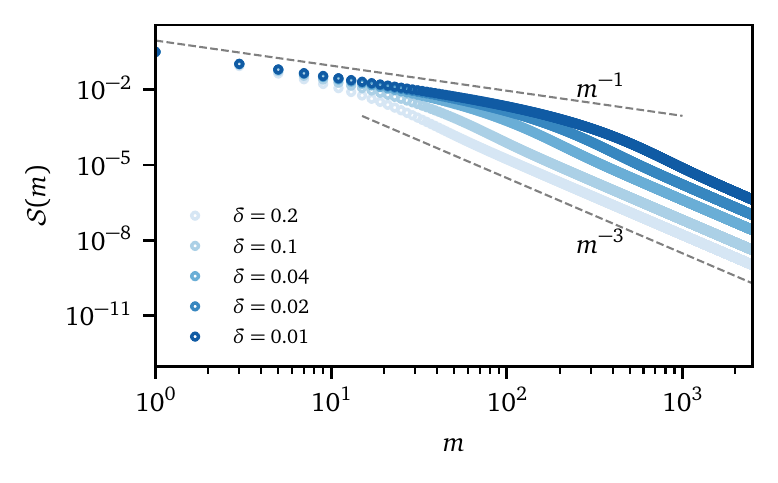}
    \caption{\textbf{Quantum critical regime of the Hermitian Green function $\bm{\mathcal{S}(m)}$} from an exact numerical treatment, for various $\delta=g - \delta\_s$, large $N = 20002$, and small $\delta\_s = 10^{-8}$ at half-filling. The exact numerical solution is compared with the approximate analytical results \cref{eq:crit_GF_scaling} indicated by the dashed lines. The crossover from intermediate-distance $m^{-1}$ to long-distance $m^{-3}$ behavior is beautifully visible for small $\delta$, and hence confirms the validity of the analytical calculation.}
    \label{fig:critical_linear_cubic_transition}
\end{figure}
\begin{figure}
    \centering
    \includegraphics{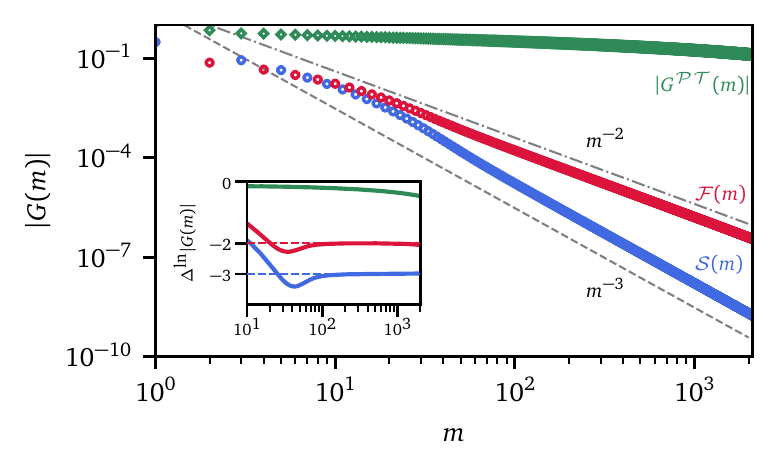}
        \caption{\textbf{Quantum critical regime of the Green functions} evaluated with the $\mathcal{PT}$-symmetric \cref{eq:correl_qc_PT} (green) and the Hermitian \cref{eq:correl_qc_PT} (red, blue) expectation value, together with the analytical predictions (dashed and dashed-dotted lines) for $\delta = g - \delta\_s = 0.2$. The inset shows the double logarithmic derivative $\Delta\^{ln}f(m) = \frac{\partial \ln\left[f(m)\right]}{\partial \ln(m)}$, of the data from the main plot, evaluated as centered differences and illustrates the power-law behavior of the Hermitian correlation function \LG{. The $\mathcal{PT}$-symmetric Green function can not be evaluated in the thermodynamic limit for $\delta_s \to 0$ (see main text). For a small but finite $\delta_s$, which provides a large distance cutoff much larger than the largest $m$ shown in the plot, the inset illustrates that it does not follow a power law. The data was obtained for $N\_{herm.} = 20 002$ and $N\_{bo} = 40 002$, with $\delta_s = 10^{-9}$ and $\delta_s = 10^{-8}$ respectively.}}
        % and in particular highlights that the  $\mathcal{PT}$-symmetric Green function does not follow a power law function. \LG{Further, it is not possible to numerically converge the results.} The data were obtained for $N\_{herm.} = 20 002$ and $N\_{bo} = 40 002$, both with $\delta_s = 10^{-9}$.}
    \label{fig:critical_td_limit}
\end{figure}

In the following, we compare the approximate analytical result with the full numerical solution of the lattice model. But more importantly, we numerically compute the $\mathcal{PT}$-symmetric correlation function \cref{eq:correl_qc_PT} to analyze if the same power-law behavior can be found in this quantity. The numerical details are explained in \cref{sec:critical_num_corr}. The quantum critical point $\delta = g$ is at the same time an exceptional point, that separates the region of broken and unbroken $\mathcal{PT}$-symmetry.  This implies, that the Hamiltonian is not diagonalizable at the quantum critical point, leading to technical difficulties. To avoid these in our numerical treatment, we consider a shift away from the quantum critical point towards the phase of unbroken $\mathcal{PT}$ symmetry, i.e. we calculate the correlation functions for $\delta = g - \delta\_s$\LG{, with a small $\delta\_s$}. The inverse of  $\delta\_s$ as well as the system size $N$ provide large-$m$ cutoff scales for the quantum critical power-law scaling \cite{Sachdev2011}. \LG{We do not want to deal with this issue and therefore aim for numerical results, converged on the scales of the plots in $\delta\_s$ and $N$.}
%We do not want to deal with this issue and thus \LG{try to} converge our numerical results in $\delta\_s$ and $N$ on the scales of the plots. 
The data shown in \cref{fig:critical_linear_cubic_transition} and \cref{fig:critical_td_limit} \LG{for the Hermitian Greens functions} does not suffer from finite $\delta\_s$ and $1/N$ corrections \LG{while we find that it is not possible to numerically converge the results for the $\mathcal{PT}$-symmetric correlation function. For a given $\delta_s$ there exists a $N(\delta_s) \sim -\ln(\delta_s)$ beyond which the data are converged in $N$. 
%that is converged, 
However, $\lim_{\delta_s \to 0} N(\delta_s) = \infty$, so that it is impossible to evaluate the thermodynamic limit of the $\mathcal{PT}$-symmetric correlation function at the quantum critical point. For completeness, in \cref{fig:critical_td_limit} we show the results for a finite but fixed $\delta_s$ (green diamonds), that allows numerical convergence in $N(\delta_s)$; see \cref{sec:critical_LR_corr} for more details.}

The transition from $m^{-1}$ to $m^{-3}$ behavior of the Hermitian Green function $\mathcal{S}(m)$ in the thermodynamic limit at half filling is illustrated in \cref{fig:critical_linear_cubic_transition}. With decreasing $\delta$, the crossover scale $J/\delta$ shifts to larger $m$. The figure nicely illustrates, that the analytical result can capture both the intermediate and long-range behavior of the system.

Hermitian and $\mathcal{PT}$-symmetric Green functions are shown in \cref{fig:critical_td_limit}. The long-distance behavior of $\mathcal{F}(m)$ and $\mathcal{S}(m)$ again agrees nicely with the analytical prediction \cref{eq:crit_GF_scaling}. This is further validated by the inset, which shows the double logarithmic derivative of the data from the main plot.  The color-matched dashed lines at large $m$ confirm the long-range power-law behavior irrevocably. The $\mathcal{PT}$-symmetric Green function on the other hand, \LG{which we can only evaluate at fixed $\delta_s$}, does not show the same behavior. In fact, it neither captures the distinction between even and odd lattice sites, nor does it show power-law decaying correlations. The latter point can be seen in the main panel, or, more clearly, in the inset. The interpretation of this quantity and the physical picture it provides is hence unclear\LG{, even ignoring the convergence issues discussed above. Note that related problems appear within the field theory (see Sect.~\ref{subsec:examples_stag}) when analytically computing the $\mathcal{PT}$-symmetric Green function.\footnote{Private communication with B.~Dora.}} 

This concludes our final argument. The $\mathcal{PT}$-symmetric correlation functions, while mathematically very convenient, does not seem to carry any physical significance. They can not capture the same critical behavior that is observed in the Hermitian correlation function. Therefore, the usage of conventional path integrals and $\mathcal{PT}$-symmetric (and biorthogonal) correlation functions seems questionable.

\section{Summary}

\label{sec:sum}

In the course of this review we have collected strong evidence that the established formalism of Hermitian quantum mechanics should be applied when theoretically studying $\mathcal{PT}$-symmetric, non-Hermitian quantum many-body systems. They are realized as open quantum systems in which this symmetry originates in the systems symmetry and in a fine tuned system-reservoir coupling; a balanced gain and loss.
%which show this symmetry and system-reservoir couplings which are compatible.
Investigating the physics of such systems requires more than the mere computation of 
% the spectrum
spectral properties of the non-Hermitian Hamiltonian. Vital physical information can only be accessed by studying observables and correlation functions.
%and the corresponding eigenstates.
Due to the non-unitarity of the time evolution the formalism propagated in the present review is, in parts, clumsy. However, it does not suffer from basic problems and in all the examples we studied, leads to physically sensible results. This has to be contrasted, to the mathematically elegant formalisms of $\mathcal{PT}$-symmetric and biorthogonal quantum mechanics, which were mainly developed to study the spectral properties of $\mathcal{PT}$-symmetric, non-Hermitian systems. They suffer from methodological peculiarities which are difficult to reconcile with our physical intuition. Examples for this are the dependence of the concept of an observable on the Hamiltonian and the lack of a direct probabilistic interpretation of what is denoted as an ``expectation value''. The density matrix considered in these formalisms, which is supposed to be the analogue of the canonical one, is non-stationary and non-Hermitian. The former shows that it does not correspond to an equilibrium ensemble, but was still employed in this sense. The latter contradicts what one expects from the 
established theory of open quantum systems. More specifically, we showed for several model Hamiltonians that the formalisms of $\mathcal{PT}$-symmetric and biorthogonal quantum mechanics lead to results for which physical interpretations can hardly be found; e.g., negative fermionic level occupancies
%being larger than 1 or even negative
and correlation functions of quantum critical systems which do not decay as a power law.     

We reviewed the ancilla approach. It allows to embed the non-Hermitian system with its non-unitary dynamics into a Hermitian one. The effective non-unitarity of the dynamics follows from a measurement of the ancilla spin and the post-selection. In contrast to other methods from open system quantum mechanics, the ancilla approach is free of approximations. It presents a transparent way to substantiate the application of the  Hermitian formalism for $\mathcal{PT}$-symmetric systems
%It forms a transparent foundation of the use of the Hermitian formalism
and is fully consistent with the general theory of open quantum systems. Using the ancilla approach we constructed statistical ensembles and their corresponding density matrices for non-Hermitian systems. For time independent, $\mathcal{PT}$-symmetric Hamiltonians with entirely real spectra, stationary, Hermitian density matrices corresponding to equilibrium ensembles, which differ from the canonical ones, can be found. 

The necessity to apply the formalism of Hermitian quantum mechanics to $\mathcal{PT}$-symmetric, non-Hermitian systems has severe consequences for quantum mechanical many-body theory. Established methods such as Green functions, functional integrals, and generating functionals, being at the heart of many of the elaborate quantum many-body methods cannot be directly employed. Ironically, they can be used within $\mathcal{PT}$-symmetric and biorthogonal quantum mechanics. Furthermore, the extension of linear response theory to non-Hermitian systems suffers from a vital drawback. It requires the computation of the $\eta_{\rm r}$-operator of the unperturbed Hamiltonian, something that is, up to rare exceptional cases, unfeasible, even for non-interacting systems. Adding insult to injury, the final expression of linear response theory still involves references to the non-conserved norm of the density matrix, which can not be recast as a response function of the unperturbed system. Hence, a reformulation only involving response functions is not possible.
%Furthermore, the standard expression of linear response theory involving response functions must be complemented by an additional term explicitly containing the equilibrium density matrix of the unperturbed system (or the corresponding eigenstate of the Hamiltonian for pure states). The response function does neither contain the operator corresponding to the observable which is addressed to investigate the response nor the one, the perturbation couples to. Instead operators transformed by the $\eta_{\rm r}$-operator of the unperturbed system have to be inserted. Only in rare examples this operator can be given in closed form which provides a serious hindrance for the application of linear response theory to non-Hermitian systems.
%
With the appearance of the norm of the state or the trace of the density matrix in the denominator of expectation values, also correlation functions, being crucial to unravel emergent many-body phenomena, have a structure different to what is known from standard many-body theory. Their efficient evaluation in a many-body context requires further theoretical progress, as shown in this review. 
% This requires different ways to evaluate correlation functions as exemplified in this review.

To wrap up we want to provide a ``cookbook'' of the methodology to investigate $\mathcal{PT}$-symmetric---or more generally, pseudo-Hermitian---, non-Hermitian Hamiltonians. 
\begin{enumerate}
    % \item Observables are given by Hermitian operators which only contain system degrees of freedom. If desired the concept of an observable can be extended to pseudo-Hermitian operators with entirely real spectra, e.g., the Hamiltonian of a  $\mathcal{PT}$-symmetric system in its symmetry unbroken phase. The possible outcome of a measurement of the observable $O$ is one of its real eigenvalues.
    % \item The pure state expectation value of an observable $O$ is defined as in Hermitian quantum mechanics and has a direct probabilistic interpretation. As the state is not necessarily normalized, and, in particular, will not stay normalized during the non-unitary time evolution, one has to divide by the norm of the state. The resulting expression is Eq.~(\ref{eq:ex_value_rewitten_nochmal}). The metric operator does not appear. Left-right state ``expectation values'' cannot be measured.    
    \item Observables are given by Hermitian operators. The state expectation value of an observable $O$ is defined as in Hermitian quantum mechanics
    \begin{equation*}
        \expval{O}_{\ket{\psi(t)}} = \frac{\mel{\psi(t)}{O}{\psi(t)}}{\ip{\psi(t)}{\psi(t)}} , 
    \end{equation*}
    has a direct probabilistic interpretation, and always leads to physically sensible results.
    \item The Heisenberg picture, as known for Hermitian Hamiltonians, is not applicable.  
    \item Density matrices are Hermitian operators which obey the generalized von Neumann equation (\ref{eq:von_Neuman}). 
    \item For time-independent, $\mathcal{PT}$-symmetric non-Hermitian Hamiltonians with real spectra stationary density matrices corresponding to equilibrium ensembles have the general form Eq.~(\ref{eq:nH_can}). The conventional (grand) canonical density matrix from Hermitian quantum mechanics does not fall into this class. 
    \item Linear response theory can be set up formally, but leads to expressions Eqs.~(\ref{eq:response_full}) and (\ref{eq:def_response}) which are much more complex than for Hermitian Hamiltonians. As the $\eta_{\rm r}$ operator enters, it is questionable if it can be employed in practical computations beyond the exceptional cases in which $\eta_{\rm r}$ is known analytically.  
    \item Correlation functions remain to be useful objects. As expectation values they contain the norm of the state or the trace of the density matrix. It is, however, not evident how to compute them from generating functionals, prohibiting the direct usage of several established many-body methods.
    %set up by employing the functional integral formalism.  
\end{enumerate}

\section*{Acknowledgements}

We thank Silas Hardt and Lukas Heinen for collaborations during an early stage of this work. We have greatly benefited from discussions with Nils Caci, Tilman Enss, Tobias Meng, Ipsika Mohanty, Maarten Wegewijs, Stefan We{\ss}el and Balázs Dóra.

We acknowledge support by the Deutsche Forschungsgemeinschaft (DFG, German Research Foundation) via  the Research Training Group 1995 ``Quantum Many-Body Methods in Condensed Matter Systems'' project number 240766775, Germany’s Excellence Strategy---Cluster of Excellence Matter and Light for Quantum Computing (ML4Q) EXC 2004/1---project number 390534769, and project number 508440990. We also acknowledge support from the Max Planck-New York City Center for Non-Equilibrium Quantum Phenomena. Simulations were performed with computing resources granted by RWTH Aachen University under project thes0823 and on the HPC system Raven at the Max Planck Computing and Data Facility.

% \appendix
% \renewcommand{\thesection}{Appendix \Alph{section}}% Adjust section printing (from here onward)
% \renewcommand{\theequation}{\Alph{section}.\arabic{equation}}
\begin{appendices}
\crefalias{section}{appendix}

\section{Ladder operators of a biorthonormal basis}
\label{sec:apdx_nH_ladder}

In this section we discuss the ladder operators, associated with a biorthonormal basis as introduced in \cref{sec:non-herm_ham} \cite{Yoshimura2020}. To start out, let us consider a single-particle Hamiltonian $\hat h$, that in second quantization, i.e.,  on the Fock space, is written as a quadratic form in terms of creation $c^\dagger_i$ and annihilation $c_i$ operators corresponding to a complete orthonormal basis $\{ \ket*{e_i} \}$ of single-particle states
\begin{equation}
    H = \sum_{ij} h_{ij} c^\dagger_i c_j.
    \label{eq:num_ham_general}
\end{equation}
Here, the ladder operators fulfill the usual (anti-) commutation relations $\comm*{c^\dagger_i}{c_j}_{\mp} = \delta_{ij}$, with the upper (lower) sign corresponding to fermions (bosons). The matrix $h_{ij} = \mel*{e_i}{\hat h }{e_j}$ is in general not Hermitian, but we assume that its biorthonormal eigensystem $\{\ket*{\rm R_\nu}, \ket*{\rm L_\nu} \}$ constitutes a basis of the single particle Hilbert space. Using the resolution of unity  \cref{eq:biorthogonal} in terms of this biorthogonal basis, the ladder operators of the biorthonormal eigenbasis can be obtained via a conventional basis transformation
\begin{align}
    c^\dagger_i
    = \sum_\nu \ip*{\textnormal{L}_\nu}{e_i} c^\dagger_{ \rm R,\nu}
    =\sum_\nu \ip*{\textnormal{R}_\mu}{e_i} c^\dagger_{ \rm L,\nu},
    \label{eq:num_ladder_trafo}
\end{align} 
with $c^\dagger_{{\rm R}, \nu} \ket{0} = \ket*{{\rm R}_\nu}$. Using \cref{eq:num_ladder_trafo} and the commutation relations of $c^\dagger_i$ and $c_i$, we can directly deduce the (anti-) commutators of the biorthogonal ladder operators
\begin{align}
    \commutator{c^\dagger_{\textnormal{X},\mu}}{c_{\textnormal{Y},\nu}}_{\mp} &= \ip*{\textnormal{Y}_\nu}{X_\mu}, \\ 
    \commutator{c^{(\dagger)}_{\textnormal{X},\mu}}{c^{(\dagger)}_{\textnormal{Y},\nu}}_{\mp} &= 0,
\end{align}
where ${\rm X}, {\rm Y} \in \{\textnormal{R}, \textnormal{L}\}$. While the (anti-) commutator between $c^\dagger_{\textnormal{R},\mu}$ and $c_{\textnormal{L},\nu}$ reduces to the conventional relation, $\commutator*{c^\dagger_{\textnormal{R},\nu}}{c_{\textnormal{R},\mu}}_\mp = \ip*{\textnormal{R}_\mu}{\textnormal{R}_\nu}$ is in general difficult to evaluate. This observation implies, that the state $\ket{\textnormal{R}_\nu}$ is created by $c^\dagger_{\textnormal{R},\nu}$ and annihilated by $c_{\textnormal{L},\nu}$ and not by $c_{\textnormal{R},\nu}$. Consequently, the role of the pair $(c^\dagger_i, c_i)$, which creates and destroys particles in state $\ket{e_i}$, is taken by $(c^\dagger_{\textnormal{R},\nu}, c_{\textnormal{L},\nu})$, which correspondingly create and destroy particles in $\ket{\textnormal{R}_\nu}$. In particular the (biorthogonal) number operator, measuring the occupation in $\ket{\textnormal{R}_\nu}$, is given by
\begin{equation}
    \hat{n}_{\textnormal{R},\nu} \ket*{\psi} = c^\dagger_{\textnormal{R},\nu} c_{\textnormal{L},\nu} \ket{\psi} = 
    \left.
    \begin{cases}
        1 & \text{if } \ket*{\textnormal{R}_\nu} \text{ occupied} \\
        0 & \text{else}
    \end{cases}
    \right\} \; \ket*{\psi}.
    \label{eq:appendix_number_operator}
\end{equation}
Inserting \cref{eq:num_ladder_trafo} into \cref{eq:num_ham_general} and using that $\ket{\textnormal{R}_\nu}, \ket{\textnormal{L}_\nu}$ are the biorthonormal eigensystem of $\hat{h}$, we attain a diagonal representation in terms of the biorthogonal number operator
\begin{equation}
    H = \sum_{ij} \sum_{\mu \nu} \ip*{\textnormal{L}_\mu}{e_i}\mel*{e_i}{\hat h}{e_j} \ip*{e_j}{\textnormal{R}_\nu} c^\dagger_{\textnormal{R},\mu} c_{\textnormal{L},\nu}
     = \sum_\nu E_\nu c^\dagger_{\textnormal{R},\nu} c_{\textnormal{L},\nu}.
\end{equation}

\section{Numerical computation of correlation functions}
\label{sec:critical_num_corr}

In this section we describe the numerical evaluation of biorthogonal and Hermitian correlation functions in many-body states. We expand on the exposition given in \cite{Yoshimura2020}. If proper operators are chosen the correlation functions become (many-body) expectation values of observables, which were discussed in Sect.~\ref{sec:obexp}. However, to be efficient we from now on speak of correlation functions only. We consider correlation functions evaluated in a given many-body state (Slater-determinant) $\ket*{\psi}$, e.g., the groundstate of the system in question. The two distinct Green functions are then defined by
\begin{align}
    \label{eq:apdx_GF}
    G(i, j) &= \frac{\mel*{\psi}{c_i^\dagger c_j}{\psi}}{\ip*{\psi}{\psi}} ,\\
    G\^{bo}(i, j) &= \mel*{\psi}{\eta\_r c_i^\dagger c_j}{\psi}.
    \label{eq:apdx_GF_PT}
\end{align}
The action of the metric operator $\eta_{\rm r}$ on a Fock space Slater-determinat $\ket*{\psi}$ must be understood as follows. On every 
%$\eta_r = \sum_\mu \dyad{\textnormal{L}_\mu}{\textnormal{L}_\mu}$ in \cref{eq:apdx_GF_PT}, acts on the 
single-particle wave function $\ket*{\psi_{\rm 1p}} = \sum_\nu c_\nu  \ket*{\textnormal{R}_\nu}$  entering the Slater-determinant it acts as
\begin{equation}
    \eta_r \ket*{\psi_{\rm 1p}} = \sum_\nu c_\nu \eta_r \ket*{\textnormal{R}_\nu}
    = \sum_\nu c_\nu \ket*{\textnormal{L}_\nu}.
\end{equation}
Therefore, \cref{eq:apdx_GF_PT} corresponds to the calculation of the expectation value in the biorthogonal sense, with the left-eigenvector as the dual vector appearing in scalar products, see \cref{sec:obexp}. In the following we assume the single-particle wave functions $\ket{\psi_{\rm 1p}}$ to be normalized as $\mel*{\psi_{\rm 1p}}{\eta\_r}{\psi_{\rm 1p}} = \sum_n \abs{c_n}^2 = 1$, which corresponds to the biorthogonal normalization \cref{eq:biorthogonal}. Accordingly, many-body wave functions are normalized as $\mel*{\psi}{\eta\_r}{\psi} = 1$.

The biorthogonal correlation function \cref{eq:apdx_GF_PT} can be obtained by transforming the ladder operators $c^\dagger_i, c_j$ to the biorthonormal basis, using \cref{eq:num_ladder_trafo}, upon which one can directly employ \cref{eq:appendix_number_operator} to obtain
\begin{align}
    G\^{bo}(i, j) = \sum_{\mu,\nu} \ip*{\textnormal{L}_\mu}{e_i}\ip*{e_j}{\textnormal{R}_\nu}
    \mel*{\psi}{\eta\_r c_{\textnormal{R},\mu}^{\dagger} c_{\textnormal{L},\nu}}{\psi} =
    \sum_{\nu \in \text{occp.}} \ip*{\textnormal{L}_\nu}{e_i}\ip*{e_j}{\textnormal{R}_\nu}.
    \label{eq:apdx_corr_PT}
\end{align}
Here the sum $\sum_{\nu \in \text{occp.}}$ runs over all occupied states, i.e. all $\nu$ with $\hat{n}_{\textnormal{R}, \nu} \ket{\psi} = \ket{\psi}$. We observe that for Hermitian $\hat h$, \cref{eq:apdx_corr_PT} reduces to the well known formula of Hermitian quantum mechanics.

The calculation for the conventional (normalized) Green function \cref{eq:apdx_GF} is more involved. We begin by rewriting the state $\ket*{\psi}$, which consists of $M$ particles occupying states $\{\epsilon_1, \dots, \epsilon_M \}$, in terms of ladder operators in some orthonormal basis $\{ \ket{e_i} \}$, using \cref{eq:num_ladder_trafo}
\begin{align}
    \ket*{\psi} = \prod_{\nu \in \{\epsilon_1, \dots, \epsilon_M\}} c^\dagger_{\textnormal{R},\nu} \ket*{0}
    = \sum_{i_1 \dots i_M} \ip*{e_{i_1}}{{\rm R}_{\epsilon_1}} \ip*{e_{i_2}}{{\rm R}_{\epsilon_2}} \dots \ip*{e_{i_M}}{{\rm R}_{\epsilon_M}} c^\dagger_{i_1} \dots c^\dagger_{i_M} \ket*{0}.
    \label{eq:appendix_psi_decomp}
\end{align}
Here $\ket*{0}$ denotes the vacuum state. To simplify the notation, we define the matrix $R_{i, \nu} = \ip*{e_i}{{\rm R}_\nu}$. For the computation  of the inner product $\ip*{\psi}{\psi}$, we need to evaluate $\mel*{0}{c_{i_M'} \dots c_{i_1'} c^\dagger_{i_1} \dots c^\dagger_{i_M}}{0}$ for arbitrary indices. This can be done directly using Wicks-theorem \cite{Bruus2003,NegeleOrland_QTCM}
\begin{align}
    \mel*{0}{c_{i_M'} \dots c_{i_1'} c^\dagger_{i_1} \dots c^\dagger_{i_M}}{0}
    = \sum_{\pi \in S_M} (\mp 1)^{N(\pi)}
    \delta_{i_1, i'_{\pi(1)}} \dots \delta_{i_M, i'_{\pi(M)}},
    \label{eq:rlm_ladder_wick}
\end{align}
with $S_M$ denoting the permutation group of size $M$ and $N(\pi)$ the number of inversions of the permutation $\pi$.

Using \cref{eq:rlm_ladder_wick}, the norm of the wave function $\ip*{\psi}{\psi}$ becomes
\begin{equation}
    \label{eq:apdx_norm}
    \ip*{\psi}{\psi}
    = \sum_{\pi \in S_M} \text{sgn}(\pi) \prod_{i \in {1 \dots M}} \left[R^\dagger R\right]_{\epsilon_i, \pi(\epsilon_i)}
    = \det_{\mu, \nu \in \text{occp.}}\left( R^\dagger R \right).
\end{equation}
In the second equation we have identified the Leibinz form of the determinant, which has to be evaluated over all occupied states, i.e. the determinant of the matrix $R^\dagger R\vert_{\text{occp.}} \equiv \left[R^\dagger R\right]_{\mu, \nu}\vert_{\mu, \nu \in \text{occp.}}$. With this we have a prescription to calculate the denominator of \cref{eq:apdx_GF}.

The expectation value of the ladder operators, i.e. the numerator of \cref{eq:apdx_GF}, can also be brought into the form of an overlap. For this we write
\begin{align}
    \mel*{\psi}{c_i^\dagger c_j}{\psi} &=
    \sum_{\mu, \nu} (R^\dagger)_{\mu, i} R_{j, \nu} \mel*{\psi}{c^\dagger_{{\rm L},\mu} c_{{\rm L},\nu}}{\psi} \\*
    \label{eq:apdx_corr_1}
    &= \sum_{\mu, \nu \in \text{occp.}} (\mp 1)^{\mu + \nu} (R^\dagger)_{\mu, i} R_{j, \nu} \ip*{\psi \setminus \textnormal{R}_\mu}{\psi \setminus \textnormal{R}_\nu} \\*
    \label{eq:apdx_corr_2}
    &= \sum_{\mu, \nu \in \text{occp.}} (\mp 1)^{\mu + \nu} (R^\dagger)_{\mu, i} R_{j, \nu} \text{minor}_{\mu,\nu}(R^\dagger R\vert_{\text{occp.}}).
\end{align}
The sign factor in \cref{eq:apdx_corr_1}  results from the Wigner string of applying the ladder operators. In \cref{eq:apdx_corr_2} we have rewritten the overlap $\ip*{\psi \setminus \textnormal{R}_\mu}{\psi \setminus \textnormal{R}_\nu}$ using \cref{eq:apdx_norm} and identified the resulting expression, which is a determinant without $\nu$-th column and $\mu$-th row, as the $\text{minor}_{\mu, \nu}$.

Combining \cref{eq:apdx_norm,eq:apdx_corr_2}, we obtain an expression for the conventional (Hermitian) correlation function
\begin{equation}
    G(i, j) =  
    \frac{1}{\det_{\mu, \nu \in \text{occp.}}\left(R^\dagger R\right)}
    \sum_{\mu, \nu} (\mp 1)^{\mu + \nu} (R^\dagger)_{\mu, i} R_{j, \nu} \text{minor}_{\mu,\nu}(R^\dagger R\vert_{\text{occp.}}).
    \label{eq:adx_GF_intermediate}
\end{equation}
While \cref{eq:adx_GF_intermediate} is a valid representation of the correlation function, it is numerically unstable and not efficient for practical implementations, since it involves the calculation of minors and determinants of potentially large matrices. Realizing that $(-1)^{\mu + \nu} \text{minor}_{\mu,\nu}(R^\dagger R\vert_{\text{occp.}})$ is just the cofactor $C_{ij}$ of the matrix $R^\dagger R\vert_{\text{occp.}}$, we can rewrite \cref{eq:adx_GF_intermediate} using Cramers rule
\begin{equation}
    C_{ij} = \det(R^\dagger R\vert_{\text{occp.}}) \left(R^\dagger R\vert_{\text{occp.}} \right)^{-1}_{ji}.
    \label{eq:apdx_cramers_rule}
\end{equation}
The determinant of \cref{eq:apdx_cramers_rule} cancels the norm of the wave-function, leading to our final, numerically stable and efficient result
\begin{equation}
    G(i, j) = \sum_{\mu, \nu \in \text{occp.}} (\pm 1)^{\mu + \nu} R_{j, \nu} (R^\dagger R\vert_{\text{occp.}})^{-1}_{\nu, \mu} (R^\dagger)_{\mu, i}
    \label{eq:apdx_GF_final}
\end{equation}
Replacing $R^\dagger \to L^\dagger$, one can also calculate the $\mathcal{PT}$-symmetric correlation function \cref{eq:apdx_GF_PT} using \cref{eq:apdx_GF_final}, which provides a convenient consistency check for numerical implementations.

\section{Numerical convergence of the \texorpdfstring{$\mathcal{PT}$}{PT}-symmetric correlation function}
\label{sec:critical_LR_corr}

\begin{figure}[t]
    \centering  
    \includegraphics{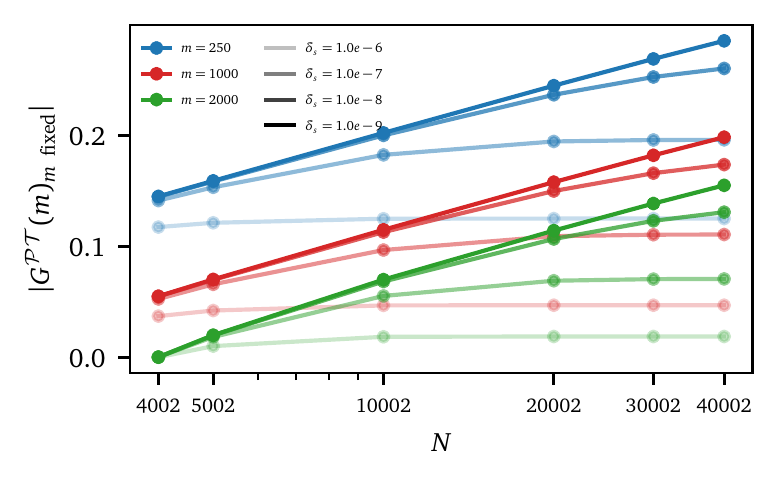}
    \caption{\textbf{Numerical convergence of the $\mathbf{\mathcal{PT}}$-symmetric correlation function} in system size $N$ and shift $\delta_s$. Distinct colors denote different $m$-arguments and the saturation's indicate various $\delta_s$. While the figure only shows $m$ even, the curves for $m$ odd appear completely analogous [also see \cref{fig:critical_td_limit}]. We observe that $N(\delta_s) \sim -\ln(\delta_s)$, implying that the thermodynamic limit is not numerically accessible.}
    \label{fig:apdx_LR_scaling}
\end{figure}

\LG{
We comment on the numerical convergence of the $\mathcal{PT}$-symmetric correlation function, discussed in \cref{subsec:critical_stag}. To assess numerical convergence in system sizes $N$ and the shift away from the quantum critical point $\delta_s$, we consider the correlation function $G^\mathcal{PT}(m)$ [see \cref{eq:correl_qc}] at fixed argument $m$ as a function of $N$ in \cref{fig:apdx_LR_scaling}.}

\LG{The larger two $m$-arguments lie well within the regime, where the Hermitian correlation functions acquire their power law behavior. We observe the system size necessary to achieve convergence depends on the shift $\delta_s$. The functional form is roughly logarithmic $N = N(\delta_s) \sim -\ln(\delta_s)$, indicating that it is impossible to numerically represent the thermodynamic limit at the quantum critical point ($\delta_s \to 0$). We are restricted to analyze the behavior of $G^\mathcal{PT}(m)$, for a given $\delta_s > 0$, for which we can achieve numerical convergence in $N(\delta_s)$. Since a finite $\delta_s$ provides a large distance cutoff, this is expected to give a valid representation of the thermodynamic limit for short and intermediate separations.}

\end{appendices}

\bibliographystyle{iopart-num}
\bibliography{references}

\end{document}